\documentclass[12pt]{iopart}

\usepackage{textcomp}
\usepackage{subcaption}
\usepackage{booktabs}
\usepackage{multirow}
\usepackage{graphicx}
\usepackage{soul}
\usepackage{bm}
\usepackage[utf8]{inputenc}
\usepackage[hidelinks]{hyperref}
\usepackage{cite}
\usepackage{eqnarray}
\usepackage{caption}
\usepackage{amssymb}
\usepackage{cancel}

\usepackage{xurl}
\usepackage{color}

\usepackage[normalem]{ulem}

\usepackage{soul}

\begin{document}

\title[Advances in laser-assisted nuclear decay and nuclear excitation]{Advances in laser-assisted nuclear decay and nuclear excitation}

\author{Q. Xiao$^1$\footnotemark[1], J. H. Cheng$^1$\footnotemark[1], Y. Y. Xu$^1$, Y. T. Zou$^1$, Z. Z. Ren$^2$, A. Ya. Dzyublik$^3$ and T. P. Yu$^1$}

\address{$^1$ College of Science, National University of Defense Technology, Changsha 410073, People's Republic of China}

\address{$^2$ School of Physics Science and Engineering, Tongji University, Shanghai 200092, People's Republic of China}
\address{$^3$ Institute for Nuclear Research, National Academy of Sciences of Ukraine, Kyiv 03680, Ukraine}

\ead{tongpu@nudt.edu.cn}
\vspace{10pt}
\begin{indented}
\item[]
\end{indented}

\footnotetext[1]{These authors contributed equally to this work}

\begin{abstract}

From the synthesis and evolution of the elements to the celestial nuclear processes of stellar explosions and neutron star mergers, nuclear physics is the foundation of our understanding of the universe. After more than a century of progress, the field of nuclear physics remains vibrant. The rapid advancement of laser technology has opened unprecedented avenues in nuclear physics, driven by the interdisciplinary convergence of laser physics, nuclear structure, plasma science, and quantum dynamics. This emerging field enables studies on laser-induced nuclear excitation, laser-assisted nuclear decay, and precision manipulation of nuclear isomers for optical clocks. This review comprehensively examines the research achievements over the past decade regarding the influence of lasers on radioactive charged-particle emissions and nuclear excitation. Regarding theoretical developments, the review details various methods used to evaluate the interactions between lasers and nuclei, including the time-dependent Schrödinger equation for $\alpha$ decay, proton radioactivity, and two-proton radioactivity and Fermi’s golden rule for nuclear excitation as well as the application and advancement of various theoretical models and approximation methods. In experimental research, the review synthesizes significant breakthroughs in laser-induced nuclear excitation experiments, particularly emphasizing the excitation of the $^{229}$Th, $^{83}$Kr, and $^{45}$Sc. These achievements provide essential groundwork for future breakthroughs in both fundamental nuclear science and its broader technological applications.
\end{abstract}

%
%
%
%
%

\maketitle
\pagestyle{plain}
\tableofcontents
\newpage


\section{Introduction}\label{intro}

The nucleus is the core of all physical particles and accounts for 99.9\% of their total mass. In terms of spatial scale, the nucleus ranges from femtoscopic protons and neutrons to macroscopic celestial systems, encompassing the currently recognized limits of the observable universe. With regard to the temporal dimension, the nucleus formed approximately one-millionth of a second after the Big Bang, and it has undergone continuous evolution for an impressive span of 13.8 billion years, corresponding to the history of the universe \cite{ma20232022,ma20242023}. The development of nuclear science began with Becquerel's uranium fluorescence experiment in 1896, which marked the first revelation of the natural radioactivity inherent to nuclei \cite{1370017279823743616}. For more than a century that followed, there has been substantial advancement in our understanding of the nucleus's fundamental properties and pivotal role within the cosmic framework. This burgeoning knowledge has been successfully translated into practical applications in various domains, notably nuclear energy \cite{Tang2021,LIU2023104772,TESTONI2021103822} and nuclear medicine \cite{KLEYNHANS2024778,POLOMSKI2022597}.

Nowadays, nuclear physics has evolved into a dynamic discipline encompassing several broad and interconnected subfields. Its research domains include but are not limited to, quantum chromodynamics \cite{Ball2022,universe7080312}, nuclear structure and reactions \cite{PhysRevLett.131.202501,PhysRevC.99.044908,PhysRevLett.130.212302,Jalili2024,CHENG2024109317,Ye2025}, nuclear astrophysics \cite{Zhou2023,PhysRevLett.130.192501,PhysRevLett.131.112701}, fundamental symmetries \cite{osti_2281639,Nishi2023}, and others \cite{Cai2023,PhysRevLett.130.062501,PhysRevLett.131.202502,PhysRevLett.134.052501}. The challenges in these areas relate to our understanding of the material world and the quantum realm, which feature interdisciplinary connections with other scientific fields and shape the modern discipline of nuclear physics. Moreover, nuclear physics seeks to unravel the nature and composition of matter and its implications for the universe's structure, achieving impressive results \cite{Duer2022,Sarmiento2023,Abdulhamid2024}. Nevertheless, numerous fundamental questions remain unresolved, indicating the ongoing journey of inquiry and discovery in nuclear physics.

In the realm of nuclear structure, understanding the nucleus largely depends on observing its spontaneous behavior, owing to the strong interactions that occur within the nucleus. There is a lack of powerful probes that interfere with the nucleus and provide new information about the nucleus. Similarly, concerning nuclear reactions, the utilization of nuclear energy is currently largely restricted to natural nuclear processes, such as fusion \cite{Maggi_2024,Joffrin_2024}, fission \cite{PhysRev.56.426,HYDE200882}, and decay \cite{Ayodele2019,D4CC03980G}, and there is a need for advanced tools for manipulating and controlling these reactions to utilize nuclear energy more comprehensively. The rapid advancement of laser technology has presented potential solutions to these challenges, thereby evolving into an essential branch of nuclear physics, known as laser nuclear physics \cite{Neugart2002,LEDINGHAM2005633,Hannachi_2007,CAMPBELL2016127,YANG2023104005}.

Laser nuclear physics investigates the interaction between lasers and nuclei, explores the fundamental laws governing these interactions, generates secondary particles and radiation sources in an efficient and compact manner, and explains how laser fields affect nuclear structure, decay, and other properties \cite{Li_2021,PhysRevC.90.054619,Popruzhenko2023,Gales_2016,ZILGES2022103903,Collaboration_2025}. In fact, the development of this field has been an ongoing process of addressing challenges and exploring new frontiers. Since the advent of high-power high-intensity lasers, while bringing new opportunities, it has also faced issues such as attaining sufficient beam intensities for nuclear astrophysics studies, understanding electron screening in extreme plasmas, and developing reliable detection techniques under intense laser irradiation \cite{10.3389/fphy.2024.1503516}. The inherent characteristics of laser—namely, its high intensity, excellent collimation, and exceptional coherence—render it a valuable tool for probing the internal structures of nuclei in nuclear physics experiments. With the ongoing efforts in experiments and the innovation of related theories, a full range of laser nuclear physics has already been explored and many others are being initiated \cite{vonderWense2016,PhysRevLett.121.253002,PhysRevC.100.064610,PhysRevC.105.054001,PhysRevC.109.064615,PhysRevC.102.044332,PhysRevC.102.024312,10.1088/978-0-7503-3199-9ch4,PhysRevLett.133.152503,Dzyublik2010,Cooke:82,PhysRevLett.128.162501,PhysRevC.110.064326,Qi2026,Yu_2025}. Among these studies, the effects of high-intensity lasers on nuclear decay and laser-induced nuclear excitations have attracted significant attention due to their broad application prospects. The application of high-intensity lasers to modulate nuclear decay pathways presents a novel potential strategy for nuclear waste management, while laser-driven nuclear excitations hold promise for achieving extreme precision in timing systems, which could revolutionize fields such as global positioning systems, fundamental physics research, and high-speed communication networks. It has been well shown that manipulation of nuclei using lasers holds great potential for practical applications, including nuclear lasers \cite{PhysRevLett.106.162501} and nuclear optical clocks \cite{Masuda2019,Seiferle2019,vonderWense2020}, etc.

This paper reviews the recent advances in the impact of high-intensity lasers on nuclear decay, especially the charged particle emission like $\alpha$ decay, proton radioactivity, and two-proton radioactivity, and laser's role in the excitation of nuclei over the past decade and provides the theoretical basis and practical foundation for the subsequent manipulation of nuclei using lasers. In the following, we commence by introducing the recent development of high-power high-intensity lasers. Then we provide a concise overview of the theoretical foundations underlying laser-nucleus interactions. Subsequently, we explore the effects of lasers on nuclear structures, classifying these influences into four primary categories: the impact of lasers on $\alpha$ decay, the effects on proton and two-proton radioactivity, and the novel phenomena of laser-induced nuclear excitation. We also summarize the latest experimental research progress and give a brief summary and outlook.

\section{High-power high-intensity laser}

Over the past decade, various models and methods \cite{Casta2011,PhysRevLett.128.043001,PhysRevC.110.014330,PhysRevLett.133.032501} have been proposed to assess the impact of high-intensity lasers on nuclear decay half-lives and excitation rates. These studies have been significantly driven by the rapid advancements in laser technology. The core physics of high-intensity lasers lies in their ability to generate extreme electromagnetic fields. Here, the laser intensity $I$ relates to the electric field $E$ through
\begin{equation}
  I = \frac{1}{2}c\epsilon_0E^2,
\end{equation}
where $c$ is the speed of light in vacuum and $\epsilon_0$ is the vacuum permittivity. For relativistic-intensity lasers ($I > 10^{18}\ \mathrm{W/cm}^2$) with the laser wavelength $\lambda_0 = 1\ \mu\mathrm{m}$, the electric field reaches
\begin{equation}
  E_{\mathrm{rel}} \approx \frac{2\pi m_e c^2}{e\lambda} \sim 10^{12}\ \mathrm{V/m},
\end{equation}
with $m_e$ the electron mass and $e$ the elementary charge. The corresponding magnetic field $B = E/c$ reaches kilotesla levels. To characterize the laser-matter interactions in this regime, a normalized laser field parameter is introduced
\begin{equation}
  a_0 = \frac{eE}{m_e c\omega} = 0.85 \times 10^{-9} \lambda_0\ [\mu\mathrm{m}] \sqrt{I\ [\mathrm{W/cm}^2]}.
\end{equation}
Here, $\omega$ is the laser angular frequency and $a_0 \geq 1$ indicates entering the relativistic regime, where electrons oscillate at the approximately speed of light in several laser cycles. In this chapter, we first review the evolution of high-power high-intensity lasers in the past decades.

In 1917, Einstein theoretically identified stimulated emission, laying the foundational principle for coherent light amplification. Nearly four decades later, in 1958, Townes and Schawlow formalized the conceptual framework for lasers (Light Amplification by Stimulated Emission of Radiation) \cite{PhysRev.112.1940}, thereby bridging Einstein's 1917 theoretical insight with practical technological design. This breakthrough was followed by the invention of the first ruby laser by Maiman \cite{MAIMAN1960} and laser technology have experienced significant progress since then. Initially, it was primarily applied in scientific research fields like physics, chemistry, and biology \cite{PhysRevLett.116.061102, doi10.1126/science.269.5221.198}. Nonetheless, the development of laser technology is multifaceted. In the field of laser materials, the exploration of novel gain media, such as crystals, glasses, and semiconductors, has been ongoing. These materials, with their diverse characteristics, have enabled lasers to fulfill diverse requirements. For instance, Nd:YAG crystal is renowned for its high gain and stability. Simultaneously, advancements in laser resonator design, including different cavity configurations like plano-concave and stable cavities, have substantially improved laser performance. The development of mode-locking \cite{10.1063/1.1754253} and Q-switching \cite{McClung62} technologies has led to a significant leap in laser peak intensity, making it reach \(10^{15}\,\mathrm{W/cm^2}\) and above.

The introduction of Chirped Pulse Amplification (CPA) technology in 1985 \cite{STRICKLAND1985447} marked a significant advancement in the field of laser physics, ultimately earning its inventors the Nobel Prize for its pioneering impact in 2018. CPA operates through a systematic process whereby a short laser pulse is temporally stretched to minimize peak power, subsequently amplified, and then recompressed. This innovative methodology has substantially enhanced the peak power of laser pulses, thereby unlocking new potential applications across diverse domains, including material processing and high-energy physics research. Furthermore, CPA has effectively mitigated nonlinear optical effects and reduced the risk of optical damage, contributing to improved system stability. Notwithstanding its transformative impact, early implementations of CPA technology exhibited certain limitations. For example, its dependence on multiple optical components introduced an inherent complexity and elevated costs associated with the system, posing great challenges in the laser maintenance. Additionally, substantial energy losses encountered during stretching, amplification, and compression resulted in relatively low energy conversion efficiencies. Moreover, the management of high-order spectral phases and the uniformity of spectral modulation presented challenges that adversely affected the stability and optimal compression of the generated laser pulse. In 1986, a notable proposal by Moulton \textit{et al.} to utilize $\mathrm{Ti:Al}_2\mathrm{O}_3$ as a gain medium \cite{Moulton86} represented a critical milestone in laser technology. The $\mathrm{Ti:Al}_2\mathrm{O}_3$ laser, distinguished by its straightforward energy level structure and broad tunability, provided several advantages in the domain of pulse generation. Subsequently, the development of Optical Parametric Chirped Pulse Amplification (OPCPA) further booted the achievable laser energy levels \cite{Audrius2023}. OPCPA is based on the principles of optical parametric amplification, involving the interaction of signal and pump photons within a nonlinear crystal. Since the 1990s, numerous high-power laser systems have emerged \cite{RevModPhys.91.030501}. The current landscape features nearly a hundred 100-TW-level (TW, 1 TW = $10^{12}$ W) systems and approximately 20 PW-level (PW, 1 PW = $10^{15}$ W) systems in operation or under construction, with some teams aiming to achieve the 100-PW and even higher level. This advance in laser technology has been driven by the need for higher power and shorter pulses in various scientific and technological applications, and continues to expand the frontiers of research and innovation.

The eight orders of magnitude increase in laser intensity over the past three decades (from $10^{15}\ \mathrm{W/cm^2}$ to $10^{23}\ \mathrm{W/cm^2}$) has established a new paradigm of laser-matter interactions. When the laser intensity is relatively low ($I < 10^{12}\ \mathrm{W/cm^2}$), the interaction between light and matter mainly involves linear optical processes such as reflection and refraction. In the regime of medium laser intensity ($10^{12}\ \mathrm{W/cm^2} \leq I < 10^{16}\ \mathrm{W/cm^2}$), nonlinear optical effects begin to emerge. Common nonlinear optical phenomena include second harmonic generation, third harmonic generation, and four-wave mixing \cite{DonaldUmstadter_2003, Rouyer93, DANSON1993392}. Nowadays, the laser intensity has reached the regime of high intensity and high power ($I \geq 10^{16}\ \mathrm{W/cm^2}$) \cite{RevModPhys.91.030501,LiChen2025}. Under the action of such lasers, highly nonlinear and extreme physical phenomena may occur such as relativistic charged-particle acceleration, high-energy photon emission and strong-field quantum electrodynamics (QED) \cite{PhysRevLett.43.267, PhysRevSTAB.10.061301,Thaury_2010,Yu2024}. 

\begin{figure}[htp]
	\centering
	\captionsetup{justification=justified, singlelinecheck=false}
	\includegraphics[width=14cm]{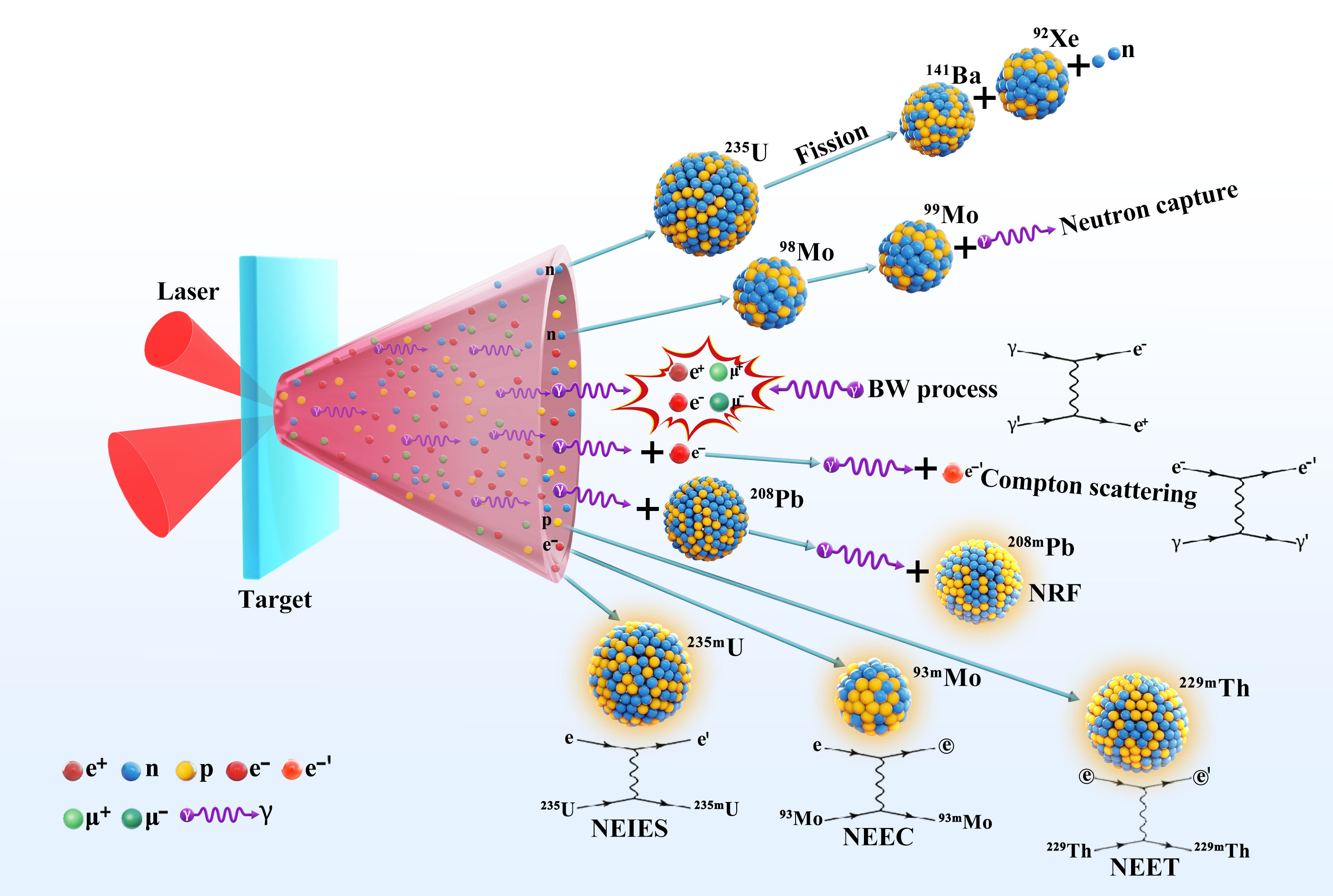}
	\caption{ Schematic diagram of secondary particles and radiation generated via laser irradiation of target material and associated nuclear reactions and excitation.}
	\label{fig2.1}
\end{figure}

The European Extreme Light Infrastructure (ELI) project is a highly anticipated European scientific research initiative aimed at developing world-class laser research facilities to investigate high-intensity light–matter interactions and their potential applications. The project plans to establish three research institutions, with the Extreme Light Infrastructure–Nuclear Physics (ELI-NP) facility being one of the most advanced globally in photonuclear physics. ELI-NP houses two state-of-the-art light sources: a 10-petawatt high-power laser system (HPLS) and a planned high-brilliance inverse-Compton $\gamma$-beam system. Full-energy 10.2 PW pulses have been demonstrated after compression and transport through the laser beam line to a beam dump \cite{RadierChalus2022}. The HPLS is designed to enable laser–matter interactions under extreme conditions, facilitating studies in High Energy Density Physics (HEDP), including QED phenomena and nuclear physics involving high-density matter. In contrast, the planned gamma-beam system is intended to enable the detailed study of various photonuclear physics topics, such as nuclear resonance fluorescence and photonuclear reactions, with sensitivities and resolutions significantly exceeding existing facilities worldwide \cite{Gales_2018,10.1063/1.5093535,doria2020,Whitebook,ELINPGammaSystem2026,ELINPGammaKickoff2026}.

Among the major large-scale laser facilities worldwide, in addition to ELI-NP, the Omega Laser Facility in the United States and the Apollon Laser Facility in France have also proposed plans to conduct laser nuclear physics research \cite{omega2012,apollon}. These plans include investigating nuclear-atomic-plasma interactions in laser-produced plasmas, such as the electron-induced excitation of nuclei in plasma and the reduction of nuclear lifetimes in hot plasma. Furthermore, the National Ignition Facility (NIF) in the United States has successfully achieved a historic series of fusion ignition \cite{JeffTollefson2024, NIF,kritcher2025progress}. The journey began on December 5, 2022, when NIF made scientific history by achieving ``target gain'': a fusion experiment produced 3.15 MJ of energy—54\% more than the 2.05 MJ laser energy delivered to the target—marking the first robustly ignited fusion plasma. Subsequent to the historic milestone, NIF conducted nine additional successful ignition experiments. On April 7, 2025, its best trial used 2.08 MJ of laser energy to generate approximately 8.6 MJ, exceeding quadruple the input and marking improved fusion energy efficiency. These experiments created target temperatures of order $10^8$ K and pressures over 100 billion times Earth’s atmosphere, enabling hydrogen atoms in the target to fuse via controlled thermonuclear reaction. On the other hand, the Shanghai Superintense Ultrafast Laser Facility (SULF) is now an operational 10-PW-class ultra-intense laser platform, with commissioning and user experiments already performed on its PW-to-multi-PW beamlines. Further optimization of focusing and pulse contrast is expected to make on-target intensities in the $10^{22}$--$10^{23}\ \mathrm{W/cm^2}$ range increasingly accessible \cite{Li18,xue2025research,LiChen2025}. In recent years, significant progress has also been made in the research of laser-driven neutron sources at existing large laser facilities \cite{PhysRevLett.131.025101,Horny2022,PhysRevLett.110.044802,Gunther2022,PhysRevLett.113.184801,pnas.2518397122,PhysRevX.13.011011,Vallieres2025,KIM2026103148,feng2026efficientgenerationneutronsbased}, which not only provides a novel tool for neutron-related research but also holds great promise for applications in fields such as nuclear energy and neutron imaging \cite{Krasilnikov_2005,WORACEK2018141,Canova_2025,Jiang_2020,Deng_2022}. A detailed schematic of the nuclear reactions induced by secondary particles and radiation interacting with the target nucleus is presented in figure \ref{fig2.1}. The interaction between laser and target first generates short-duration secondary particles that initiate various nuclear reactions with nuclei, including nuclear fission, neutron capture, the Breit–Wheeler (BW) process, Compton scattering, nuclear resonance fluorescence (NRF), nuclear excitation via inelastic electron scattering (NEIES) and electron transition (NEET), as well as nuclear excitation through electron capture (NEEC). High-power lasers, such as the ELI-NP, Omega Laser Facility, Apollon Laser Facility, NIF, and SULF, offer important possibilities for the exploration of the nucleus. These lasers can be utilized independently or in conjunction with matter and accelerators, thereby providing unique opportunities for a more complete understanding of nuclear structure and properties.

\section{General formalism of laser-assisted nuclear decay and excitation}

Nuclear decay processes play a fundamental and indispensable role in the realm of nuclear physics. Among them, the charged particle emission, e.g., $\alpha$ decay, proton radioactivity, and two-proton radioactivity stand as three of the significant decay modes. The controlled manipulation of these decay mechanisms holds promising applications in the management of nuclear waste materials, which is vital for ensuring environmental safety and preventing potential hazards to human health and ecological systems. In $\alpha$ decay, the emission of an $\alpha$ particle, composed of two protons and two neutrons, is a common mechanism through which unstable nuclei seek to attain greater stability. Proton radioactivity, on the other hand, occurs when a proton-rich nucleus ejects a single proton, providing crucial insights into the structure and behavior of such nuclei \cite{Wang:2025bbn}. Two-proton radioactivity, although relatively rare, offers a unique insight into the extreme conditions of nuclear instability. From a theoretical perspective, $\alpha$ decay, proton radioactivity, and two-proton radioactivity are electromagnetic decay processes dominated by strong interactions, for which theoretical modeling (e.g., the time-dependent Schrödinger equation) has been well-established. By contrast, $\beta$ radioactivity fundamentally involves weak interaction processes with quark-level transformations (e.g., d→u), which differ inherently from the electromagnetic interactions of laser fields. Therefore, this review focuses on the charged particle emission like $\alpha$ decay, proton radioactivity, and two-proton radioactivity. For details on $\beta$ decay can be found in the review \cite{RevModPhys.15.209}.

The decay half-life serves as a crucial parameter in describing the decay rate of a radioactive nucleus. It is precisely defined as the time interval during which half of the initial number of radioactive nuclei decay. For instance, $^{238}$U, a well-known isotope, has a remarkably long half-life of approximately 4.5 billion years \cite{Kondev_2021}, emphasizing its relatively stable nature. In contrast, $^{212}$Po has an extremely short half-life of only about 0.3 microseconds \cite{Kondev_2021}, exemplifying the highly unstable characteristics of certain nuclei. These examples vividly illustrate the vast range of half-life values and their profound implications for understanding the diverse properties of nuclei. Similarly, nuclear excitation processes are characterized by the mean lifetime of excited nuclear states, which dictates the temporal evolution of energy release through $\gamma$-ray emission or internal conversion. These excitation lifetimes provide insights into nuclear structure and support technological developments such as the nuclear clock \cite{vonderWense2016}.

Given the significance of these processes, it is critical to investigate how external factors, such as high-intensity lasers, may impact these natural processes. To evaluate the impact of high-intensity lasers on nuclear decay and excitation, the models and methods that have emerged in the past decade can be classified in different ways, based on how lasers interact with nuclei and whether lasers are treated as fields or particles. In this Chapter, we briefly outline the general formalism of laser-assisted charged particle emission and low-energy nuclear excitation.

\begin{figure*}[htb]\centering
\captionsetup{justification=justified, singlelinecheck=false}
	\includegraphics[width=16cm]{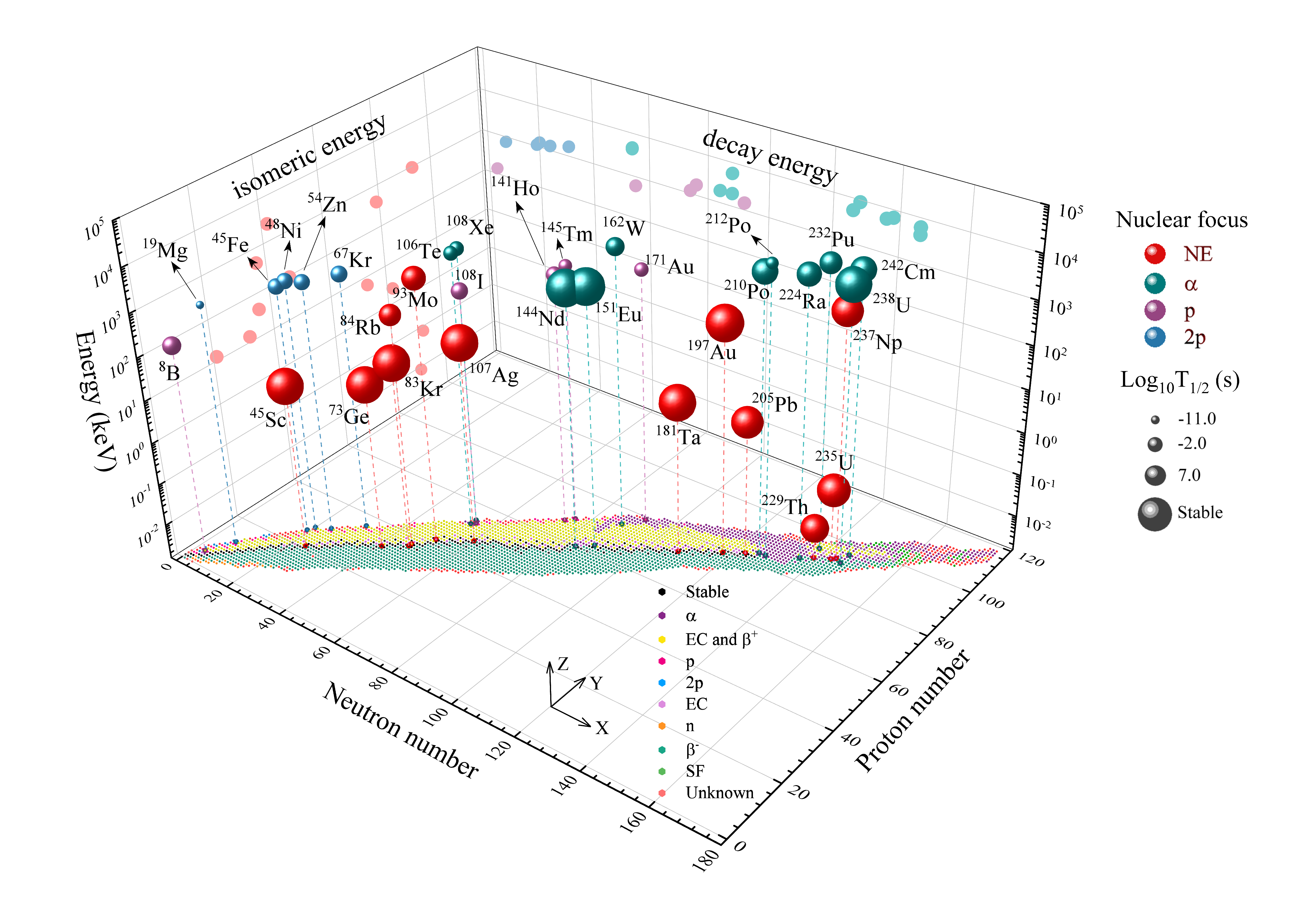}
	\caption{A 3D visualization of nuclear excitation and decay characteristics, with the X-axis representing neutron number, Y-axis proton number, and Z-axis isomeric or decay energy in keV. The colored spheres denote different decay processes: nuclear excitation (red), $\alpha$ decay (purple), proton radioactivity (green), and two-proton radioactivity (blue), while sphere size reflects the relative nuclear stability. Projections onto the Y-Z and X-Z planes reveal energy trends within nuclear decay and excitation, and the base plane displays a comprehensive nuclide chart categorized into nine decay modes.}
	\label{fig2.2}
\end{figure*}

The radioactivity of nuclei has long been a major interest in nuclear physics, which is undergoing a renaissance experimentally and theoretically with the rapid development of laser technology towards ultra-high intensity, ultra-short duration, and ultra-fast process. Figure \ref{fig2.2} presents a detailed visualization of the extensively studied nuclei in the dimensions of neutron number, proton number and energy, focusing on nuclear excitation, $\alpha$ decay, proton radioactivity, and two-proton radioactivity. The size of each sphere reflects the relative nuclear stability, with larger spheres indicating more stable nuclei. The base plane displays a comprehensive nuclide chart that categorizes all known nuclei into nine decay types based on their dominant decay mechanism: $\alpha$ decay, electron capture (EC), $\beta^{+}$ decay, proton radioactivity, two-proton radioactivity, neutron emission, $\beta^{-}$ decay, spontaneous fission, and unknown or unclassified types. These categories provide a clear framework for understanding the diverse decay behaviors across nuclei. Projections onto the Y-Z and X-Z planes further reveal the energy trends within the nuclear decay and nuclear excitation. The projection of spheres onto the base plane also reveals that the nuclei for nuclear excitation study are relatively stable. Moreover, two unique nuclei, $^{229}$Th and $^{235}$U, exhibit excitation energies of only a few eV and tens of eV, respectively, highlighting their significance in current nuclear excitation.

\begin{table}
\captionsetup{justification=justified, singlelinecheck=false}
\caption{\label{jlab1}Distribution and information of nuclides in laser-assisted nuclear decay and excitation research*.}
\footnotesize
\setlength{\tabcolsep}{8.5pt} 
\renewcommand{\arraystretch}{1.0} 

\begin{tabular}{@{}llllll@{}} 
\br
Decay mode & Nuclide & $T_{1/2}$ & $Q/E_{is}$(keV) & $J^{\pi}$ & References \\
\mr
\multirow{12}{3cm}{$\alpha$ decay}
    & $^{106}$Te	& 78 $\mu$s	& 4290	& $0^{+}$	& \cite{CASTANEDACORTES2013401,Miicu_2013,Miicu_2016,PhysRevC.111.034602,PhysRevLett.124.212505}	\\
    & $^{108}$Xe	& 72 $\mu$s	& 4570	& $0^{+}$	& \cite{Xiao2024,CHENG2024138322}	\\
    & $^{144}$Nd	& 2.29 Py	& 1901.3	& $0^{+}$	& \cite{Xiao2024,CHENG2024138322,PhysRevLett.124.212505,PhysRevC.99.044610}	\\
    & $^{151}$Eu	& 4.6 Ey	& 1964	& ${5/2}^{+}$	& \cite{cjh2025}	\\
    & $^{162}$W	& 1.19 s	& 5678.3	& $0^{+}$	& \cite{PhysRevLett.124.212505}	\\
    & $^{210}$Po	& 138.376 d	& 5407.53	& $0^{+}$	& \cite{Kis_2018}	\\
    & $^{212}$Po	& 294.4 ns	& 8954.19	& $0^{+}$	& \cite{PhysRevLett.119.202501,BAI201823,Bai_2018,PhysRevC.99.044610,PhysRevC.101.044304,PhysRevC.111.034602,PhysRevLett.124.212505}	\\
    & $^{224}$Ra	& 3.6316 d	& 5788.92	& $0^{+}$	& \cite{PhysRevC.99.044610}	\\
    & $^{232}$Pu	& 33.7 m	& 6716	& $0^{+}$	& \cite{PhysRevLett.119.202501,BAI201823,Bai_2018,PhysRevC.99.044610,PhysRevLett.124.212505}	\\
    & $^{238}$Pu	& 88.7 y	& 5593.27	& $0^{+}$	& \cite{PhysRevC.111.034602,PhysRevLett.124.212505}	\\
    & $^{238}$U	& 4.463 Gy	& 4269.9	& $0^{+}$	& \cite{PhysRevLett.124.212505}	\\
    & $^{242}$Cm	& 162.8 d	& 6215.63	& $0^{+}$	& \cite{BAI201823}	\\
\mr
\multirow{5}{3cm}{Proton radioactivity}	
    & $^{8}$B	& 771.9 ms	& 136.4	& $2^{+}$	& \cite{PhysRevC.106.064610}	\\
    & $^{108}$I	& 26.4 ms	& 597	& $1^{+}$	& \cite{PhysRevC.105.024312}	\\
    & $^{141}$Ho	& 4.1 ms	& 1177	& ${7/2}^{-}$	& \cite{Misicu2019}	\\
    & $^{145}$Tm	& 3.17 $\mu$s	& 1736	& ${11/2}^{-}$	& \cite{Misicu2019}	\\
    & $^{171}$Au	& 22.3 $\mu$s	& 1448	& ${1/2}^{+}$	& \cite{Misicu2019}	\\
\mr
\multirow{5}{3cm}{Two-proton radioactivity}	
    & $^{19}$Mg	& 5 ps	& 760	& ${1/2}^{-}$	& \cite{Zou_2024}	\\
    & $^{45}$Fe	& 2.5 ms	& 1800	& ${3/2}^{+}$	& \cite{Zou_2024}	\\
    & $^{48}$Ni	& 2.8 ms	& 2390	& $0^{+}$	& \cite{Zou_2024}	\\
    & $^{54}$Zn	& 1.8 ms	& 2280	& $0^{+}$	& \cite{Zou_2024}	\\
    & $^{67}$Kr	& 7.4 ms	& 2890	& ${3/2}^{-}$	& \cite{Zou_2024}	\\
\mr
\multirow{12}{3cm}{Nuclear excitation}	
& $^{45}$Sc    & STABLE    & 12.4    & ${7/2}^{-}$    & \cite{Shvydko2023} \\
& $^{73}$Ge    & STABLE    & 13.2845    & ${9/2}^{+}$    & \cite{10.1063/5.0212163,PhysRevLett.128.212502} \\
& $^{83}$Kr    & 32.41 h    & 41.56    & ${7/2}^{+}$    & \cite{PhysRevLett.128.052501,pnas2413221121} \\
& $^{84}$Rb    & 32.82 d    & 463.59    & $2^{-}$    & \cite{PhysRevC.109.054327,PhysRevC.96.024604} \\
& $^{93}$Mo    & 4.0 Ky    & 2424.95    & ${5/2}^{+}$    & \cite{PhysRevLett.122.212501,chiara2018isomer,PhysRevLett.128.242502,Guo2021} \\
& $^{107}$Ag    & STABLE    & 93.125    & ${1/2}^{-}$    & \cite{10.1063/5.0212163} \\
& $^{181}$Ta    & STABLE    & 6.237    & ${7/2}^{+}$    & \cite{10.3389/fphy.2023.1203401,PhysRevLett.133.132501} \\
& $^{197}$Au    & STABLE    & 409.15    & ${3/2}^{+}$    & \cite{PhysRevC.88.054616} \\
& $^{205}$Pb    & 17.0 My    & 2.329    & ${5/2}^{-}$    & \cite{PhysRevLett.133.032501} \\
& $^{229}$Th    & 7.916 Ky    & 0.00836    & ${5/2}^{+}$    & \cite{PhysRevLett.134.113801,vonderWense2016,Masuda2019,PhysRevLett.128.043001,PhysRevC.110.014330,PhysRevLett.109.262502,PhysRevA.81.042516,PhysRevLett.105.182501,Bilous_2018,PhysRevC.100.044306,PhysRevC.102.024604,PhysRevC.106.064608,PhysRevLett.124.192502,PhysRevLett.125.032501,PhysRevLett.130.112501,PhysRevLett.124.242501,PhysRevC.106.044604,PhysRevC.110.064621,10.3389/fphy.2023.1166566,PhysRevLett.118.212501,PhysRevLett.122.162502,PhysRevC.103.014313,PhysRevC.105.064313,PhysRevC.100.024315,Seiferle2019,Kraemer2023,PhysRevLett.132.182501,PhysRevLett.133.013201,Zhang2024,Zhang20242,Yamaguchi2024,PhysRevC.76.054313,KARPESHIN1999579,PhysRevC.95.034310,PhysRevC.110.064326,PhysRevResearch.7.L022036,PhysRevResearch.7.013052,PhysRevA.111.L041103,PhysRevC.110.054307,PhysRevC.111.054316,8jhh-ktfy,Lan2025Photonuclear} \\
& $^{235}$U    & 704 My    & 0.0767    & ${7/2}^{-}$    & \cite{PhysRevC.106.064604,PhysRevC.59.2462,PhysRevC.93.034610,PhysRevC.110.L051601} \\
& $^{237}$Np    & 2.144 My    & 945.2    & ${5/2}^{+}$    & \cite{Izosimov01092008} \\    

\br
\end{tabular}\\
$\ast$The nuclear half-life, excitation energy, and spin-parity are extracted from reference  \cite{Kondev_2021} (except that the excitation energy of the $^{229}$Th is taken from reference \cite{PhysRevLett.132.182501}), and the decay energy are extracted from reference \cite{Wang_2021}.
 
\end{table}
\normalsize

Table \ref{jlab1} compiles key nuclear parameters for the isotopes visualized in Figure \ref{fig2.2}, including decay mode, half-life, and decay energy $Q$ (or isomeric energy \(E_{is}\) for nuclear excitation). Additionally, the table also shows the spin-parity (\(J^{\pi}\)) of the corresponding nuclei, a parameter that plays a critical role in determining the decay or excitation channels of nuclei. This table includes representative isotopes from each decay category, such as \(^{229}\)Th (nuclear excitation, \(E_{is} = 8.36\) eV), \(^{212}\)Po ($\alpha$ decay, \(T_{1/2} = 294.4\) ns), \(^{8}\)B (proton radioactivity, \(T_{1/2} = 771.9\) ms), and \(^{45}\)Fe (two-proton radioactivity, \(T_{1/2} = 2.5\) ms). This compilation facilitates quantitative correlation between the visualized nuclear properties in Figure \ref{fig2.2} and their corresponding physical parameters, enabling ones to link graphical representations with experimental and theoretical datasets discussed throughout the review.

It is interesting to see that in figure \ref{fig2.2}, these decay modes in heavy, superheavy, and extremely neutron-deficient nuclei are mostly distributed at the periphery of the nuclide chart, which contains rich nuclear structure information crucial for synthesizing new nuclides. Compared with spontaneous fission and proton radioactivity, $\alpha$ decay is more widely distributed and has a larger half-life span, and nuclei with a low neutron-to-proton ratio undergo $\alpha$ decay to increase the ratio, making its structure more stable. The earliest theoretical research on $\alpha$ decay was independently proposed by Condon, Gurney, and Gamow by utilizing the quantum tunneling effect \cite{Gamow1928,Gurney1928}. This theoretical breakthrough offered a new perspective on explaining the mechanism of nuclear decay and contributed to the early development of quantum mechanics. In the ensuing a hundred years, the theory of $\alpha$ decay has been continuously developed, giving rise to many debates and disputes, while most of these theories are almost forgotten. For a review of this, see reference \cite{JACKSON1977151}.

\subsection{Time-dependent Schrödinger equation}

\begin{figure*}[htb]\centering
\captionsetup{justification=justified, singlelinecheck=false}
	\includegraphics[width=16cm]{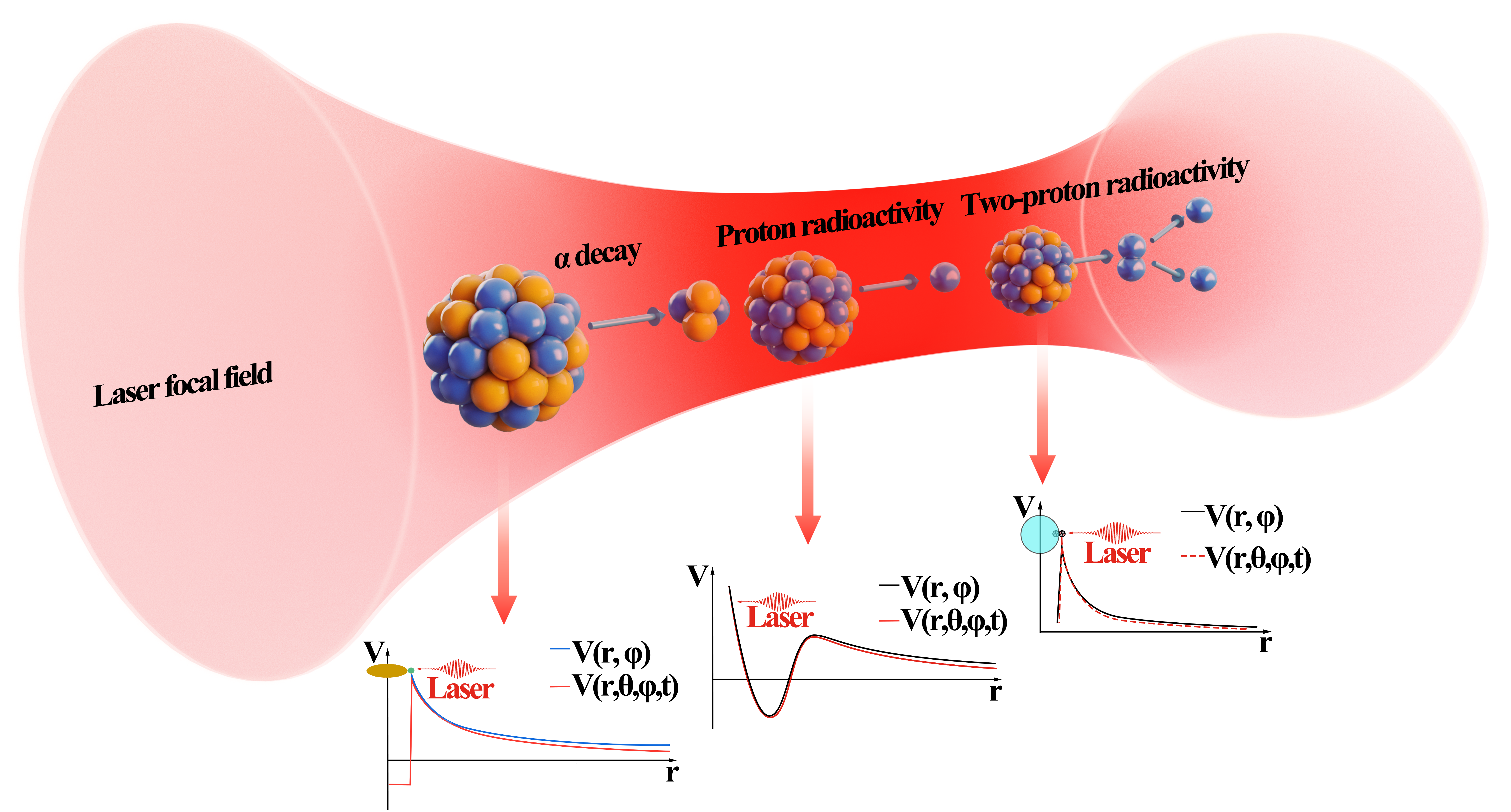}
	\caption{Schematic of the influence of lasers on $\alpha$ decay, proton radioactivity, and two-proton radioactivity. The red arrows highlight the schematics from previous studies \cite{Xiao2024,PhysRevC.105.024312,Zou_2024}, illustrating the influence of high-intensity laser focal fields on the emitted particle and the daughter nucleus interaction potential for different nuclear decay modes.}
	\label{fig2.2.1}
\end{figure*}

In contrast to the continuous updating of $\alpha$ decay theory mentioned in the previous section, the current understanding of the influence of lasers on the emissions of radioactive charged particles—$\alpha$ decay, proton radioactivity, and two-proton radioactivity, as illustrated in figure \ref{fig2.2.1}—is predominantly based on a shared theoretical framework: the time-dependent Schrödinger equation. This equation fundamentally describes the relative motion of the emitted particle and the daughter nucleus within the context of an electromagnetic field

\begin{equation}
  i\hbar\frac{\partial\Psi(\mathbf{r},t)}{\partial t}=\left[\frac{1}{2\mu}\left(\mathbf{P}-\frac{Z_{\mathrm{eff}}}{c}\mathbf{A}(t)\right)^2+V(\mathbf{r})\right]\Psi(\mathbf{r},t),
  \label{eq1.1}
\end{equation}
where $\hbar$ is the reduced Planck constant, $\mu$ is the reduced mass of the emitted particle and the daughter nucleus, $Z_{\mathrm{eff}}$ is the effective charge \cite{Miicu_2013}, $\mathbf{A}(t)$ is the time-dependent radiation field and $V(\mathbf{r})$ represents the interaction potential between the emitted particle and the daughter nucleus. $\Psi(\mathbf{r},t)$ and $\mathbf{P}$ are the total wave function and total momentum of the system, respectively. While the interaction Hamiltonian can, in principle, be written in terms of the nuclear current density interacting with the electromagnetic vector potential, the present work adopts an alternative approach based on the Henneberger transformation \cite{PhysRevLett.21.838}. This method leads to an effective potential that consistently incorporates the influence of the oscillating laser field by averaging the nuclear potential over a classical, laser-driven displacement of the emitted particle. This treatment maintains gauge consistency at the nonrelativistic level within a quantum mechanical framework. However, it does not include gauge invariance as formulated in full quantum field theory at the level of one-loop corrections. Addressing this remains a valuable open direction for future studies.

This equation underlies the fundamental principle of nucleus-laser interaction under intense laser irradiation, while the solution method is crucial for quantifying laser effects on nuclear decay. Depending on whether the laser field $\mathbf{A}(t)$ described in equation (\ref{eq1.1}) varies with time, the solving process can be divided into two different approaches, the results of which are widely debated. In cases where the time variation of the laser field during the particle emission is neglected, a quasistatic process is employed to calculate the change in nuclear decay half-life, where the most widely used quasistatic process is the Wenzel-Kramers-Brillouin (WKB) approximation for calculating the penetration probability. Conversely, when considering the time variation of the laser field, the Kramers-Henneberger (KH) transformation, commonly used in atomic physics, is introduced for calculating decay processes in nuclear physics. The discrepancy between results from these two approaches can exceed an order of magnitude under current laser intensities \cite{PhysRevLett.119.202501,PhysRevLett.124.212505}. These two approaches reflect the different theoretical frameworks developed to address the temporal mismatch between laser pulse durations and nuclear decay timescales. The discrepancies in calculations show the remaining challenge of modeling laser-nucleus interactions across multiple timescales, and the validity of the two theories needs to be verified experimentally to end this debate.

While the interaction Hamiltonian in equation (\ref{eq1.1}) provides the theoretical foundation for understanding laser-induced nuclear processes, practical investigations in this field have historically been constrained by technological limitations. In particular, the electric field strengths produced by early laser systems were far below the threshold required to influence nuclear decay or excitation, due to the extremely short timescales and strong forces inherent to nuclear dynamics. As a result, experimental and theoretical studies of laser–nucleus interactions remained scarce for many years. This situation has significantly been improved with the breakthroughs in laser technology over the last two decades. This will provide a crucial foundation for laser-nucleus experiments \cite{RevModPhys.84.1177,10.1063/1.5093535}. The advances in laser technology have not only enabled investigations into the effects of high-intensity lasers on $\alpha$ decay but have also been applied to studying proton radioactivity and two-proton radioactivity, which involve analogous physical mechanisms with $\alpha$ decay. Although there are limited studies in this domain, these studies still hold significant importance for exploring nuclear structures. In the research of $\alpha$ decay and proton radioactivity, the current primary conclusion is that high-intensity lasers affect the nuclei's half-life by changing the penetration probability of the emitted particles. Based on the WKB approximation, the penetration probability $P$ considering the laser field can be expressed as
\begin{equation}
P(\theta,\varphi)=\exp\! \left[- \frac{2}{\hbar} \int_{R_{in}}^{R_{out}} \sqrt{2\mu (V(r,\theta,\varphi,t)-Q)}\, \rm{dr}\right],
\label{eq2}
\end{equation}
where $R_{in}$ and $R_{out}$ are the classical turning points, $Q$ is the energy of the relative motion of the emitted particle and the daughter nucleus (decay energy), and $V(r,\theta,\varphi,t)$ represents the total interaction potential between the emitted particle and the daughter nucleus in the laser field. $\theta$ represents the angle between the laser field direction and the emitted particle, and $\varphi$ is the emitted particle's orientation angle concerning the daughter nucleus's symmetry axis.

The WKB approximation, as a semiclassical method to solve the Schrödinger equation, has been successfully applied to calculate the tunneling probability of the emitted particles, such as $\alpha$ particles, protons, and clusters, penetrating the Coulombic potential barrier around the nucleus \cite{Ren:2004ah,PhysRevC.105.024307,Xu_2022,Xu2023,Zou_2021,Cheng2020,Cheng_2022,Xiao_2023,PhysRevC.97.044322,PhysRevC.74.014304,PhysRevC.74.017304,QI2019214,PhysRevC.78.044310,Ren:2012zza,Poenaru:2006dp,PhysRevC.73.031301,Seif:2023jcv,PhysRevC.105.034311,PhysRevC.92.014602,w6wq-mj9b,czvc-ymfl,Xiao2024NPA}. This is essential for understanding the structure and dynamics inside the nucleus. Importantly, the accuracy of the WKB approximation depends on factors including the barrier shape, the emitted particle energy, and the wave function properties. Beyond the WKB approximation, one can use the density-dependent cluster model and the multi-channel cluster model to describe the $\alpha$ decays of deformed nuclei by solving the quasi-bound Schrödinger equations \cite{Ni:2013tba,Delion:2018rrl,Ni:2011zzb,Wang:2022yaj,Wang:2022axn,Wang:2024hmt}. Besides, the superfluid tunneling model can also describe the $\alpha$ decays of deformed nuclei \cite{Clark:2023xrh}. Consequently, uncertainties in laser-modified decay calculations do not originate solely from the external-field coupling. They also reflect the underlying description of the unperturbed tunnelling problem, including the choice of nuclear potential, deformation treatment, preformation factor, and boundary conditions. The laser field should therefore be regarded as a perturbation applied to a model-dependent nuclear tunnelling framework, rather than to a uniquely defined decay baseline.

The interpretation of laser-induced modifications becomes even more delicate when one considers the timescale hierarchy of the problem. The tunnelling process in $\alpha$ decay is usually associated with a characteristic timescale much shorter than an optical laser cycle. This observation motivates quasistatic treatments and is central to the debate over the range of validity of KH-based reductions. Accordingly, while equation (\ref{eq1.1}) provides a common formal starting point, the appropriate reduction of this equation depends on the frequency and coherence regime of the external field. For optical and near-infrared laser pulses currently available in high-power laser facilities, the separation between the sub-attosecond tunnelling timescale and the femtosecond optical period naturally supports a quasistatic treatment. By contrast, the large enhancements obtained in KH-based calculations arise from a different high-frequency, time-averaged reduction of the same Hamiltonian, and their applicability to present optical-laser conditions remains experimentally unverified. This regime distinction will be discussed in detail in section \ref{4.3}.

Beyond $\alpha$ decay, the same formal framework has also been extended to proton radioactivity and two-proton radioactivity. In proton radioactivity, the lower emitted-particle mass and different effective charge alter the barrier sensitivity to the laser field. In two-proton radioactivity, the situation is more complex because one must also consider proton--proton correlations and competing emission mechanisms. These extensions show the versatility of the time-dependent Schrödinger equation as a unifying formalism for charged-particle emission, but they should not be taken to imply that all such predictions are equally close to future experimental verification.

\subsection{Fermi's Golden Rule}

Similar to the study of the nuclear decay using the time-dependent Schrödinger equation, one of the most crucial tools for calculating transition rates between nuclear states is Fermi's Golden Rule. This rule provides a way to calculate the transition rate from an initial state to a final state when the system is subject to perturbations, such as the interaction with an external electromagnetic field. In the context of nuclear excitation, Fermi's Golden Rule is used to determine the rate at which a nucleus undergoes a transition between the different energy levels due to the interaction between the nucleus and the laser field \cite{PhysRevLett.129.140402,PhysRevB.104.184302,Bransden2000}. The equation giving the general form of Fermi's Golden Rule is as follows
\begin{equation}
	W_{i \rightarrow f} = \frac{2 \pi}{\hbar} \left|\langle f | V | i \rangle\right|^2\rho(E_f),
	\label{eq111}
\end{equation}
where $W_{i \rightarrow f}$ is the transition rate from the initial state $|i\rangle$ to the final state $|f\rangle$. $V$ is the interaction Hamiltonian describing the coupling between the nuclear system and the external field, and $\langle f| V|i\rangle$ is the matrix element of the interaction Hamiltonian between the initial and final states, which measures the strength of the coupling between the states. The term $\rho\left(E_f\right)$ represents the density of final states at the energy $E_f$, accounting for the number of available final states into which the system can transition.

Fermi's Golden Rule assumes that the perturbation is weak and that the transition probability per unit time remains constant, which is valid when the initial state population is not significantly depleted over time. This assumption holds for many nuclear processes, including laser-induced nuclear excitation, where the electromagnetic interaction with the nucleus is relatively weak compared to other nuclear forces. The transition probability is determined by the matrix element $\langle f |V| i \rangle$, which characterizes the coupling strength between the initial and final nuclear states. The matrix element depends on the type of interaction involved, such as electric dipole ($E1$), magnetic dipole ($M1$), or higher-order multipole transitions. The density of final states $\rho(E_f)$ plays a crucial role in determining the transition rate, as it accounts for the number of available states into which the system can transition. In cases where the final states form a continuous spectrum, such as in inelastic electron scattering, $\rho(E_f)$ must be integrated over the relevant energy range.

Although Fermi’s Golden Rule provides a fundamental framework for calculating nuclear transition rates, its practical applicability varies among different excitation mechanisms. Direct laser excitation is known to have extremely low efficiency due to the narrow nuclear transition widths and the difficulty of achieving resonance with the available laser sources. In contrast, mechanisms such as nuclear excitation by electron capture (NEEC) \cite{GOLDANSKII1976393,CUE198925,PhysRevC.47.323,PhysRevC.59.2462,PhysRevA.73.012715,PhysRevLett.112.082501,chiara2018isomer,PhysRevLett.122.212501,PhysRevLett.128.242502} and inelastic electron scattering (NEIES) \cite{PhysRevLett.124.242501,CPC10.1088/1674-1137/ac9f0a,PhysRevC.106.044604,PhysRevC.106.064604,PhysRev.92.978,PhysRevC.79.014604,PhysRevC.85.044612,PhysRevC.105.014608,Ya.Dzyublik2013} can exhibit higher transition rates because they involve electronic degrees of freedom that enhance the coupling to the nucleus. In NEEC, a free electron is resonantly captured into a bound state of an ion while simultaneously exciting the nucleus, which can lead to enhanced effective excitation probabilities when the electronic capture channel matches the nuclear transition energy. In NEIES, inelastic scattering of electrons transfers energy to the nucleus via Coulomb interaction, with a transition probability that depends strongly on the scattering angle and initial electron energy distribution. The differences between these mechanisms show the importance of interaction strength and density of final states in determining nuclear transition probabilities.

Fermi's Golden Rule remains a fundamental tool in the theoretical understanding of nuclear excitation, particularly in laser-induced processes, and it forms the basis for many calculations in this area of nuclear physics. Its application extends beyond direct laser excitation to other excitation mechanisms, providing a unified framework for understanding nuclear transitions under external perturbations.

\section{$\alpha$ decay in laser fields}\label{third}

$\alpha$ decay is one of the earliest and most debated topics in laser--nucleus interaction studies. Because it is governed by tunneling through a Coulomb-dominated barrier, even a small external-field modification may alter the calculated penetration probability. However, no direct experiment has yet demonstrated a measurable laser-induced change of an $\alpha$-decay half-life under presently accessible optical high-power laser conditions.

The earliest exploration regarding the impact of lasers on nuclear $\alpha$ decay can be traced back to the work conducted by Cortés et al \cite{CASTANEDACORTES2013401}. By developing a formalism based on the imaginary time method and the strong-field approximation, they characterized the tunneling effect of $\alpha$ particles traversing the Coulomb barrier of the nucleus in laser field. The imaginary time method was originally put forward in multiphoton ionization of atoms driven by intense lasers. It has been verified to be directly applicable to the domain of intense monochromatic laser radiation \cite{Popov2005}. In the more microscopic realm of nuclei, the imaginary time method ought to be integrated with the strong-field approximation approach to characterize the tunneling probability of $\alpha$ particles moving along complex trajectories under the influence of the long-range Coulomb potential and the laser electric field \cite{CastaedaCorts2012}.

\begin{figure*}[htb]\centering
\captionsetup{justification=justified, singlelinecheck=false}
	\includegraphics[width=12cm]{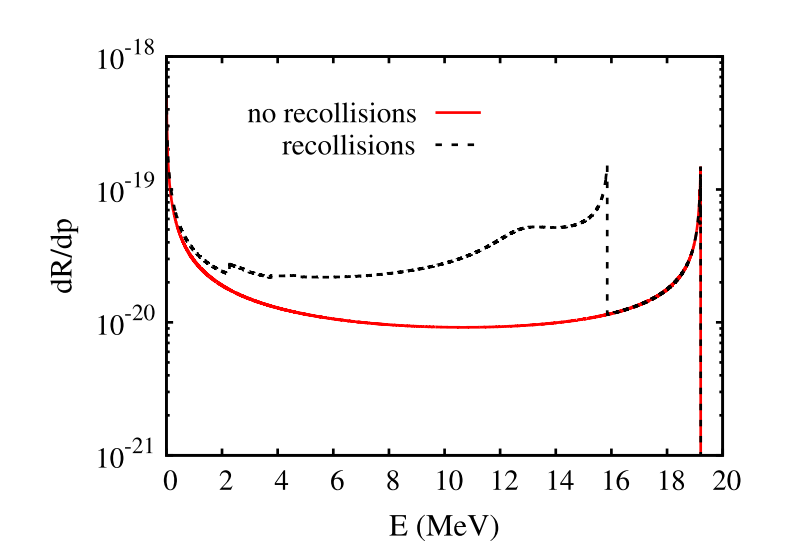}
	\caption{Predicted energy spectrum of the laser-assisted $\alpha$ decay of $^{106}$Te at a laser intensity of $7.9\times10^{22}$ W/cm$^{2}$. The dashed black line represents the recollision contribution to the $\alpha$ decay spectrum, while the solid red line indicates the spectrum without recollisions \cite{CASTANEDACORTES2013401}.}
	\label{fig3}
\end{figure*}

It has been ascertained that at a laser intensity of $10^{22}\ \mathrm{W/cm^{2}}$, the influence of the laser on $\alpha$ decay is negligible \cite{CASTANEDACORTES2013401}, with the relative change of half-live ranging from $10^{-7}$ to $10^{-8}$, exhibiting a linear dependence on the laser intensity. This indicates that even though the relative change in the half-life of $\alpha$ decay is rather small, the influence of the laser on $\alpha$ decay still occurs in the high-intensity laser field. Moreover, the laser field can significantly modify the emitted $\alpha$ spectrum. Figure \ref{fig3} shows the predicted energy spectrum of laser-assisted $\alpha$ decay of $^{106}$Te at a available laser intensity of $7.9\times10^{22}\ \mathrm{W/cm^{2}}$ \cite{Li18,Lureau2020,Yoon21}. Specifically, in the recollision process, the $\alpha$ particle initially tunnels through the nuclear barrier almost undisturbedly. Subsequently, with specific initial phases of the laser field, the $\alpha$ particle is capable of acquiring sufficient energy to recollide with the Coulomb barrier of the daughter nucleus. Figure \ref{fig3} also shows that the recollision contribution to the $\alpha$ decay spectrum commences at low energies, and reaches at a recollision plateau at around 16 MeV, serving as a distinct indication of the occurrence of recollisions. It is also worth noting that similar laser-driven recollisions might occur in the context of laser-assisted proton emission \cite{CASTANEDACORTES2013401}, which could reveal the proton-daughter nucleus interaction dynamics during recollision. The study by Cortés et al shows that the laser-driven nuclear recollisions in $\alpha$ decay are infrequent but detectable at current laser intensities, thereby allowing investigation of nuclear reactions and the relaxation dynamics of the daughter nucleus on a femtosecond scale.

Subsequently, Mi\c{s}icu and Rizea explored the influence of laser pulse characteristics on the penetration probability and half-life of $\alpha$ decay by solving the time-dependent Schrödinger equation \cite{Miicu_2013,Miicu_2016}. With the $\alpha$ decay of $^{106}$Te serving as an illustrative example, they discovered that short pulses consisting of odd half-cycles possess the ability to more effectively accelerate the $\alpha$ decay of the nucleus and increase the decay rate by several orders of magnitude when the electric field strength is as high as $10^{16}$ V/cm. Their research focused on the qualitative understanding of $\alpha$ decay under the influence of laser fields rather than quantitative evaluation. These idealized pulse-shape predictions later became part of the broader controversy over whether large laser-induced $\alpha$-decay enhancements can survive under realistic optical-laser conditions.

\subsection{Kramers-Henneberger transformation and its limitations}

In 1911, Geiger and Nuttall discovered an intriguing phenomenon: For a given isotope, a clear linear correlation exists between the logarithm of the $\alpha$ decay half-life, $\log_{10}T_{1/2}$, and the negative square root of the $\alpha$ decay energy, $Q_{\alpha}^{-1/2}$ \cite{Geiger01101911}. This relationship is subsequently designated as the Geiger-Nuttall law and can be mathematically represented as
\begin{equation}
  \mathrm{log}_{10}T_{1/2}=a\chi+b,
\end{equation}
where $a$ and $b$ are fitting parameters determined by fitting the experimental data. $\chi = 4e^{2}Z_{d}/(\hbar \sqrt{2Q_{\alpha}/\mu})$ represents the Coulomb-Sommerfeld parameter and \(Z_d\) is the charge number of the daughter nucleus. The Geiger-Nuttall law plays an important role in nuclear physics, which establishes a link between the decay half-life and the decay energy of nuclei, thus providing crucial support for understanding the nuclide stability, exploring the characteristics of exotic nuclei, predicting nuclear half-lives, and further improving theoretical decay models \cite{VIOLA1966741,PhysRevC.46.811,PhysRevC.96.034619,PhysRevC.92.064301,PhysRevLett.103.072501,PhysRevC.80.044326}.

In traditional nuclear physics, the Geiger–Nuttall law has been generalized into various forms, including more flexible semi-empirical formulations (e.g., the Royer formula \cite{GRoyer2000,Royer2010,PhysRevC.101.034307}), which introduce explicit mass and charge dependence beyond the original \(Z/\sqrt{Q_\alpha}\) structure ($Z$ is the proton number of parent nuclei). Additionally, there exists the new Geiger–Nuttall law \cite{Ren:2012zza,PhysRevC.78.044310}, which retains the original framework containing the \(\chi\) parameter while incorporating additional terms. These different formulations vary in their mathematical expressions but describe the same \(\alpha\)-decay phenomenon from different angles. In the interface of laser-nuclear physics interactions, the exploration of the Geiger-Nuttall law within strong laser fields, carried out by Delion and Ghinescu, has attracted significant attention \cite{PhysRevLett.119.202501}. They have proposed an approach based on the unitary Henneberger transformation \cite{PhysRevLett.21.838} to characterize the $\alpha$-decay process in a strong electromagnetic laser field, which involves the following theoretical formulation
\begin{equation}
  \Omega=\exp\left[\frac i\hbar\int_{-\infty}^tH_{\mathrm{int}}(\tau)\rm{d\tau}\right],
\end{equation}
where $H_{\mathrm{int}}$ is the perturbation Hamiltonian. By incorporating the classical trajectory of the emitted particle $\mathbf{S}(t)$, equation (\ref{eq1.1}) can be reformulated as 
\begin{equation}
  i\hbar\frac{\partial\Phi(\mathbf{r},t)}{\partial t}=\left[\frac{1}{2\mu}\mathbf{P}^{2}+V\Big(\mathbf{r}-\mathbf{S}(t)\Big)\right]\Phi(\mathbf{r},t),
\end{equation}
where $\Phi=\Omega\Psi $ represents the novel wave function. For continuous high-energy lasers, the impact of the external laser electric field on the nuclear decay can be more effectively analyzed via the Kramer-Hanneberg approach, which undertakes a fundamental approximation by replacing the transformed time-dependent Coulomb potential with its static component, as illustrated below
\begin{equation}
  V_0(\mathbf{r})=\frac1T\int_0^TV\Big(\mathbf{r}-\mathbf{S}(t)\Big)\rm{dt}.
\end{equation}
Taking into account the axial deformation of the nucleus, the Coulomb potential can be expressed as
\begin{equation}
  V_0(r,\theta,\varphi,S_0)=\frac{2Z_de^2}r\xi(r,\theta,\varphi,S_0).
\end{equation}
Here, $\xi(r,\theta,\varphi,S_0)$ denotes the screening function and $S_0$ represents the spatial amplitude of the classical trajectory driven by the laser—that is, the maximum displacement attained by a charged particle oscillating under the influence of the laser field (with the unit of length, typically femtometers in nuclear physics). The penetration probability in equation (\ref{eq2}) can be rewritten as
\begin{equation}
P(\theta,\varphi)=\exp\! \left[- \frac{2}{\hbar} \int_{R_{in}}^{R_{out}} \sqrt{2\mu \Big(V_0(r,\theta,\varphi)-Q_\alpha \Big)}\, \rm{dr}\right].
\end{equation}
Additionally, the dimensionless parameter $\eta= S_0/R_0$ is introduced to more precisely characterize and analyze the influence of the laser electromagnetic field on the penetration probability of the $\alpha$ decay, where $R_0$ represents the nuclear radius. They showed that the angular dependence of the $\alpha$-decay penetration probability in the spherical nucleus $^{212}$Po becomes increasingly anisotropic with the laser parameter $\eta$, changing from a weak effect at $\eta=1/\sqrt{10}$ to an equatorially concentrated enhancement exceeding six orders of magnitude at $\eta=\sqrt{10}$, and found a similar $\eta$ dependence in the deformed nucleus $^{232}$Pu, where a quadrupole deformation of $\beta_2=0.3$ further enhances the penetration probability by about one order of magnitude. Moreover, in the strong laser fields, due to the significant alteration in the $\alpha$ decay half-life and the obvious anisotropy of the emission process, the Geiger-Nuttall law has been modified. Such results stimulated considerable interest because they suggest multi-order enhancements under sufficiently strong fields. Therefore, this implies that the laser affecting nuclei becomes feasible, which has led to extensive discussions in the community of laser nuclear physics \cite{BAI201823,Kis_2018,Bai_2018,Rehman2022,PhysRevC.111.034602}.

However, these predictions must be interpreted with caution. First, the KH transformation is borrowed from atomic strong-field physics and assumes that the emitted particle experiences a coherent periodic displacement that can be meaningfully averaged over the emission process. For $\alpha$ decay under present optical femtosecond lasers, however, the under-barrier dynamics are much faster than the laser cycle, and the required coherence conditions are not established experimentally. For this reason, the large KH-based enhancements should be regarded as model-dependent extrapolations rather than as robust expectations for current experiment conditions.

Subsequently, Bai \textit{et al.} \cite{BAI201823} extended the KH-type description by deriving the transformed wave equation from the time-dependent Schrödinger equation and showing that the static component of the Henneberger potential dominates in a high-frequency alternating field, leading to results for laser-assisted $\alpha$ decay similar to those of Delion \textit{et al.} \cite{PhysRevLett.119.202501}; meanwhile, Kis and Szilvasi \cite{Kis_2018} further examined three-dimensional $\alpha$-particle tunneling in laser-modified potentials and found that circularly polarized fields may enhance the decay probability and shorten the lifetime of quasi-bound states under specific model assumptions. Nevertheless, these studies inherit the same time-averaged assumptions and therefore do not remove the underlying conceptual objection. Predictions such as a laser-driven change of the dominant decay mode, for example from $\alpha$ decay to proton emission, remain highly speculative and should not be presented as near-term experimental prospects.

\subsection{Quasistatic approximation}

Another prevalent method for investigating the laser-nucleus interaction is the quasistatic approximation. Since one optical cycle of the high-power laser is considerably longer than the time required for the emitted particle to penetrate the potential barrier, $\alpha$ decay in strong electromagnetic fields can be approximated as a quasistatic process under these conditions, where laser field variations are negligible \cite{PhysRevC.99.044610,PhysRevC.102.064629}. Consequently, under this scenario, the magnetic component is commonly neglected, and the electric-field contribution becomes the dominant term in the relative motion. Note that the notion ``tunneling time'' is not clear itself, since under the barrier the velocity of the $\alpha$ particle becomes imaginary and respectively the tunneling time also takes complex value. Hence, conclusion that $10^{-20}$ s is very approximate and we are planning additional work to get more reliable estimations of time delay in tunneling of $\alpha$ particles.

Within this framework, the effect of laser light on the $\alpha$ decay is investigated through an additional potential energy. The expression for $V(r,\theta,\varphi,t)$ in equation (\ref{eq2}) can thus be rewritten as
\begin{equation}
  V(r,\theta,\varphi,t)=V_N(r,\varphi)+V_l(r)+V_C(r,\varphi)+V_i(r,\theta,t).
\end{equation}
Here, $V_N(r,\varphi)$, $V_l(r)$ and $V_C(r,\varphi)$ respectively denote the nuclear, centrifugal, and Coulomb potentials. The electric dipole term $V_i(r,\theta,t)$ is expressed as \cite{Miicu_2013}
\begin{equation}
  V_i(r,\theta,t)=-Z_{\mathrm{eff}}\vec{r}\cdot\vec{E}(t)=-Z_{\mathrm{eff}}rE(t)\cos\theta.
\label{eq:vic}
\end{equation}
For $\alpha$ decay, the effective charge can be formulated as
\begin{equation}
  Z_{\mathrm{eff}}=\frac{Z_{\alpha}A_d-A_{\alpha}Z_d}{A_d+A_{\alpha}},
\label{eqzeef}
\end{equation}
where $A_d$ represents the mass number of the daughter nucleus. $A_{\alpha}$ and $Z_{\alpha}$ denote the mass number and charge number of the $\alpha$ particle, respectively. As seen from equation (\ref{eqzeef}), when the mass number of the daughter nucleus is twice its charge number, the effective charge is zero. This signifies that the laser electric field does not affect the relative motion of the daughter nucleus and the $\alpha$ particle, e.g., $^{108}$Xe \cite{CHENG2024138322,Xiao2024}.

In 2018, Bai \textit{et al.} also investigated the influence of the strong laser field on the penetration probability of $\alpha$ decay via the quasistatic approximation \cite{Bai_2018}. The calculations show that the expected angular asymmetry: emission along the electric field is favored, whereas the antiparallel direction is suppressed, and nuclei with larger effective charge respond more strongly. Later in 2019, Qi \textit{et al.} explored the influence of a strong laser field on the nuclear $\alpha$ decay process by utilizing a quantitative nuclear potential model \cite{PhysRevC.99.044610}. The results indicate that lower-$Q_\alpha$ nuclei are more sensitive because their longer tunneling paths amplify the external-field contribution. In addition to the quantitative revelations, Qi \textit{et al.} also put forward a practical approach to observe the response of $\alpha$ decay to the laser field of an elliptically polarized laser pulse. In this sense, the quasistatic approach offers a more experimentally consistent baseline: it predicts small effects, but it also suggests differential observables that could be more realistic than a direct half-life measurement.

In contrast to claims of significant laser-induced acceleration of $\alpha$ decay, moving on to 2020, Pálffy and Popruzhenko's study \cite{PhysRevLett.124.212505} introduced additional considerations that challenge earlier predictions. Employing semiclassical methods including the WKB approximation and imaginary time method, their analysis identified a characteristic electric field threshold (\(E_{\rm{eff}} \sim 10^{16}\)--$10^{18}\,\rm{V/cm}$) required to modify the decay rates, which exceeds current laser capabilities (\(E \sim 3 \times 10^{14}\,\rm{V/m}\) at $I = 10^{23}\,\rm{W/cm}^2$). Their calculations for a variety of $\alpha$ emitters showed that the laser-induced corrections are below the level of experimental detectability even at the predicted extreme intensities. The difference from the previous statement is attributed to the adiabatic approximation inherent in the Kramer-Henneberg transformation, which assumes that the laser field is stationary and inconsistent with the sub-attosecond nuclear decay timescale (\(\tau_{\alpha} \sim 10^{-20}\,\rm{s}\)). Additionally, Pálffy and Popruzhenko emphasized the impracticality of observing such effects in realistic experimental settings, where the number of decay events in the laser focal volume (\(\Delta N \leq 10^{-3}\) per shot for $^{212}\rm{Po}$) is orders of magnitude below detection thresholds. This result is the main quantitative counterpoint to the large KH-type enhancement scenario and explains why the absence of high-power-laser evidence is physically significant rather than accidental.

Nevertheless, in the same year, Delion \textit{et al.} countered this assertion \cite{PhysRevC.101.044304}. They argued that under continuous laser field conditions, the Kramers-Henneberger transformation can be effectively utilized to calculate the $\alpha$ particle half-life. To support their assertion, they conducted a detailed coupled-channel analysis of $\alpha$ decay in a strong electromagnetic field. This work represents the alternative KH-based viewpoint, but it does not remove the timescale and coherence objections raised for optical femtosecond experiments. It is important to distinguish between laser-induced deformation of the effective Coulomb barrier and the intrinsic nuclear deformation parameters (\(\beta_2, \gamma\)). The former refers to the transient distortion of the external potential experienced by the \(\alpha\)-particle due to the presence of an external electromagnetic field. In contrast, the latter are structural parameters governed by the quantum state of the nucleus \cite{Cline1986}. Currently, there is no theory suggesting that changes in nuclear deformation can be achieved through any means other than exciting the nucleus. Laser fields, unless reaching regimes capable of direct nuclear transitions (e.g., via NEEC or direct photoexcitation), cannot modify \(\beta_2\) or \(\gamma\) directly. Laser-induced nuclear excitation and its impact on nuclear structure will be discussed in later sections.

\subsection{Controversy over laser influence on $\alpha$ decay}
\label{4.3}

The theoretical discrepancies between the Kramers-Henneberger transformation and the quasistatic approximation in modeling laser-assisted $\alpha$ decay arise from fundamentally different assumptions about the temporal relationship between the laser fields and nuclear dynamics. In the KH approach, which was originally developed in atomic strong-field physics, the field is incorporated through a laser-driven displacement and an effective time-averaged potential. This treatment can yield large barrier distortions, strong angular anisotropies, and dramatic enhancement of penetrability in certain parameter regimes. In contrast, the quasistatic approximation treats the laser as a slowly varying external perturbation during the tunnelling event. Under current optical laser conditions, it generally predicts much smaller changes, often of order $10^{-5}$ or less for the relative half-life variation, depending on the nucleus and model assumptions.

It should be noted that under presently accessible optical and near-infrared strong-field conditions, the quasistatic approximation provides a more conservative baseline for direct laser-assisted $\alpha$ decay. This is mainly because the estimated $\alpha$-tunnelling timescale is much shorter than an optical laser cycle, while KH-based large enhancements rely on a high-frequency, coherence-sensitive time-averaged description whose applicability to realistic femtosecond laser pulses remains unsettled. In the absence of direct experimental confirmation, such KH-based enhancements should therefore be treated as regime-dependent theoretical predictions rather than established experimental expectations.

The controversy is also sharpened by experimental feasibility. A realistic $\alpha$-decay measurement in a high-intensity laser environment must cope with intense electromagnetic flash backgrounds, secondary particles generation and emission, limited source number in the focal volume, and the extremely small number of decays occurring during a single laser shot. For short-lived isotopes in particular, source preparation, survival, timing, and detection geometry become severe constraints. Therefore, the existence of a theoretically nonzero laser effect should be distinguished from the feasibility of measuring a laser-induced half-life change, since the latter requires a much more demanding assessment of signal size, background suppression, and detection statistics.

This controversy is nevertheless scientifically productive. It identifies a crucial open problem for the field: the development of a unified multiscale framework capable of connecting direct strong-field coupling, tunnelling dynamics, and realistic experimental conditions without extending approximations beyond their physically justified domain. Future progress will likely depend not only on more sophisticated theory but also on the identification of observables more realistic than direct half-life modification, such as anisotropic emission signatures, laser-modified spectra, or environment-mediated effects. 

\subsection{Determinants of laser-influenced $\alpha$ decay}

Since direct experimental validation is still lacking, it is useful to identity which nuclear properties most strongly influence the predicted laser modification of $\alpha$ decay within conservative theoretical estimates. Cheng \textit{et al.} investigated the effect of $10^{23}\ \mathrm{W/cm^{2}}$ laser pulses on 190 deformed even-even nuclei's $\alpha$ decay half-life change rate \cite{CHENG2024138322}, providing insights into the understanding of the complex interactions between lasers and nuclear $\alpha$ decay.

In their study, accounting for the nucleus deformation, the total penetration probability $P(\theta)$ is averaged from $P(\theta,\varphi)$ over particle emission directions \cite{PhysRevC.73.041301}
\begin{equation}
  P(\theta)=\frac{1}{2}\int_0^\pi P(\theta,\varphi)\sin\varphi \rm{d\varphi}.
  \label{eq2.48}
\end{equation}
The change rates $\Delta T$ and $\Delta P$ of the $\alpha$ decay half-life and penetration probability in the high-intensity laser field were defined respectively by
\begin{equation}
  \Delta T=\frac{T(E,\theta)-T(E=0,\theta)}{T(E=0,\theta)},
  \label{eq3.8}
\end{equation}
\begin{equation}
  \Delta P=\frac{P(E,\theta)-P(E=0,\theta)}{P(E=0,\theta)}.
\end{equation}

\begin{figure*}[h]
\centering
\captionsetup{justification=justified, singlelinecheck=false}
\includegraphics[width=12cm]{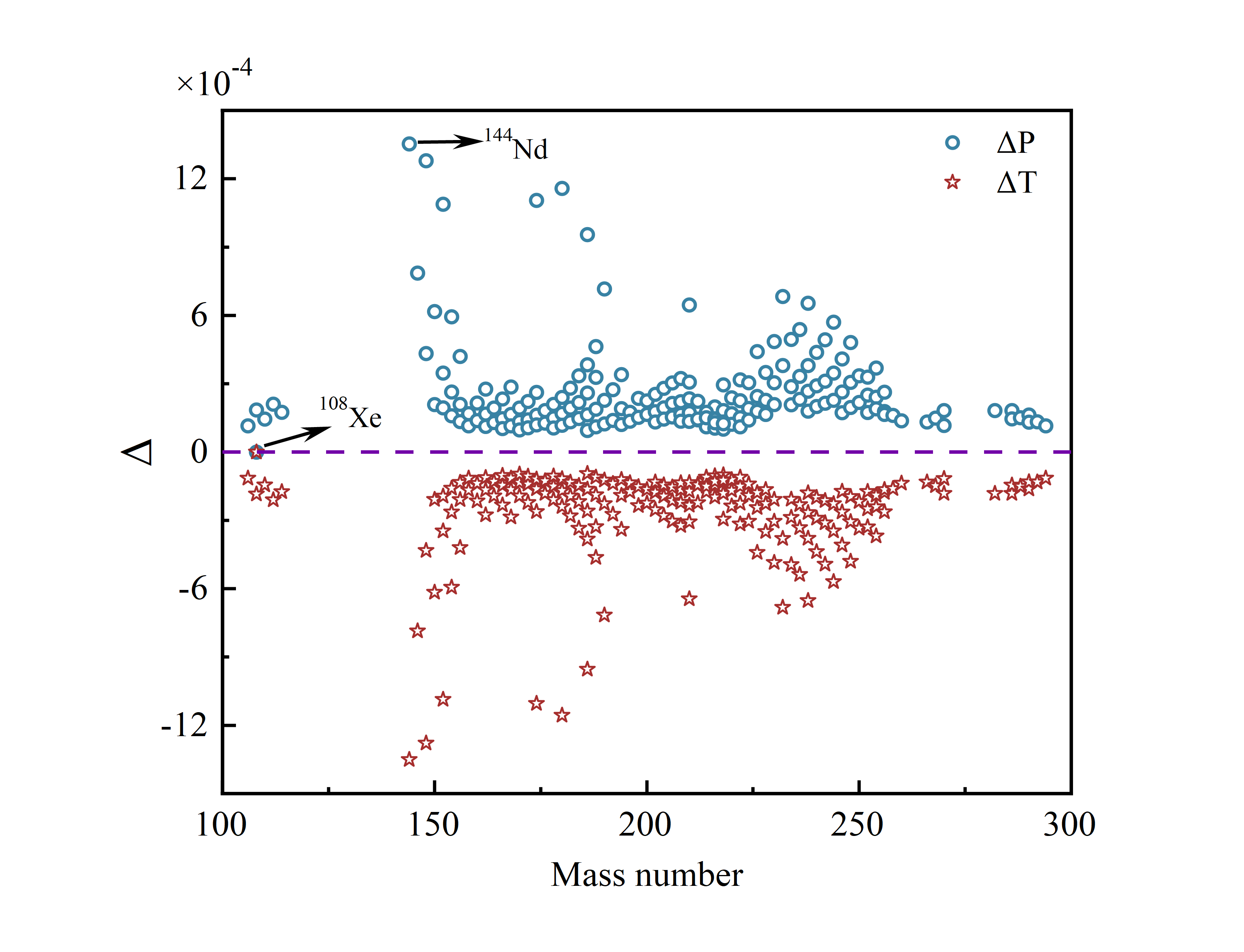}
\caption{ The change rates $\Delta T$(red star) and $\Delta P$ (blue circle) for the 190 deformed ground state even-even nuclei with the laser intensity of $10^{23}\ \mathrm{W/cm^{2}}$ \cite{CHENG2024138322}.}
\label{fig 33}
\end{figure*}

The influence of the high-intensity laser field on the $\alpha$ decay of deformed even-even nuclei is shown in figure \ref{fig 33}, assuming that the particle emission occurs along the laser electric field direction. There are significant differences in the rates of change of decay penetration probability and half-life among different parent nuclei, ranging from 0.0009\% to 0.135\%. As expected, for $^{108}\mathrm{Xe}$, $\Delta T$ and $\Delta P$ are zero because $A_d = 2Z_d$ ($Z_{\mathrm{eff}} = 0$) and the daughter nucleus and the emitted particle move together in the laser field. The largest response occurs for $^{144}$Nd, reflecting the enhanced sensitivity of low-$Q_\alpha$ nuclei while still remaining at the sub-percent level.

Cheng \textit{et al.} analytically derived the change rate of the $\alpha$ decay penetration probability \cite{CHENG2024138322}, and obtained the analytical equation as follows
\begin{equation}
	\Delta P\approx B_1Q_\alpha^{-5/2}+B_2Q_\alpha^{-2}+B_3Q_\alpha^{-1}+C_1Q_\alpha^{-9/2},
	\label{eq4.22}
\end{equation}
where $B_{1}$, $B_{2}$, $B_{3}$ and $C_{1}$ are $Q_\alpha$ independent parameters. The inverse powers of $Q_\alpha$ explain why low-energy $\alpha$ emitters are systematically more sensitive in quasistatic estimates.

For odd-A nuclei that have distinct structural traits \cite{Koura2005, PhysRevC.93.044321, PhysRevC.60.051301,PhysRevLett.81.3599,Yarman2016,QI2016, PhysRevC.77.054318, PhysRevC.76.027303, PhysRevC.93.011306,Sun2017,Yang2022}, calculations show that the change in their $\alpha$-decay penetration probability within extreme laser fields is influenced by shell effects and odd-even staggering \cite{cjh2025}. This finding helps identify comparatively sensitive nuclei within the quasistatic framework, although the predicted absolute change remains small. Additionally, radium and thorium isotopes exhibit odd-even staggering in the variation of their penetration probabilities when they are away from the magic number. Thus, these calculations mainly provide selection criteria for future tests rather than evidence that a direct measurement is already feasible.

Recently, Zou \textit{et al.} investigated $\alpha$-decay of actinide nuclei in bichromatic laser fields \cite{czvc-ymfl}. Bichromatic fields introduce additional degrees of freedom for controlling the nuclear $\alpha$-decay process, including parameters such as the frequency ratio, field amplitude ratio, phase difference, and temporal delay of the two lasers.  These parameters can be varied to optimize or modulate nuclear $\alpha$ decay. Their results show that a fundamental–second harmonic ($\omega$-$2\omega$) bichromatic field can enhance the time-averaged modification by one to two orders of magnitude. This work provides new insights into nuclear $\alpha$-decay manipulation from the perspective of laser technology optimization.

\subsection{$\alpha$ decay via electron screening in laser-heated clusters}
 
Directly manipulating the nuclear $\alpha$-decay rates by laser fields continues to pose significant challenges, as current approaches typically demand ultra-high laser intensities or suffer from limited control efficiency. Beyond direct laser-nucleus interaction, the electrons in deeply bound atomic orbitals can also be affected by the electromagnetic fields of lasers. During the laser-matter interactions, an intense femtosecond laser pulse first irradiates a target, leading to the ionization of the constituent atoms. Theoretically, the ionized free electrons can be precisely controlled by the laser fields to manipulate the nuclear processes via electron-nucleus interactions.

\begin{figure*}[h]
\centering
\captionsetup{justification=justified, singlelinecheck=false}
\includegraphics[width=9cm]{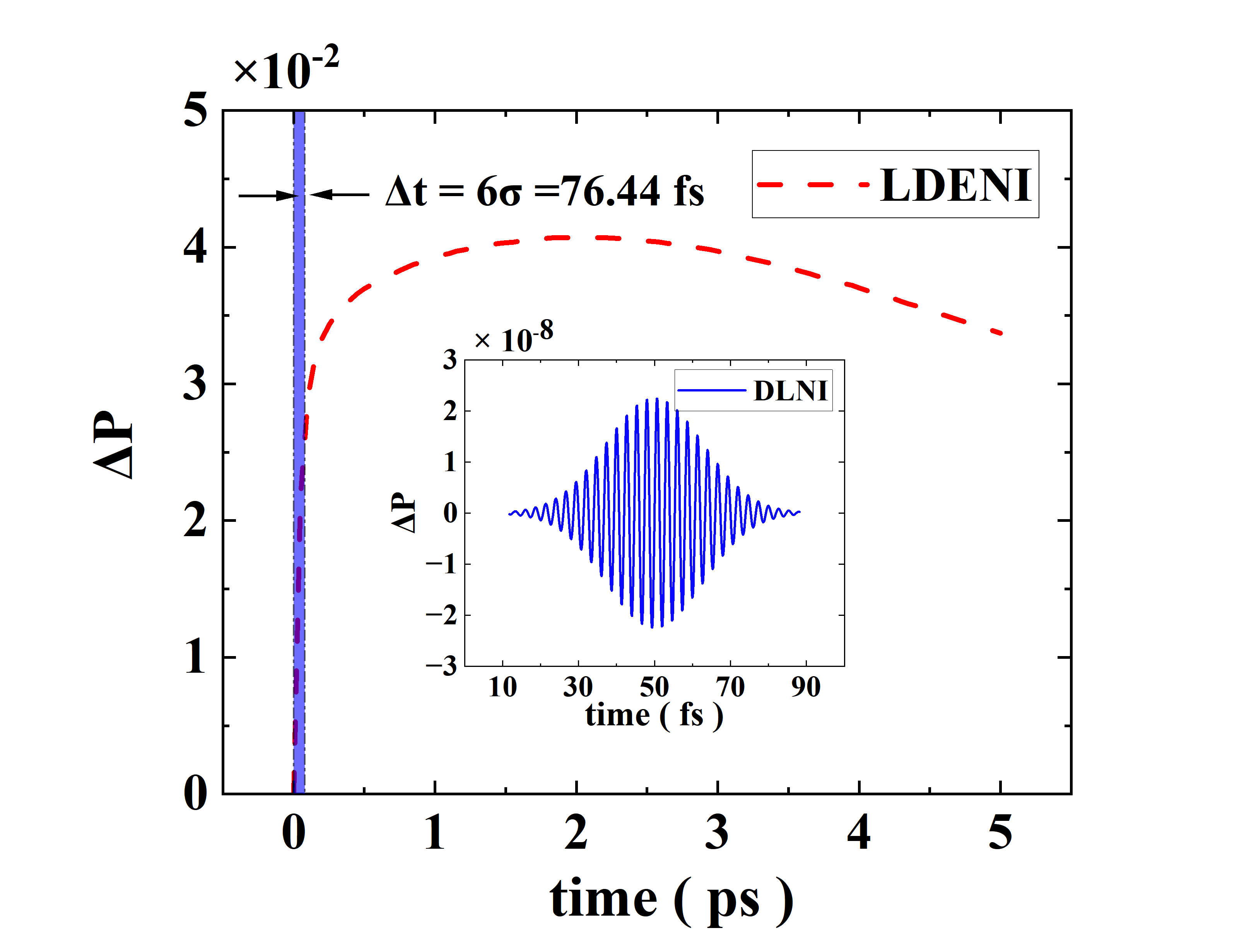}
\caption{Penetration rate change of a Gaussian laser field on $\alpha$ decay by laser-driven electron-nucleus interactions (LDENI) and direct laser-nucleus interactions (DLNI) under the condition of laser intensity $1.0 \times 10^{15}\rm{ W/cm^{2}}$, laser wavelength 800 nm, and pulse width 30 fs \cite{ZOU2026140362}.}
\label{fig445}
\end{figure*}

The electron-nucleus interaction is fundamentally dominated by electromagnetic interactions, which mediate nuclear environment modulation. These electromagnetic interactions exert key effects including electron screening, Coulomb excitation, radiative transition, and weak interaction-induced electron capture, collectively serving as a critical means to regulate nuclear processes such as nuclear decay, nuclear excitation, nuclear reactions, and nuclear structure evolution. In laser-plasma scenarios, the temporal dynamics of electron density and temperature, coupled with laser-controlled electron energy and spectrum distribution, render electrons a pivotal medium for indirect manipulation of nuclear processes. Recently, Zou \textit{et al.} proposed a novel method to accelerate the nuclear $\alpha$-decay rate by electron screening in laser-heated warm dense cluster plasma \cite{ZOU2026140362}. In this scheme, a nanocluster composed of $\alpha$-emitting nuclei is initially irradiated by a femtosecond laser pulse, inducing atom-ionization and forming the warm dense matter \cite{PhysRevLett.125.085001}. After the laser pulse's irradiation, the nanocluster starts to expand and lasts for a few picoseconds. During the expansion process, the confined electrons within the cluster dynamically screen the Coulomb barrier suffered by the $\alpha$ particle, thereby enhancing the tunneling probability. Especially, they developed a gradient-corrected screened model to evaluate the effects of electron screening on the $\alpha$ decay within the warm dense nanocluster plasma. Combined with the particle-in-cell simulations, their results show that the cluster expansion maintains a high-density core while reducing electron temperature, significantly enhancing the effects of electron screening on $\alpha$ decay. Using the deformed two-term proximity model \cite{PhysRevC.26.1969}, they predicted that the rate of change in the penetrability of $^{226}\rm{Ra}$ can be enhanced by up to $4\%$ under the laser intensity of $1.0\times 10^{15}\ \mathrm{W/cm^{2}}$, as shown in figure \ref{fig445}. This modification is extremely larger than that from direct laser field acceleration by six orders of magnitude under this laser intensity. In the subsequent optimization of the scheme, they also promised the extension by employing high repetition$-$rate lasers interacting with multi$-$cluster targets, which can further prolong the effective interaction timescale and amplify cumulative effects. These findings offer new theoretical insight for future experimental efforts aimed at controlling the $\alpha$-decay rate in warm dense matter assisted by lasers. In this sense, the warm-dense-matter route shifts the problem from direct nucleus-field coupling to laser-controlled electronic environments, offering a conceptually distinct and potentially more experimentally tractable pathway for testing laser-assisted modifications of $\alpha$ decay \cite{ZOU2026140362}.

Notably, this scheme differs from the direct laser field approach in several key aspects. First, indirect laser-driven electron-nucleus interaction (LDENI) achieves more effective modulation than direct laser-nucleus interaction (DLNI) under same laser intensities, with the picosecond-scale expansion of the nanocluster supporting longer interaction timescales. Secondly, using laser-heated nanoclusters relaxes the alignment requirement between the laser field and the $\alpha$ particle’s emission direction, a feature that may facilitate its feasibility in practical experimental setups. Additionally, employing high-repetition-rate lasers with multiple clusters may help mitigate constraints related to rapid electron diffusion, which could contribute to improved experimental viability in future work. Furthermore, this method links nuclear physics and laser-driven plasma science, with potential extension to other $\alpha$-emitters and charged-particle decay modes, thus providing a possible reference for developing advanced nuclear decay control strategies.

In summary, this chapter has systematically reviewed the theoretical progress in understanding laser-assisted $\alpha$ decay, highlighting the interplay between extreme electromagnetic fields and nuclear structure. The competing frameworks of the Kramers-Henneberger transformation and the quasistatic approximation have illuminated the temporal and energetic mismatches inherent in laser-nucleus interactions, underscoring the need for hybrid models that bridge decay dynamics with femtosecond-scale laser pulses. While current laser intensities remain insufficient to observe measurable half-life changes, advancements in high-intensity laser facilities like ELI-NP and SULF offer unprecedented opportunities to test these predictions. Besides that, the new regulation schemes, like $\alpha$-decay rate altered by laser-driven electron screening and by laser-induced nuclear excitation of isomeric states ($\alpha$-decay rate in the excited state of certain nuclei is significantly higher than that observed in the ground state, e.g., $^{212}$At, $^{212}$Bi, etc.), provide scalable solutions to modulate the $\alpha$-decay rate by lasers. By means of the laser-induced isomeric state excitation, a representative example is $^{212}$At. Its ground state exhibits an $\alpha$-decay half-life of 314 ms, while pumping the nucleus into its $9^-$ isomeric state reduces the half-life to 121 ms \cite{NNDC2025}. This significant enhancement in decay rate is primarily attributed to the elevation of the parent nucleus's decay energy $Q_{\alpha}$ upon excitation, and a critical reduction in angular momentum hindrance—stemming from a change in the orbital angular momentum carried by the emitted $\alpha$-particle. Although substantial modulation of nuclear $\alpha$-decay rates remains challenging under present laser facilities, these frontier explorations offer pioneering attempts and shed new light on the laser-based control of nuclear $\alpha$ decay.

\section{Proton radioactivity in laser fields}\label{fourth}

Despite many studies on $\alpha$ decay in high-intensity laser fields, limited research exists on proton radioactivity in such fields. One possible reason is that the decay energy levels and decay dynamics of proton radioactivity are more challenging to assess and control with current laser technology compared to those of $\alpha$ decay. However, recent studies have started to explore this new area. This chapter summarizes the strong X-ray laser-induced proton radioactivity in halo nuclei and the effect of high-intensity laser fields on proton radioactivity in deformed nuclei, together with some new insights from these recent research efforts.

\subsection{Proton radioactivity from halo nuclei}
Proton radioactivity in laser fields is a research area of great significance since it provides crucial insights into nuclear structure and the interaction of nuclei with intense electromagnetic fields. This may enable new approaches for investigating proton-rich nuclei in astrophysical environments, such as those encountered in X-ray bursts and supernovae \cite{PARIKH2013225}. In this regard, Wu and Liu conducted a detailed study on the proton emission from halo nuclei induced by intense X-ray lasers \cite{PhysRevC.106.064610}. They modeled the laser field as an arbitrarily polarized, monochromatic wave of frequency $\omega$ with the vector potential $A(t)=\frac{A_{0}}{\sqrt{1+\delta^{2}}}\left[\cos(\omega t)\vec{e}_{z}+\delta\sin(\omega t)\vec{e}_{y}\right].$ Here, $A_{0}$ is the maximal amplitude and $\delta$ is the laser ellipticity. They focused on one-proton halo nuclei, specifically $^{8}$B, and separated the two-body total Hamiltonian of the halo nuclei in the laser field into a center-of-mass part and a relative-motion part under the dipole approximation. By using the Dyson expansion \cite{REISS19921,Joachain2012} and strong-field approximation, the transition amplitude $T_{p}^{\mathrm{SFA}}$ was derived. Subsequent detailed calculations and approximations led to the analytical derivation of the scattering amplitude $S_{p}^{\mathrm{SFA}}$ matrix, which provided the basis for calculating both the differential and total proton emission rates.

It is also shown that both the Coulomb factor (CF) and the polarization of the laser fields had a substantial influence on the total rates of proton emission, since the CF had a profound impact on the relationship between the total rate and the laser intensity of different frequencies. It not only reduced the total rate but also modified the slope of the rate-intensity curve and caused blueshifts of the multiphoton transition frequency. At the same time, the polarization effects manifested as variations in the total rates with the laser ellipticity $\delta$, which varied with frequency. By analyzing the laser ellipticity corresponding to the maximum total rates as a function of $\omega/E_{b}$, a clear transition from perturbative to nonperturbative proton emission was identified. The comparison between the power-law approximation in the perturbative regime and the nonperturbative S-matrix theory results, along with the distributions of partial rates $R_{n}$, further validated the nonperturbative signatures. This clearly shows the limitations of traditional perturbative models and emphasizes the need for more innovative theoretical frameworks. Future studies could expand on this work by extending the scope to a wider range of halo nuclei \cite{PhysRevC.53.R572,Ren:1999fcl}. For example, recently, the nuclear masses of $^{23}$Si, $^{26}$P, $^{27}$S, and $^{31}$Ar were precisely measured, and new experimental evidence supporting the existence of proton halos in these nuclei has been reported through the analysis of mirror symmetry breaking effects \cite{Yu:2024ieb}.

\subsection{Symmetry of the rate of change}

\begin{figure*}[h]
\centering
\captionsetup{justification=justified, singlelinecheck=false}
\includegraphics[width=16cm]{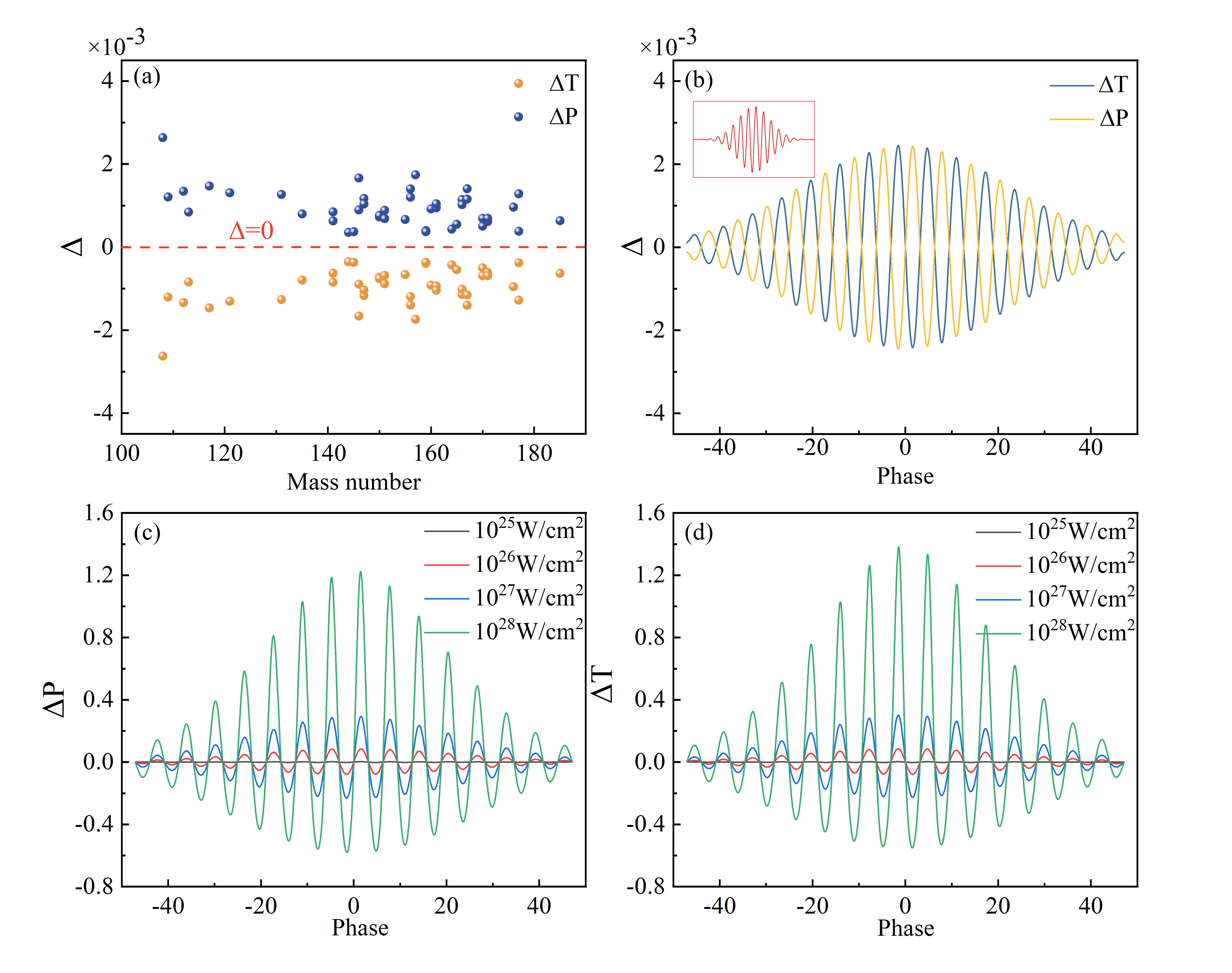}
\caption{Analysis of proton radioactivity-related parameters (rate of change of penetration probability $\Delta P$ and half-life $\Delta T$) for different mother nuclei under laser fields of different intensities: (a) representation of $\Delta T$ and $\Delta P$ at $I=10^{23}\,\mathrm{W/cm^{2}}$, (b) oscillation of $\Delta T$ and $\Delta P$ of $^{108}\mathrm{I}$ at $I=10^{23}\,\mathrm{W/cm^{2}}$, (c) comparison of $\Delta P$ of $^{108}\mathrm{I}$ under $I=10^{25}$--$10^{28}\,\mathrm{W/cm^{2}}$, (d) same as (c) but for $\Delta T$ \cite{PhysRevC.105.024312}.}
\label{fig4.1.1}
\end{figure*}

Recently, Cheng \textit{et al.} have found that under the current laser conditions, \(\Delta P\) and \(\Delta T\) have symmetry for different nuclei (under the same laser electric field conditions) and for the same nuclei in the same pulse of different electric field intensity \cite{PhysRevC.105.024312}. Figure \ref{fig4.1.1}(a) and (b) shows such a symmetry with regard to \(\Delta\) = 0 for different nuclei under same conditions and the same nucleus under different conditions.

To elucidate this phenomenon, they introduced another important parameter $P'(E, \theta)= P(E, \theta)-P(E = 0, \theta)$ to quantify the difference in proton penetration probability with and without a laser electric field, providing a fundamental variable for deriving the relationship between the \(\Delta T\) and the \(\Delta P\). Substituting it into equation (\ref{eq3.8}), this yields
\begin{equation}
\Delta T = -\frac{P'(E, \theta)}{P'(E, \theta) + P(E, \theta)}.
\end{equation}
At the current laser electric field intensity, $P'(E, \theta) \ll P(E, \theta),$ thus the equation (\ref{eq3.8}) could be approximated as $\Delta T \approx -P'(E, \theta)/P(E, \theta) \approx -\Delta P.$

Analogous to $\alpha$ decay, the change rate of proton radioactivity penetration probability can also be expressed as two terms
\begin{equation}
\Delta P_{\varphi} = \chi_{\varphi}^{(1)} + \chi_{\varphi}^{(2)}.
\label{eq3.24}
\end{equation}
Here, $\chi_{\varphi}^{(1)}$ is related to the electric field strength, and $\chi_{\varphi}^{(2)}$ is related to the square of the electric field strength, since the square of the electric field strength is always positive, this symmetry will be progressively disrupted as the laser intensity steadily increases. To validate the aforesaid conjecture, they took \(^{108}\mathrm{I}\) as an exemplar and calculated the proton radioactivity penetration probability change rate \(\Delta P\) under the influence of four distinct laser pulses with the intensity ranging from \(10^{25}\,\mathrm{W/cm^{2}}\) to \(10^{26}\,\mathrm{W/cm^{2}}\), \(10^{27}\,\mathrm{W/cm^{2}}\) and \(10^{28}\,\mathrm{W/cm^{2}}\). It is shown that with the increase of the laser intensity, the symmetry of \(\Delta P\) and \(\Delta T\) with respect to \(\Delta = 0\) is progressively disrupted. The root cause lies in the fact that \(\chi_{\varphi}^{(2)}\) in equation (\ref{eq3.24}) is directly proportional to the square of the electric field, and its rate of growth in tandem with the escalating electric field surpasses that of \(\chi_{\varphi}^{(1)}\), which is only proportional to the electric field itself. Once the increase in the laser field exceeds a specific threshold, \(\chi_{\varphi}^{(2)}\) can no longer be ignored, thus breaking the symmetry of \(\Delta P\) and \(\Delta T\).

Given that the values of \(\chi_{\varphi}^{(2)}\) are always positive and will not take negative values as \(\chi_{\varphi}^{(1)}\) does when the electric field turns negative, the variation trend exhibited in figure \(\ref{fig4.1.1}\)(c) and figure \(\ref{fig4.1.1}\)(d) are, in general, displaced upward in the positive direction. It is foreseeable that when the laser intensity reaches extremely high threshold, e.g., $I=10^{25}\,\mathrm{W/cm^{2}}$, nearly the entire distribution profiles of the \(\Delta P\) and \(\Delta T\) become entirely positive. This implies that as the laser intensity increases, the behavior of proton radioactivity in the laser field undergoes a significant transformation. It could potentially lead to new approaches for controlling or manipulating proton radioactivity with lasers, which may have far-reaching consequences in fields like nuclear energy and medical imaging. However, further investigations are required to fully understand and harness these effects.

\subsection{Asymmetric chirped laser pulses}

As the intensity of the laser electric field oscillates with the phase changes, the effect of \(\chi_{\varphi}^{(1)}\) on proton radioactivity penetration probability mostly nullifies, making the average penetration probability change rate within a laser period rather feeble. Mi\c{s}icu and Rizea suggested using a rectangular short laser pulse with an odd number of half-cycles to increase the proton radioactivity decay rate by three orders of magnitude \cite{Misicu2019}. However, realizing half-electric field laser pulses is still difficult with current technology. Qi \textit{et al.} put forward an elliptically polarized laser field experimental scenario \cite{PhysRevC.99.044610}, yet only factored in the angle \(\theta\) between \(\vec{E}(t)\) and \(\vec{r}\) on the average penetration probability change rate and overlooked the laser frequency shifts. It is clearly indicated that the existing methods still have significant limitations in effectively enhancing and precisely controlling the proton radioactivity in laser fields. There is an urgent need for more innovative approaches that can comprehensively consider multiple factors such as laser phase, frequency, and polarization to achieve a more substantial and reliable manipulation of proton radioactivity. 

Building on this, researches aimed to boost the average change rate of proton radioactivity penetration probability by introducing chirp to break the laser electric field symmetry \cite{PhysRevC.105.024312}. To achieve this goal, Cheng \textit{et al.} utilized specific pulse sequences within the chirped pulse amplification framework. The electric field of a Gaussian chirped pulse sequence can be described by \cite{Zhang2012}
\begin{equation}
E(p,b)=E_{0}\exp\left(-\frac{p^{2}}{4\pi^{2}x^{2}}\right)\sin\left(p+\frac{b\times p^{2}}{2\pi}\right),
\end{equation}
where \(b\) is the chirp parameter (\(b < 0\) for negative chirp and \(b > 0\) for positive chirp) and $p=\omega t$. They quantifies \(\Delta P_{avg}\) values for diverse chirp values with the aimed laser intensity of \(I_0 = 10^{25}\,\mathrm{W/cm^{2}}\) in ELI-NP \cite{RevModPhys.84.1177,10.1063/1.5093535} and the Station of Extreme Light (SEL) in Shanghai \cite{10.34133/2022/9894358,Li2023}. 

\begin{figure*}[h]
\centering
\captionsetup{justification=justified, singlelinecheck=false}
\includegraphics[width=16cm]{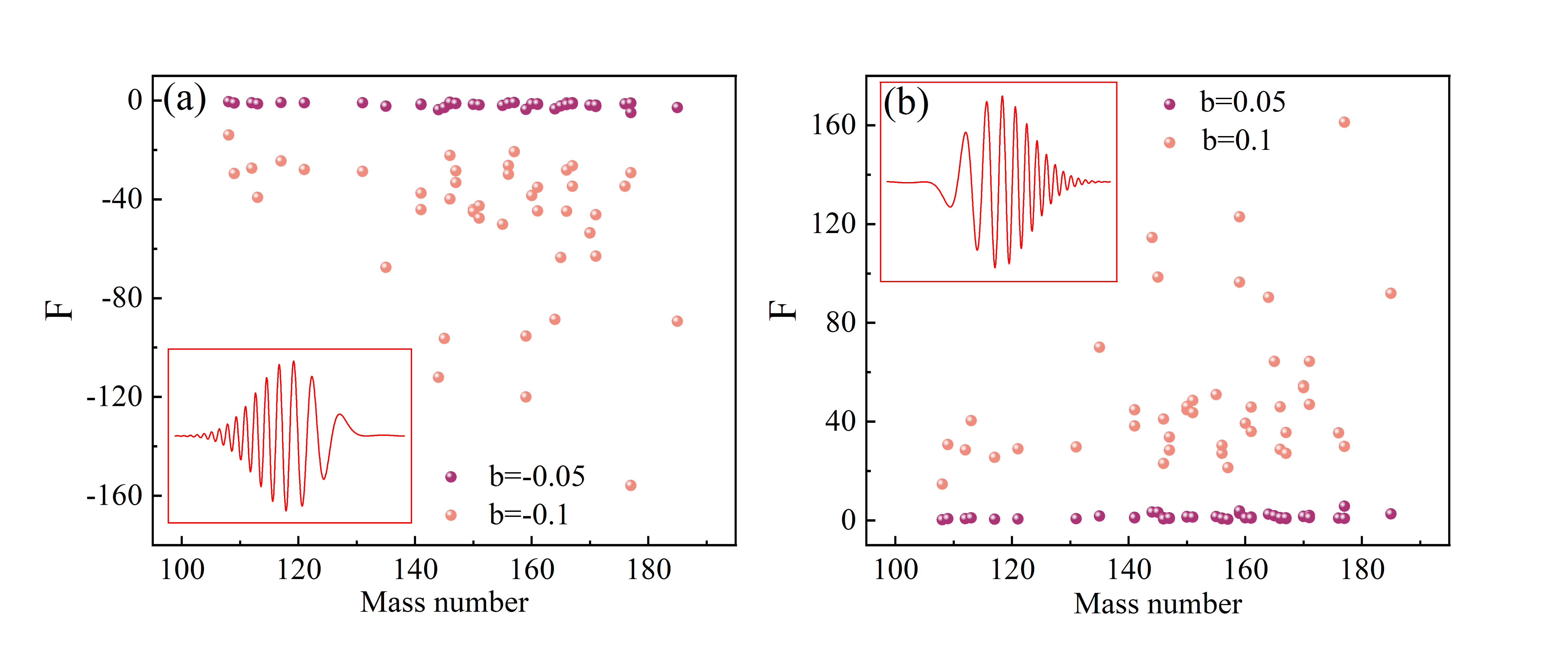}
\caption{The averaged change rate ($\Delta P_{avg}$) of proton radioactivity penetration probability at $I_{0}=10^{25}\,\mathrm{W/cm^{2}}$: (a) with different negative chirp values, (b) with different positive chirp values \cite{PhysRevC.105.024312}.}
\label{fig4.2.1}
\end{figure*}

To compare symmetric and chirped pulses clearly, a new parameter \(F = (\Delta P_{\mathrm{avg}}-\Delta P_{\mathrm{avg}}^{b = 0}) / \Delta P_{\mathrm{avg}}^{b = 0}\) is defined, representing the relative enhancement factor of the average penetration probability change rate of the asymmetric chirped pulse compared to the symmetric pulse without chirp. Figures \(\ref{fig4.2.1} \)(a) and \(\ref{fig4.2.1} \)(b) display that \(F\) grows with \(b\), indicating that the asymmetric chirped pulses exhibit superior performance in attaining a larger \(\Delta P_{avg}\) as compared to the pulse without chirp. With symmetric pulses, the \(\chi_{\varphi}^{(1)}\) cancellation effect is suppressed by asymmetric chirped pulses, increasing \(\Delta P_{avg}\) by several times. For example, with \(b = 0.1\), \(F\) can reach 160, which is equivalent to increasing the laser intensity by two orders of magnitude. It shows the potential of asymmetric laser pulses to enhance the rate of proton radioactivity and emphasizes the need for optimizing chirp parameters and pulse shaping techniques.

\section{Two-proton radioactivity in laser fields}\label{Fifth}
Two-proton ($2p$) radioactivity is a rare but fascinating mode of nuclear decay occurring in proton-rich nuclei beyond the proton drip line. First predicted by Goldansky in the early 1960s, this process involves the simultaneous emission of two protons from an unstable nucleus, offering valuable insights into nuclear structure and the nuclear interplay forces at the limits of stability \cite{GOLDANSKY1960482}. The study of $2p$ radioactivity has not only advanced our understanding of nuclear decay mechanisms but also holds significant implications for nuclear astrophysics and the synthesis of heavy elements. The advent and advancement of laser technology have engendered novel opportunities for the examination of $2p$ radioactivity within exotic nuclei. Particularly, the laser-assisted $2p$ radioactivity is closely associated with the ($2p$, $\gamma$) and ($\gamma$, $2p$) processes of stellar evolution, which are crucial for exploring the nuclear properties at waiting points in stellar nucleosynthesis \cite{Fisker_2004}. However, the influence of intense laser fields on $2p$ radioactivity remains a relatively unexplored area.

\subsection{Three mechanisms of two-proton radioactivity}

Over the past few decades, both experimental and theoretical progress has significantly enhanced our understanding of $2p$ radioactivity. Experimentally, the $2p$ radioactivity phenomenon was derived from extremely short-lived light nuclei such as $^{6}\rm{Be}$\cite{PhysRev.150.836}, $^{12}\rm{O}$\cite{PhysRevC.86.011304,PhysRevLett.103.152503} and $^{16}\rm{Ne}$\cite{PhysRevC.17.1929}. This type of nuclei is characterized by $Q_{2p}$ $>$ 0 and $Q_{p}$ $>$ 0, also called the not true $2p$ radioactivity because the $1p$ radioactivity is energetically allowed, and the effect of paring energy of emitted two protons is ignored. However, owing to their short lifetimes and broad intermediate states, the decay mechanisms of these nuclei are not fully understood. The other type is the true $2p$ radioactivity whose $Q_{2p}$ $>$ 0 and $Q_{p}$ $<$ 0, which is featured in that the $1p$ radioactivity is strongly forbidden or substantially suppressed, and the energy level of the $2p$ emitting channel is lower than that of $1p$ radioactivity\cite{RevModPhys.84.567,Blank_2008,BLANK2008403}. Here, $Q_{p}$ and $Q_{2p}$ represent the released energy of $1p$ radioactivity and $2p$ radioactivity, respectively. The true $2p$ radioactivity phenomenon was first reported in 2002, when $^{45}\rm{Fe}$ decayed into $^{43}\rm{Cr}$ with the emission of two protons, in two independent experiments conducted at GANIL\cite{PhysRevLett.89.102501} and GSI\cite{Edvice297}. Later on, the true $2p$ radioactivity of $^{19}\rm{Mg}$\cite{PhysRevLett.99.182501}, $^{48}\rm{Ni}$\cite{PhysRevC.72.054315}, and $^{54}\rm{Zn}$\cite{PhysRevLett.94.232501} was also observed in different experiments. Recently, in an experiment conducted with the BigRIPS separator at the RIKEN Nishina Center, the radioactivity of $^{67}\rm{Kr}$ was observed, showing a good agreement with the predictions of possible radioactivity candidates by theoretical models\cite{PhysRevLett.117.162501}. Moreover, the $2p$ radioactivity of the long-lived isomer $^{94m}\rm{Ag}$, whose parent nucleus has a very large deformation, was observed by Mukha in an experiment at GSI\cite{NatureAg94}. Experimental breakthroughs have enabled the observation of $2p$ radioactivity in various nuclei, validating theoretical predictions and allowing for detailed studies of decay properties such as half-lives and emission energies. 

\begin{figure*}[htb]\centering
\captionsetup{justification=justified, singlelinecheck=false}
	\includegraphics[width=9cm]{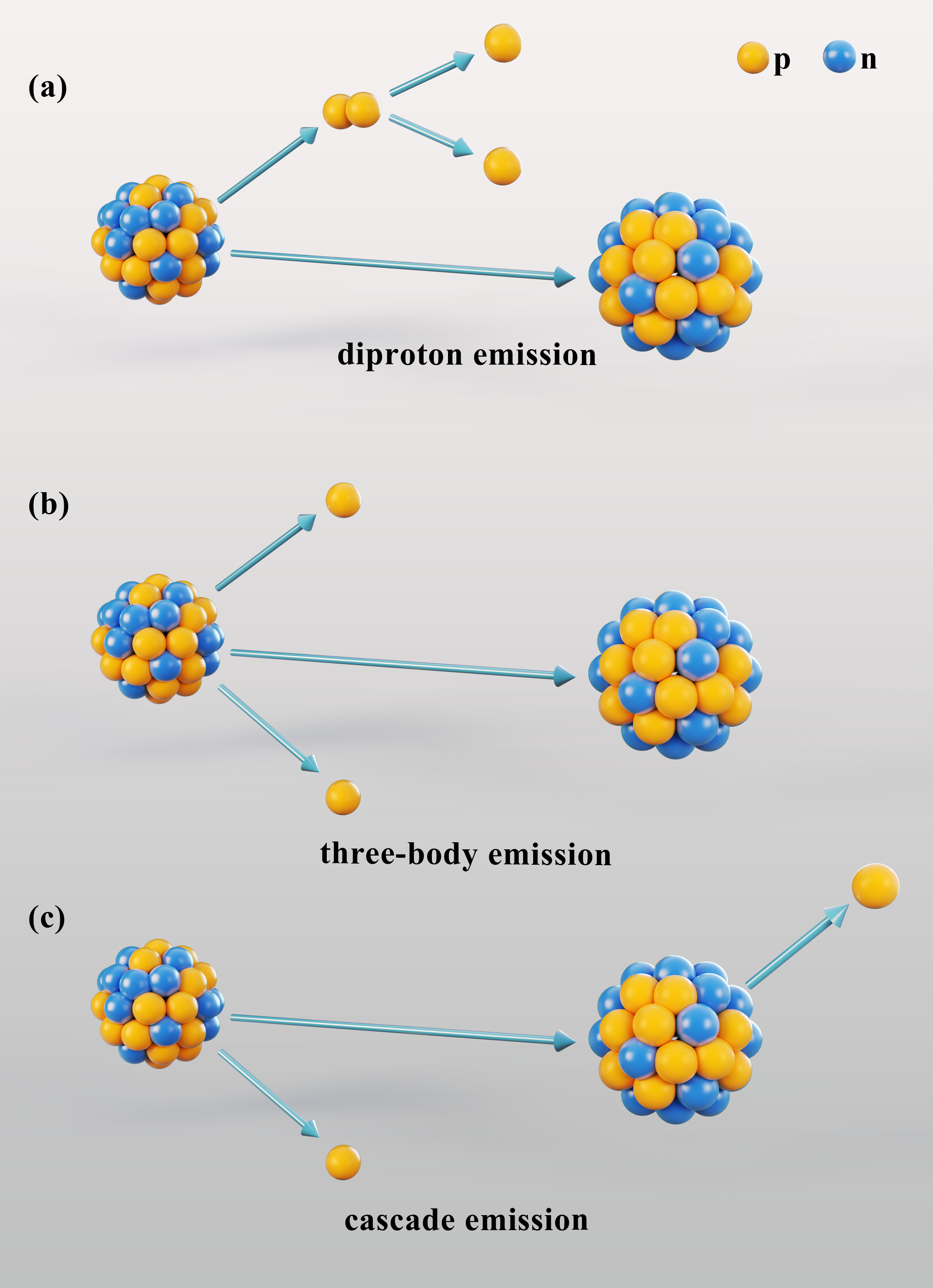}
	\caption{ Schematic diagram of the three different $2p$ emission mechanisms: (a) diproton emission, (b) three-body emission, (c) cascade emission.}
	\label{fig5.1.1}
\end{figure*}

Theoretically, the emission mechanism of $2p$ radioactivity can be divided into the following three types \cite{Fang-Deqing}, see figure \ref{fig5.1.1}: (a) diproton emission, (b) three-body emission, and (c) cascade emission. In the diproton emission, the emitted two protons are strongly correlated between each other and form a quasi-bound configuration with a notably short lifespan. This quasi-bound state then gets separated quickly due to the dominance of the Coulomb repulsion after escaping from the barrier. In the three-body emission, the emitted two protons and the daughter nucleus separate simultaneously, and the valence protons are located distantly in coordinated space and released with wide opening angles. For the cascade emission, the initial nucleus first decays to an intermediate state via $1p$ radioactivity. The final state is then formed by emitting the other proton from this intermediate state. Therefore, cascade emission can be viewed as a double sequential emission. Owing to the involvement of nucleon wave function configurations and nucleon-nucleon interaction in the two correlated protons, the first two mechanisms attract the interest of researchers. Based on this, a large number of diproton emission models have been proposed to study $2p$ radioactivity, including the density-dependent cluster model \cite{cone}, generalized liquid drop model \cite{ctwo}, unified fission model \cite{cthree}, Gamow-like model \cite{cfour}, three-body emission model \cite{cfive}, etc. These models aim to capture the complex dynamics between the emitted protons and the daughter nucleus, accounting for factors such as proton-proton correlations and nuclear deformation.

\subsection{Parameter dependence of laser-assisted diproton emission}

For the laser-assisted $2p$ radioactivity, the electric dipole term as described by the equation (\ref{eq:vic}) can be directly incorporated to modify the potential barrier of $2p$ emission. The electric dipole term encompasses important information on laser parameters, including the laser intensity and time profile evolution. Therefore, a normalized method for controlling variables can be employed to investigate how varying the laser parameters influences the laser-assisted $2p$ radioactivity.

\begin{figure*}[h]
\centering
\captionsetup{justification=justified, singlelinecheck=false}
\includegraphics[width=16cm]{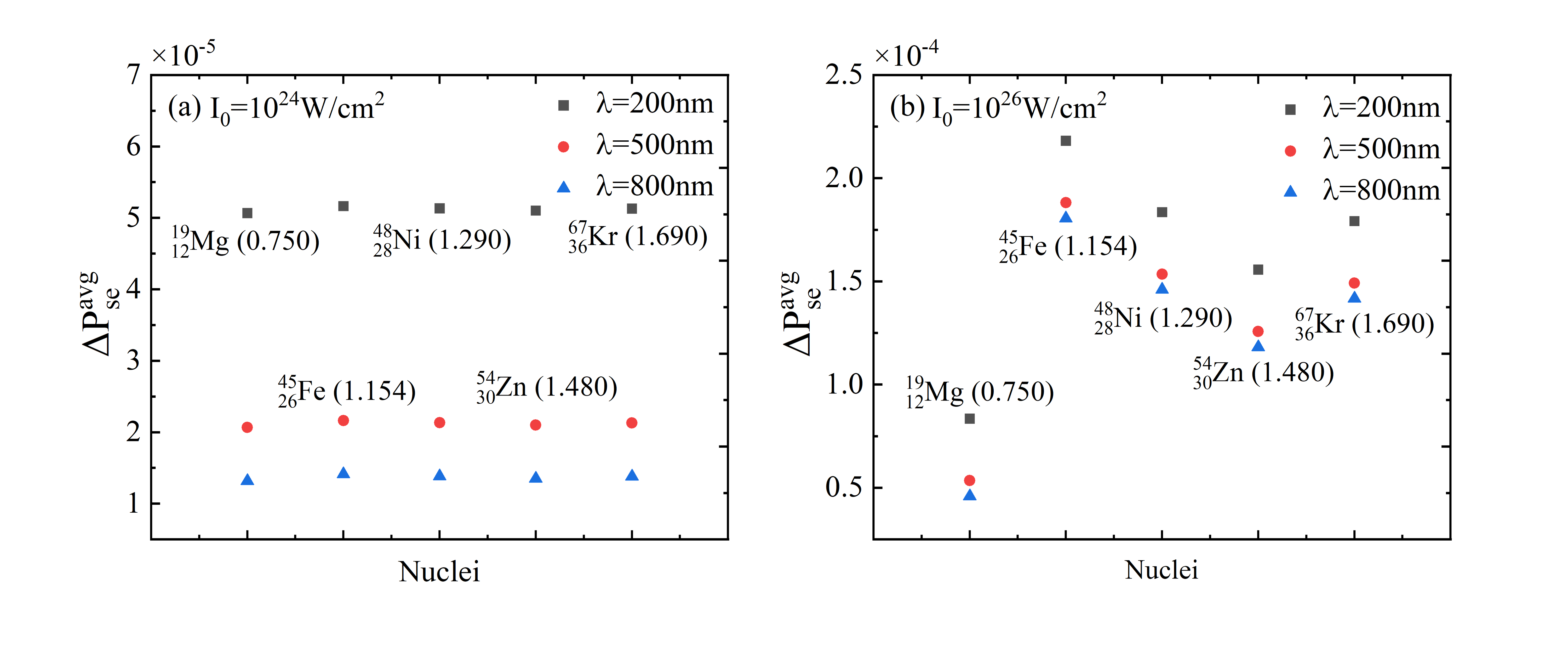}
\caption{ The influences of a laser pulse on $2p$ radioactivity penetration probability for different laser wavelength $\lambda_{0}$ at $I = 10^{24}\ \rm{W/cm^2}$ and $I = 10^{26}\ \rm{W/cm^2}$, respectively. The values provided in parentheses represent the experimentally determined $2p$ decay energy, measured in $\rm{MeV}$ \cite{Zou_2024}.}
\label{fig 5.2}
\end{figure*}

Recently, Zou \textit{et al.} systematically studied laser-assisted $2p$ radioactivity using a linearly polarized Gaussian laser within a deformed one-parameter nuclear calculation model \cite{Zou_2024}. Their findings indicated that the high-intensity laser field can not only affect the tunneling probability of $2p$ radioactivity but also influence the preformation of the $2p$-pair. The change in the half-life during the pulse duration was mainly attributed to the laser field’s effect on the tunneling probability. With $I$=$10^{ 24}\rm{\ W/cm^{2}}$ and $\lambda_{0}$=800\ $\rm{nm}$ laser, the maximum relative rate of change in half-life was reduced by 0.4\% for $^{45}\rm{\rm{Fe}}$ with $Q_{2p}$=1.100 MeV and the existing true $2p$ radionuclides are more susceptible to laser fields than not true $2p$ radionuclides, due to the longer tunneling path for true $2p$ radionuclides. In addition, they also investigated the effects of varying laser parameters on the laser-assisted $2p$ radioactivity. To characterize this effect, they introduced the average rate of change over a single pulse as
\begin{equation}
S_{2p}^{avg} = \frac{1}{\Delta t}\int_{t_1}^{t_2} \Delta S_{2p}(t)\rm{dt} , 
\label{eq6.23}
\end{equation}
\begin{equation}
P_{se}^{avg} = \frac{1}{\Delta t}\int_{t_1}^{t_2} \Delta P_{se}(t)\rm{dt} \ ,
\label{eq6.24}
\end{equation}
where $\Delta t = t_2-t_1$ is the pulse width, typically on the order of femtoseconds or picoseconds in high-intensity laser experiments. The calculations are presented in figure \ref{fig 5.2}. The results indicated that $\Delta P_{se}^{avg}$ is more pronounced with lasers of shorter wavelengths and higher intensities over a laser pulse duration. For the preformation probability, despite the fact that the effect of the laser field is not a determining factor in changing the half-life, the preformation probability encapsulates critical information about the nuclear structure, including the strong nuclear force governing nucleon-nucleon interactions at the femtometer scale \cite{PhysRevC.110.044321,PhysRevC.91.014322}, as well as shell closure effects \cite{zhao2018alpha,deng2016alpha}. The limited direct influence of laser field on the nuclear structure arises from the competition between the strong nuclear force and electromagnetic interactions. Thus, the authors also investigated the $\Delta S_{2p}^{avg}$ under different laser wavelengths and peak intensities over a pulse duration. Their results show that the shorter wavelength laser is more likely to affect the preformation probability of $2p$-pair, and the rate of average change in preformation probability seems to be irrelevant to the laser peak intensity. It is worth noting that the laser intensity utilized currently is far below the threshold for modulating intranuclear interactions, and the frequency detuning between the laser field and nuclear transitions precludes effective perturbative coupling. These two factors jointly render dynamic Stark splitting insufficient to significantly modify the $2p$-pair preformation probability, thus, the static $S_{2p}$ assumption remains valid. These findings indicate the regulation of nuclear structures by different laser wavelengths. More recently, Kaur \textit{et al.} studied the laser-assisted $2p$ radioactivity of nuclei $^{52}\rm{\rm{Zn}}$, $^{42}\rm{\rm{Cr}}$ and $^{34}\rm{\rm{Ca}}$ using a realistic short-range nuclear potential \cite{innet}. It is shown that for $^{42}\rm{\rm{Cr}}$ greater modifications are found in this nucleus when compared to other ones since $^{42}\rm{\rm{Cr}}$ has the lowest decay energy among them. It is worth noting that previous studies on laser-assisted $2p$ radioactivity have primarily considered the emission as diproton emission. However, in actual emission modes, all three mechanisms can co-exist. Therefore, the research on laser-assisted $2p$ radioactivity can serve as an excellent means to explore proton-proton correlations and the structure of proton-rich nuclei. Nevertheless, further investigation into the three-body emission in the laser field is still required.
\par Despite these theoretical advancements, experimentally verifying the effects of laser-assisted $2p$ radioactivity remains challenging. The technical difficulties of synchronizing high-intensity laser pulses with radioactive nuclei and detecting subtle changes in decay properties present significant obstacles. Moreover, comprehensive theoretical frameworks that accurately predict and describe laser-induced modifications in $2p$ radioactivity are still in development.

\section{Theory of nuclear excitation with lasers}\label{Sixth}

High power lasers have revolutionized our ability to investigate nuclear excitation behavior by providing a highly controllable and intense electromagnetic field environment. These fields enable the creation of extreme plasma environments and laser-driven secondary particles, thereby activating various nuclear excitation mechanisms, especially electron-mediated pathways. In special cases, such as the low-lying isomeric transition in $^{229}$Th, direct laser excitation of nuclear states may also become possible. Theoretical studies in this domain aim to bridge the gap between atomic and nuclear physics, exploring how the interplay of laser intensity, frequency, and duration, as well as laser-driven secondary particles, especially electrons and $\gamma$-photons influence the nuclear transitions. This chapter delves into the relevant theoretical frameworks, computational approaches, and fundamental mechanisms that underpin our understanding of nuclear excitation in laser fields, covering both direct and electron-mediated pathways.

The energy level structure of nuclei  plays a vital role in understanding laser-induced nuclear excitation mechanisms, as depicted in figure \ref{figlevel}, which illustrates key nuclides like \(^{83}\)Kr, \(^{45}\)Sc, \(^{93}\)Mo, and \(^{229}\)Th with excitation energy, spin-parity, and half-life. These parameters define transition feasibility and efficiency under laser fields: the low-lying isomeric state (8.36 eV) of \(^{229}\)Th enables precise coherent manipulation critical for nuclear clock applications, whereas the higher-energy isomeric state of $^{93}$Mo illustrates the relevance of electron-mediated mechanisms for accessing nuclear transitions far beyond the optical-energy range.

This energy level information is essential for constructing theoretical frameworks, in which the transition probability between nuclear states is determined by the properties of these levels. It bridges abstract theoretical models and specific nuclear properties, facilitating a deeper understanding of how laser fields induce nuclear excitations via various pathways and enabling systematic exploration of the interaction between laser parameters and nuclear state responses during excitation.

\begin{figure}[p]
    \centering
    	\captionsetup{justification=justified, singlelinecheck=false}
    \begin{subfigure}[b]{\textwidth}
        \centering
        \includegraphics[width=14cm]{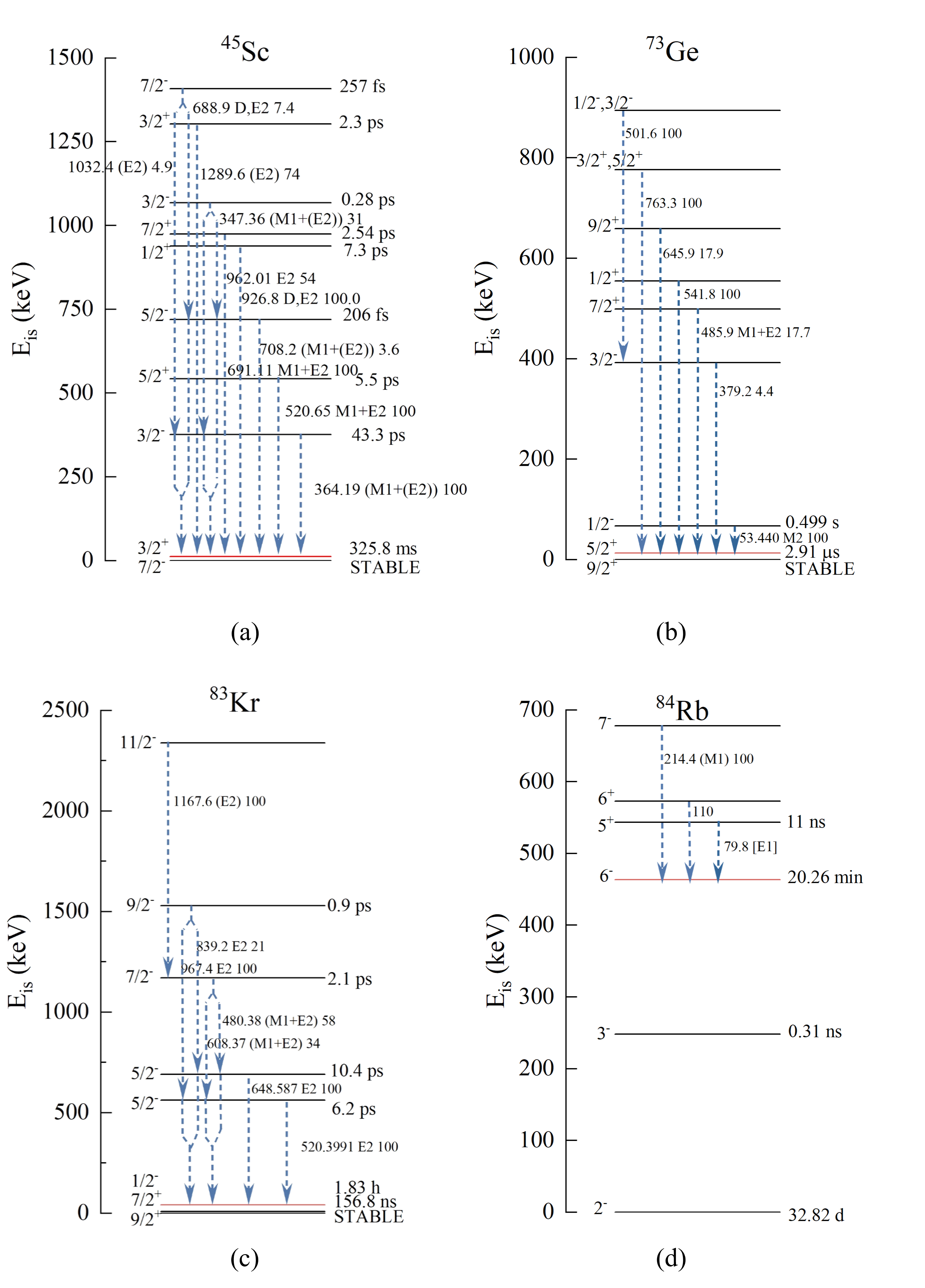}
        \label{fig:subfig1}
    \end{subfigure}
    \label{xx}
\end{figure}

\begin{figure}[p]
    \ContinuedFloat
    \centering
        	\captionsetup{justification=justified, singlelinecheck=false}
    \begin{subfigure}[b]{\textwidth}
        \centering
        \includegraphics[width=14cm]{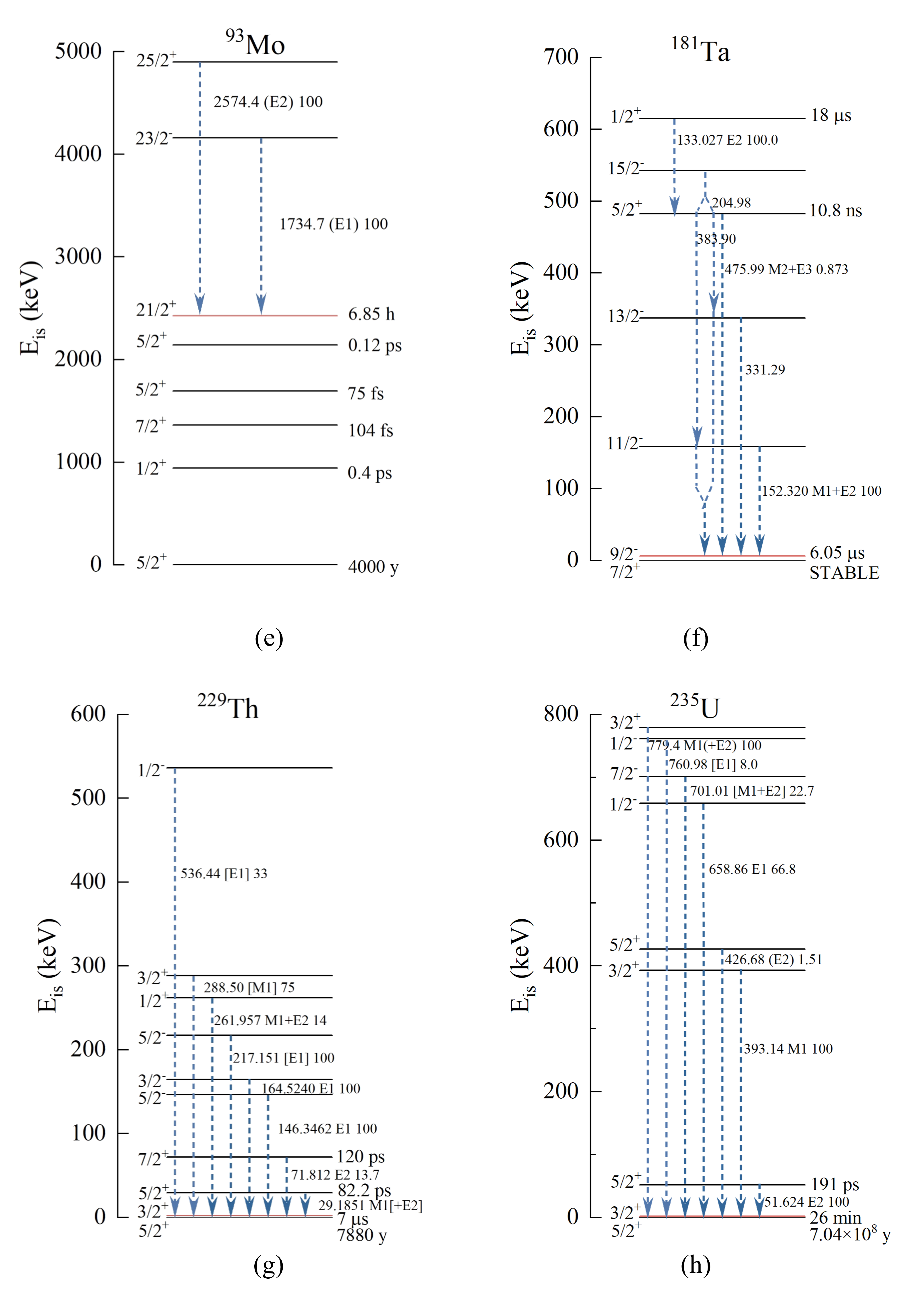}
        \label{fig:subfig2}
    \end{subfigure}
    \caption[]{The partial relevant energy levels of the excited states of different nuclei. The red lines denote the nuclear isomer states. Spin-parity is presented on the left side of the energy levels, while the half-life of some energy levels is shown on the right side. The blue dashed arrows indicate the decay of higher excited states into the isomer state, with the adjacent numerical values indicating the $\gamma$-ray energy, transition type, and relative intensity, respectively. All data available are from reference \cite{NNDC2025}, except for the $^{229}$Th partial energy level half-life from reference \cite{PhysRevC.104.024306}.}
    \label{figlevel}
\end{figure}

\subsection{Theoretical models and computational approaches}

The study of nuclear excitation in laser fields necessitates robust theoretical models and computational approaches capable of capturing the relativistic and quantum nature of these processes. These models and methods play a crucial role in predicting experimental observations and guiding future research, encompassing those from the description of the electronic environment in highly charged ions to the modeling of the complex dynamics of electron-nucleus interactions. In this subsection, two pivotal computational approaches are introduced: the Dirac Hartree-Fock-Slater (DHFS) method \cite{PhysRevA.36.467,EPJDAN,CMPAN} and the Dirac distorted wave Born approximation (DWBA) \cite{PhysRevC.106.044604,PhysRevC.106.064604}. The DHFS method is primarily employed to obtain the static electronic structure and potentials of atomic systems, which in turn serve as essential inputs for DWBA calculations that describe the dynamic scattering processes involving electrons and nuclei. Each offers unique insights into different aspects of laser-induced nuclear excitation. It should be noted that the theoretical models introduced in this subsection are primarily applied to electron-mediated excitation mechanisms.

\subsubsection{Dirac Hartree-Fock-Slater method}

{\ }

{\ }

\noindent The DHFS method is a foundational approach for solving the relativistic Dirac equation in systems where relativistic effects significantly influence atomic and nuclear properties. It is particularly useful for heavy atoms and ions, where relativistic effects associated with high nuclear charge become significant. The DHFS method is built upon the central-field independent-electron approximation, where electrons move in a common central potential $V_{\mathrm{DHFS}}(r)$ \cite{PhysRevA.36.467}. This potential combines the contributions from the nuclear field, electron density, and an approximate exchange interaction, enabling the efficient computation of binding energies, wavefunctions, and electron distributions \cite{EPJDAN,CMPAN}.

The electronic states in the DHFS framework are represented by single Slater determinants, constructed from central-field orbitals of the form
\begin{equation}
\psi_{n \kappa m}(\mathbf{r})=\frac{1}{r}\left(\begin{array}{c}
P_{n \kappa}(r) \Omega_{\kappa  m}(\hat{\mathbf{r}})\\
\mathrm{i} Q_{n \kappa}(r) \Omega_{-\kappa m}(\hat{\mathbf{r}})
\end{array}\right),
\end{equation}

where $P_{n \kappa}(r)$ and $Q_{n \kappa}(r)$ are the radial wavefunctions of the upper and lower components of the Dirac spinor, respectively. The functions $\Omega_{\kappa m}(\hat{\mathbf r})$ and $\Omega_{-\kappa m}(\hat{\mathbf r})$ represent the spherical spinors associated with the angular part of the wavefunction, where $\kappa$ is the relativistic quantum number, and $m$ is the magnetic quantum number. These spherical spinors encapsulate both the orbital and spin angular momentum of the electron. These orbitals satisfy the DHFS equations, given by
\begin{equation}
\left[
c\bm{\alpha}\cdot\mathbf p
+
(\beta-1)m_e c^2
+
V_{\mathrm{DHFS}}(r)
\right]
\psi_{n\kappa m}(\mathbf r)
=
E_{n\kappa}\psi_{n\kappa m}(\mathbf r),
\end{equation}
where $E_{n \kappa}$ are the energy eigenvalues of the system, representing the discrete energy levels associated with the quantum states characterized by the quantum numbers $n$ and $\kappa$. The term $c \bm{\alpha} \cdot \mathbf{p}$ corresponds to the relativistic kinetic energy operator, with $\bm{\alpha}$ being the vector of Dirac matrices and $\mathbf{p}$ being the momentum operator. The term $(\beta-1)m_e c^2$ accounts for the relativistic mass-energy contribution after subtracting the electron rest energy, where $\beta$ is the Dirac matrix associated with the time-like component of the Dirac equation. This potential is derived self-consistently by replacing the non-local Hartree-Fock exchange potential with the Slater approximation by assuming a local, density-dependent exchange interaction inspired by the Thomas-Fermi model \cite{PhysRev.81.385}.

The DHFS potential is composed of three terms: the nuclear potential $V_{\mathrm {nuc}}(r)$, the electronic potential $V_{\mathrm{el}}(r)$, and the exchange potential $V_{\mathrm{ex}}(r)$. The nuclear potential accounts for the interaction of the electron with the spherically symmetric nuclear charge distribution. For a finite-sized nucleus, the Fermi distribution describes the proton density $\rho_{\mathrm{p}}(r)$, parameterized by nuclear radius $R_n=1.07 A^{1/3}\ \mathrm{fm}$ and diffuseness $z=0.546\ \mathrm{fm}$ \cite{PhysRev.101.1131,SALVAT2019165}. The corresponding nuclear potential is calculated numerically by  $V_{\mathrm{nuc}}(r)
=
-\frac{e^2}{r}
\int_0^r \rho_{\mathrm p}(r^{\prime})4\pi r^{\prime 2}\,\mathrm d r^{\prime}
-
e^2
\int_r^\infty \rho_{\mathrm p}(r^{\prime})4\pi r^{\prime}\,\mathrm d r^{\prime}$.
The electronic potential reflects the interaction between the electron at $r$ and the surrounding electron cloud, given by
$
V_{\mathrm{el}}(r)
=
\frac{e^2}{r}
\int_0^r \rho(r^{\prime})4\pi r^{\prime 2}\,\mathrm d r^{\prime}
+
e^2
\int_r^\infty \rho(r^{\prime})4\pi r^{\prime}\,\mathrm d r^{\prime},
$
where $\rho(r)$ is the electron density.
Finally, the exchange potential approximates electron-electron correlations using the Thomas-Fermi model \cite{PhysRev.81.385}
$
	V_{\mathrm{ex}}(r)=C_{\mathrm{ex}} V_{\mathrm{ex}}^{(\mathrm{TF})}(r),
$
where $C_{\mathrm{ex}}=1.5$ and $V_{\mathrm{ex}}^{(\mathrm{TF})}(r)$ is the exchange potential for a free-electron gas, defined as
$ V_{\mathrm{ex}}^{(\mathrm{TF})}(r)=-e^2\left(\frac{3}{\pi}\right)^{1 / 3}[\rho(r)]^{1/3} .
$
To ensure the correct asymptotic behavior of 
$V_{\mathrm {DHFS}}(r)$, a correction known as the Latter tail is applied \cite{PhysRev.99.510}. This correction ensures that the potential remains continuous and enforces physical consistency at large distances, which is crucial for accurately describing the long-range interactions in atomic systems. By incorporating these approximations and iterative self-consistency, the DHFS method provides an efficient and robust approach for calculating key atomic properties, such as electronic structure, binding energies, and transition probabilities. This facilitates investigations into the nuclear excitation, inelastic scattering, and other complex processes \cite{PhysRevC.110.064621}.

\subsubsection{Dirac distorted wave Born approximation}
{\ }

{\ }

\noindent The DWBA is a robust theoretical framework widely used in nuclear and atomic physics to investigate relativistic scattering and excitation processes \cite{PhysRevC.106.044604,PhysRevC.106.064604}. As a natural extension of the Born approximation \cite{https://doi.org/10.1155/2010/367180}, the DWBA accounts for the distortion of wavefunctions caused by the central potential of the target nucleus or atom, providing a more accurate description of the interaction. This approach is particularly suitable for high-energy particles, where relativistic effects such as spin-orbit coupling and fine-structure splitting become critical. The relativistic formulation of the DWBA ensures consistency with the Dirac equation, making it indispensable for systems involving heavy nuclei or high-energy projectiles.

In the DWBA, the motion of the incident particle is described by a distorted wavefunction, which is a solution to the Dirac equation in the presence of an effective central potential. The potential 
 $V(r)$ includes the Coulomb interaction for charged particles and, when relevant, additional nuclear potentials to account for the short-range interactions. The distorted wavefunctions $\psi_i^{(+)}(\mathbf r)$ and $\psi_f^{(-)}(\mathbf r)$, corresponding to the initial and final states, incorporate the influence of $V(r)$, ensuring that the scattering process is accurately modeled. The transition amplitude is given by the matrix element
\begin{equation}
T_{fi}
=
\int
\left[\psi_f^{(-)}(\mathbf r)\right]^{\dagger}
V_{\mathrm{int}}(\mathbf r)
\psi_i^{(+)}(\mathbf r)
\,\mathrm d^3 r,
\end{equation}
where $V_{\mathrm{int}}(\mathbf r)$ is the interaction potential responsible for the excitation or scattering process. This interaction potential is treated as a perturbation, and the distorted wavefunctions encapsulate the non-perturbative effects of the central potential. By using the distorted wavefunctions, DWBA provides a more accurate description of the scattering process than the plane wave Born approximation, especially in the presence of strong or long-range potentials.

The DWBA is particularly advantageous for relativistic systems, as it naturally incorporates spin-orbit coupling and relativistic corrections that arise in high-energy or heavy-ion scattering. For example, it has been successfully applied to calculate excitation cross sections in electron-nucleus scattering, where low-lying nuclear states, such as the isomeric state in $^{229}\mathrm{Th}$, are of interest \cite{PhysRevC.106.044604}. Additionally, DWBA has been used to model nuclear excitation processes involving heavy ions, providing insights into the influence of relativistic effects on the transition probabilities and differential cross sections \cite{PhysRevC.106.064604}.

Despite its capacity, the DWBA's limitations are obvious. The accuracy of the method relies on the validity of the distorted wavefunctions, which are solutions to the Dirac equation with the chosen central potential. For systems with highly deformed targets or strongly coupled states, the distorted wavefunctions may not fully capture the complexity of the interactions. Furthermore, the perturbative treatment of $V_{\mathrm{int}}(\mathbf r)$ assumes that this interaction is sufficiently weak, which may not hold for certain strong coupling scenarios. In such cases, alternative approaches such as coupled-channel methods \cite{GRANDE1997264} or non-perturbative techniques \cite{KMomberger1995} may be required. In conclusion, the DWBA offers a robust framework for studying relativistic scattering and excitation processes, particularly in systems dominated by relativistic effects. By accurately modeling wavefunction distortions and incorporating relativistic corrections, it provides reliable predictions for transition amplitudes and cross sections. These predictions have been critical in analyzing electron-nucleus scattering and heavy-ion collisions. Despite its limitations in strongly coupled or highly deformed systems, the DWBA remains a key tool in theoretical studies, bridging experimental observations and theoretical models in nuclear and atomic physics.

\subsection{Photon-mediated nuclear excitation}

Laser-induced nuclear excitation involves several different mechanisms, each leveraging different aspects of the interaction between the laser field, electrons, and nuclei. These mechanisms include not only direct photon absorption by the nucleus but also more complex processes mediated by electron transitions or collisions. Understanding these mechanisms is essential for designing experiments aimed at probing nuclear states or inducing controlled transitions. This subsection focuses on two key pathways: direct laser excitation, which relies on the resonance between the nuclear transitions and laser photons, and photoexcitation via blackbody radiation and bremsstrahlung, which explores broader spectral contributions in dynamic environments.

\subsubsection{Direct laser excitation}

{\ }

{\ }

\noindent Direct laser excitation is a fundamental mechanism in nuclear excitation, wherein photons directly interact with nuclei to induce transitions between nuclear energy levels. This process exploits the interaction between the electromagnetic field of a laser and the nucleus, enabling precise control over nuclear states. Unlike electronic transitions, which occur at much lower photon energies, nuclear transitions typically involve $\gamma$-ray or vacuum ultraviolet (VUV) photons due to the significant energy scales of nuclear level spacings \cite{Zhang2024}. Recent advances in laser technology, particularly in high-intensity and high-power lasers, have opened new avenues for exploring direct laser excitation in nuclei.

The interaction between the high-intensity laser field and the nucleus is governed by the multipole expansion of the electromagnetic interaction Hamiltonian. The transition probability between initial and final nuclear states is determined by the matrix elements of $V$, which depend on the laser's frequency, polarization, and intensity. In particular, the resonance condition requires that the laser photon energy $\hbar\omega$ matches the nuclear transition energy

\begin{equation}
	\hbar\omega=\hbar\omega_0=E_f-E_i,
\end{equation}
where $E_i$ and $E_f$ are the energies of the initial and final nuclear states, respectively. In the case of a large detuning $\Delta=\omega-\omega_0$ between the laser frequency and the nuclear transition frequency, the excitation probability is strongly suppressed. This suppression can be partially compensated by using high-intensity lasers, although the resonance condition remains the most favorable case \cite{Dzyublik2007}. For a laser field described by the vector potential $A(\mathbf r,t)$, the interaction Hamiltonian can be further expressed as \cite{Dzyublik2010}

\begin{equation}
V(t)=\frac{1}{c}\int \mathbf j_n(\mathbf r)\cdot \mathbf A(\mathbf r,t)\,\mathrm d^3 r,
\end{equation}
where $\mathbf j_n(\mathbf r)$ is the nuclear current density. The Hamiltonian governing laser-nucleus interaction can be expanded into electric and magnetic multipole components, depending on the transition type. The nuclear transition probability is governed by the electromagnetic multipolarity 
$\lambda$, photon wave vector $\mathbf{k}$, nuclear spin configurations, and the reduced transition probabilities $B\left(M\lambda\right)$ or $B\left(E\lambda\right)$. These matrix elements can be evaluated using standard angular momentum algebra and are well-documented in the literature \cite{PhysRevC.77.044602}.

The amplitude of the nuclear transition from the ground state $|i\rangle=\left|I_i M_i\right\rangle$ to the isomeric state $|f\rangle=\left|I_f M_f\right\rangle$ under first-order perturbation theory is

\begin{equation}
	b_{fi}(t)
	=
	-\frac{i}{\hbar}
	\int_0^t
	\left\langle I_f M_f\right|
	V(t^{\prime})
	\left|I_i M_i\right\rangle
	e^{i\omega_0 t^{\prime}}
	\,\mathrm d t^{\prime}.
\end{equation}
By averaging over the initial states and summing over final states we get the total excitation probability.
\begin{equation}
	P_{i\to f}(t)
	=
	\frac{1}{2I_i+1}
	\sum_{M_i M_f}
	\left|b_{fi}(t)\right|^2 .
\end{equation}

Direct laser excitation efficiency depends on the laser's intensity, bandwidth, and coherence, as well as the interaction cross section of the nuclear transition. High-intensity lasers are characterized by strong electric fields, enhancing coupling to nuclear multipole moments. Narrow-bandwidth lasers with high spectral purity are essential for resonant excitation, minimizing off-resonant effects and maximizing transition probabilities. However, the feasibility of direct laser excitation is limited by the typically low interaction cross section and the challenge of achieving the necessary photon energies. For a femtosecond laser pulse with a wavelength of 800 nm , an intensity of $10^{14} \mathrm{~W} / \mathrm{cm}^2$, and a Gaussian temporal profile with a FWHM of 30 fs , the probability of exciting a single nucleus is estimated to be on the order of $10^{-14}$ \cite{PhysRevLett.130.112501}. This probability is negligible compared to mechanisms like NEEC or NEIES. The latter can achieve excitation probabilities on the order of $10^{-7}$.

Despite these challenges, direct laser excitation remains critical for nuclei like $^{229}\mathrm{Th}$ with low-lying isomeric states. Ongoing advancements in laser technology, particularly in free-electron lasers (FEL) and high-intensity VUV sources, are improving the practicality of this mechanism, paving the way for future applications in nuclear clocks and precision measurements \cite{Zhang2024}.

\subsubsection{Photoexcitation via blackbody radiation and bremsstrahlung}

{\ }

{\ }

\noindent Photoexcitation via blackbody radiation and bremsstrahlung involves the interaction of a broad spectrum of photons, generated by thermal or dynamic processes, with nuclei to induce transitions between nuclear energy states. This mechanism relies on the copious number of photons produced, which can occasionally match the energy difference required for nuclear excitation. While these processes are less efficient than NEEC or NEIES, they are still relevant in specific contexts, such as laser-heated clusters \cite{PhysRevLett.130.112501}.

In a laser-heated cluster, photons are generated through both blackbody radiation and bremsstrahlung, which possess broad spectral distributions. Blackbody radiation follows the Planck distribution, where the photon flux $\phi_\gamma^{\mathrm {bb}}\left(\varepsilon, T_e\right)$ is given by

\begin{equation}
	\phi_\gamma^{\mathrm{bb}}(\varepsilon,T_e)\,\mathrm d\varepsilon
	=
	\frac{\varepsilon^2\,\mathrm d\varepsilon}
	{\pi^2 c^2\hbar^3\left(e^{\varepsilon/T_e}-1\right)},
\end{equation}
where $T_e$ is the electron temperature expressed in energy units, and $\varepsilon$ is the photon energy. As the system heats up, the spectrum shifts toward higher photon energies, increasing the overlap with nuclear transition energies, particularly for low-energy nuclear states. Bremsstrahlung, on the other hand, arises from the deceleration of electrons in the Coulomb field of ions. The photon flux for bremsstrahlung is determined by the interaction cross section and the electron distribution \cite{PhysRevE.97.063205}

\begin{equation}
	\phi_\gamma^{\mathrm {bt}}\left(\varepsilon, T_e, n_e\right) \mathrm{d} \varepsilon=\int_{E_e} \frac{\mathrm{d} \sigma_{\mathrm {bt}}}{\mathrm{d} \varepsilon} \phi_e\left(E_e, T_e, n_e\right) \mathrm{d} E_e \mathrm{d} \varepsilon,
\end{equation}
where $n_e$ is the electron density, $f\left(E_e\right)$ is the electron energy distribution with $E_e$ denoting the electron kinetic energy, $v_e\left(E_e\right)$ is the electron velocity, and $\phi_e\left(E_e, T_e, n_e\right)=n_e f\left(E_e\right) v_e\left(E_e\right)$ is the electron flux.

For blackbody radiation, substituting the Planck distribution into the photon flux yields the photoexcitation rate \cite{RevModPhys.31.920} 

\begin{equation}
	W_\gamma^{\mathrm {bb}}=\frac{2 \pi^2}{k^2} \frac{2 I_f+1}{2 I_i+1} \Gamma_\gamma^{f \rightarrow i} \phi_\gamma^{\mathrm {bb}}\left(\Delta E, T_e\right) .
\end{equation}
For bremsstrahlung, the rate is similarly expressed, with the photon flux derived from the electron-ion interactions

\begin{equation}
	W_\gamma^{\mathrm {bt}}=\frac{2 \pi^2}{k^2} \frac{2 I_f+1}{2 I_i+1} \Gamma_\gamma^{f \rightarrow i} \phi_\gamma^{\mathrm {bt}}\left(\Delta E, T_e, n_e\right) .
\end{equation}
During the expansion and cooling of the laser-heated cluster, the electron temperature decreases, leading to a time-dependent photoexcitation rate. The total excitation probability for a nucleus during this process is given by

\begin{equation}
	P_{i\to f}(t)
	\simeq
	\int_0^t W_\gamma(t^{\prime})\,\mathrm d t^{\prime},
\end{equation}
which accounts for the contributions from both blackbody radiation and bremsstrahlung. However, these contributions are minimal due to the relatively low photon flux and the mismatch between the photon energy spectrum and the nuclear transition energy. Taking a laser pulse with a peak intensity of I=$10^{14} \mathrm{~W} / \mathrm{cm}^2$ for example, the photoexcitation probabilities range between $10^{-17}$ and $10^{-16}$, approximately 10 orders of magnitude lower than the probabilities achieved via NEEC or NEIES. This stark contrast illustrates the inefficiency of photoexcitation via blackbody radiation and bremsstrahlung under typical experimental conditions.

While these mechanisms are generally negligible compared to more efficient electron-mediated excitation processes, their broad spectral distributions make them relevant in environments such as astrophysical settings or laser-driven plasmas \cite{PhysRevLett.130.112501}. In such contexts, blackbody radiation and bremsstrahlung provide complementary tools for studying nuclear excitation, offering a more comprehensive understanding of nuclear interactions in complex systems. Their relevance may extend further as experimental environment and laser technology continue to evolve, potentially enabling more precise investigations of nuclear transitions via direct laser excitation.

\subsection{Electron-coupled nuclear excitation}

\begin{figure*}[htb]
\captionsetup{justification=justified, singlelinecheck=false}
  \centering
	\includegraphics[width=9.5cm]{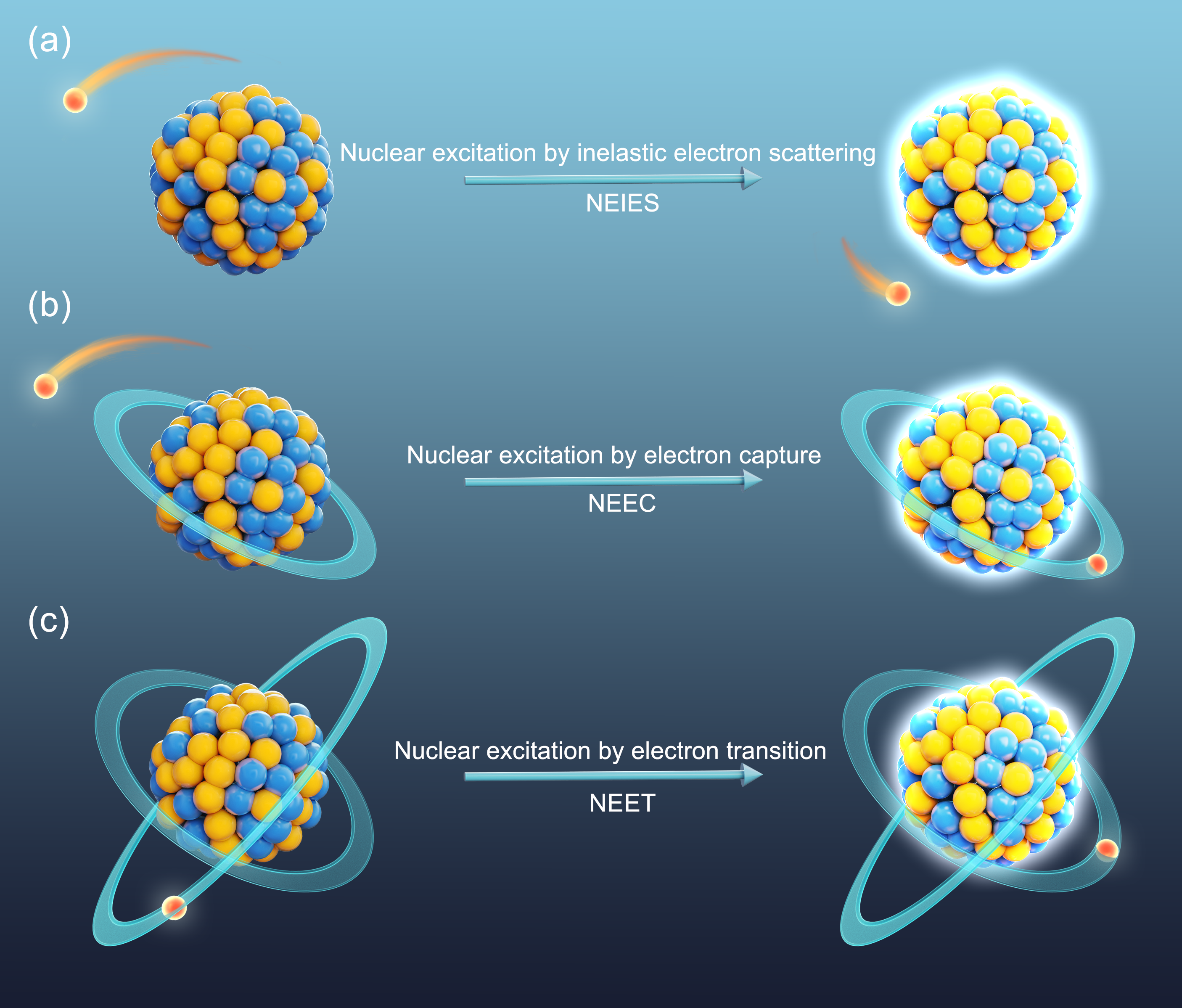}
	\caption{Schematic representation of NEET, NEEC, and NEIES processes. NEET involves bound-bound electronic transitions, NEEC entails free-bound transitions, and NEIES operates through inelastic free-free electron scattering. In each process, the nucleus is excited from an initial nuclear state to a higher-lying excited state, such as an isomeric state, through energy transferred from the electronic system.}
	\label{fig7.3.1}
\end{figure*}

Electron-coupled nuclear excitation offers an indirect yet highly efficient pathway to nuclear transitions, leveraging the interactions between electrons and nuclear states. These processes, as shown in figure \ref{fig7.3.1}, include NEEC, NEIES, and NEET, and are particularly significant in scenarios where direct photon-nucleus coupling is inefficient or impractical. By utilizing the energy transfer facilitated by electron dynamics, these mechanisms substantially broaden the parameter space for nuclear excitation, especially under controlled conditions such as laser-heated clusters and ion traps \cite{PhysRevLett.130.112501,10.3389/fphy.2023.1203401}. While other electron-mediated processes such as the electronic bridge (EB) \cite{PhysRevA.81.042516,PhysRevLett.105.182501,Bilous_2018,PhysRevC.100.044306,PhysRevC.102.024604,PhysRevC.106.064608,PhysRevLett.124.192502,PhysRevLett.125.032501,PhysRevA.111.L041103,PhysRevC.110.054307} also hold promise, this subsection focuses on NEEC, NEIES, and NEET because of their well-established theoretical formulations and their relevance to laser-driven plasma and atomic-nuclear coupling scenarios. Beyond their computational accessibility, these mechanisms reveal novel pathways for manipulating nuclear states with high selectivity and energy resolution. Their ability to facilitate resonance-matched nuclear transitions through atomic electron dynamics makes them a promising platform for both experimental verification and technological application. For instance, these processes enable precise determination of nuclear matrix elements, lifetimes, and branching ratios—quantities essential for nuclear structure studies. Moreover, in suitable atomic-nuclear systems, related electron-mediated coupling mechanisms may provide sensitive probes for testing fundamental symmetries, such as parity violation and time-reversal invariance, offering possible insights into beyond-standard-model physics \cite{PhysRevLett.128.052501,pnas2413221121}.

\subsubsection{Nuclear excitation by electron capture (NEEC)}

{\ }

{\ }

\noindent NEEC is a resonant process first proposed by Goldanskii and Namiot in the 1970s \cite{GOLDANSKII1976393}. It involves the capture of a free electron into a bound atomic state, where the electron transfers its energy directly to the nucleus, exciting it to a higher energy state. Unlike photon-mediated nuclear excitation, NEEC relies on the interaction between the electron, nucleus, and the quantized radiation field, without the emission of real photons. The electron provides the necessary energy for the nuclear transition via virtual photon exchange, making NEEC a powerful mechanism for coupling atomic and nuclear systems \cite{GOLDANSKII1976393,CUE198925,PhysRevC.47.323,PhysRevC.59.2462,PhysRevA.73.012715,PhysRevLett.112.082501,chiara2018isomer,PhysRevLett.122.212501,PhysRevLett.128.242502,Yang2025,xu2025chargestateregulationnuclear}.

The system's evolution during NEEC is governed by the interaction Hamiltonian $V$, which drives the transition from the initial state $|i\rangle=\left|I_i M_i\right\rangle \otimes\left|\phi_i\right\rangle \otimes|0\rangle$ to the final state $|f\rangle=$ $\left|I_f M_f\right\rangle \otimes\left|\phi_f\right\rangle \otimes|0\rangle$. Here, $\left|\phi_i\right\rangle$ is the free electron state, $\left|\phi_f\right\rangle$ is the bound atomic state, and $|0\rangle$ represents the vacuum state of the radiation field. It should be noted that $|f\rangle$ is the final state only for the initial energy-transfer process between the electron and the nucleus. Subsequently, the excited nuclear level $|I_fM_f\rangle$ may decay by emitting a $\gamma$ quantum, and the whole system reaches the final state $|b\rangle$. The complete NEEC process is then described by the transition matrix
	
\begin{equation}
		T_{bi}
		=
		\sum_f
		\frac{V_{bf}^{r}V_{fi}}
		{E_i-E_f+i\Gamma_f/2},
\end{equation}
\noindent where $V_{bf}^{r}$ is the interaction operator describing the radiative decay of the excited nuclear state, $V_{fi}$ defines the strength of the energy transfer from the electron to the nucleus, and $E_f$ and $\Gamma_f$ are the energy and total width of the resonant state $|f\rangle$, respectively. Substituting this expression into the standard formulas of scattering theory gives the transition rate from the state $|i\rangle$ to the resonant state $|f\rangle$ as

\begin{equation}
\omega_{i\rightarrow f}
		=
		\frac{2\pi}{\hbar}
		\left|V_{fi}\right|^2
		\frac{\Gamma_f/2\pi}
		{(E_i-E_f)^2+\Gamma_f^2/4}.
\end{equation}

In Coulomb gauge, the interaction Hamiltonian $V_r$ includes contributions from nuclear currents, electronic currents, and the Coulomb interaction between the nucleus and the electron

\begin{equation}
V_r
=
-\frac{1}{c}
\int
\left[
\mathbf j_n(\mathbf r)+\mathbf j_e(\mathbf r)
\right]\cdot \mathbf A(\mathbf r)\,
\mathrm d^3 r
+
\int
\frac{\rho_n(\mathbf r)\rho_e(\mathbf r')}
{|\mathbf r-\mathbf r'|}
\,\mathrm d^3 r\,\mathrm d^3 r',
\end{equation}
where $\mathbf j_n$ and $\mathbf j_e$ are the nuclear and electronic current densities, respectively, and $\rho_n$ and $\rho_e$ are the corresponding charge densities. $V_{f i}$ is expanded in terms of electric $(E)$ and magnetic $(M)$ multipole components of the field \cite{RevModPhys.28.432}

\begin{equation} V_{f i} = \sum_{\lambda \mu} \frac{4 \pi}{2 \lambda+1}(-1)^\mu \big[ \mathcal{T}_E-\mathcal{T}_M \big], 
\end{equation}

\begin{equation} \mathcal{T}_E = \langle\phi_f| \mathcal{N}(E \lambda, \mu)|\phi_i\rangle \langle I_f M_f| \mathcal{M}(E \lambda,-\mu)|I_i M_i\rangle, 
\end{equation}

\begin{equation} \mathcal{T}_M = \langle\phi_f| \mathcal{N}(M \lambda, \mu)|\phi_i\rangle \langle I_f M_f| \mathcal{M}(M \lambda,-\mu)|I_i M_i\rangle. 
\end{equation}
where $\mathcal{M}(T \lambda, \mu)$ and $\mathcal{N}(T \lambda, \mu)$ represent nuclear and electronic multipole operators, respectively, with $T=E, M$.

For an isolated resonance and a monoenergetic incident electron, the NEEC cross section may be written in a Lorentzian form as
\begin{equation} \sigma_{\mathrm{NEEC}}\left(E_i\right) = \frac{4 \pi^2}{c^2} \frac{E_i + m_e c^2}{p_i^3}\frac{\Gamma_{\mathrm{NEEC}}}{\left(E_i-E_r\right)^2 + \Gamma_{\mathrm{NEEC}}^2 / 4} \mathcal{S}, 
\end{equation}
\begin{equation} \mathcal{S} = \sum_{\mathcal{T} \lambda} B\left(\mathcal{T} \lambda ; I_i \rightarrow I_f\right) \frac{k^{2 \lambda+2}}{(2 \lambda+1)!!^2}\sum_{\eta_i} \left(2 j_i+1\right) \left(C_{j_f 1 / 2 \lambda 0}^{j_i 1 / 2}\right)^2 \left|M_{f i}^{\mathcal{T} \lambda}\right|^2, 
\end{equation}
where $p_i$, $B\left(\mathcal{T} \lambda ; I_i \rightarrow I_f\right)$ and $C_{j_f 1 / 2 \lambda 0}^{j_i 1 / 2}$ are the electron momentum, the reduced nuclear transition probability, and the Clebsch-Gordan coefficient, respectively. $\Gamma_{\mathrm{NEEC}}$ denotes the total resonance width of the state formed after electron capture. $k=\Delta E / \hbar c$ is the wave number associated with the nuclear transition energy $\Delta E$.
The resonance strength, integrating over all electron energies, is $S=\int \sigma_{\mathrm{NEEC}}\left(E_i\right) \mathrm{d} E_i $.

Since NEEC is a highly resonant process, its efficiency depends sensitively on the overlap between the nuclear transition energy and the available electronic capture channels. This feature makes NEEC particularly relevant for low-energy nuclear transitions and isomer-related scenarios, where suitable electronic capture channels may resonantly match the nuclear transition energy. The low-lying transition in $^{229}\mathrm{Th}$ is a prominent example, although the narrow resonance condition, small excitation probability, and strong competing backgrounds make the unambiguous observation of NEEC experimentally challenging \cite{PhysRevC.59.2462,PhysRevA.73.012715,PhysRevLett.112.082501,chiara2018isomer,PhysRevLett.122.212501,PhysRevLett.128.242502}. Recent theoretical and experimental studies have substantially improved our understanding of the NEEC mechanism and of the conditions required for its observation; however, an unambiguous and generally accepted experimental confirmation of NEEC remains elusive \cite{PhysRevLett.122.212501,PhysRevLett.128.242502,PhysRevLett.130.112501,Guo2021,Yang2025}.

Given these experimental difficulties, the history of NEEC research has been shaped not only by continuous theoretical development but also by long-standing controversy over its experimental identification. The most widely discussed example is the beam-based depletion of the $^{93m}$Mo isomer.   In the original beam-based experiment, Chiara \textit{et al.} reported isomer depletion during the slowing-down of highly charged $^{93m}$Mo ions in a solid target and interpreted the extracted excitation probability, $P_{\mathrm{exc}} = 0.010(3)$ per ion, as evidence for NEEC \cite{chiara2018isomer}. However, the observed isomer depletion was not a channel-specific signature of NEEC. Its interpretation therefore relied on the assumption that other depletion mechanisms and background contributions during the slowing-down process could be excluded, which made the claim sensitive to the underlying atomic-physics modeling and background analysis. A major challenge to the original claim was provided by the theoretical reanalysis of Wu \textit{et al.}, who calculated the beam-based NEEC probability using state-of-the-art atomic structure and stopping-power models \cite{PhysRevLett.122.212501}. Their results were approximately nine orders of magnitude smaller than the value inferred in Ref.~\cite{chiara2018isomer}, indicating that NEEC is far too weak in the experiment, under the reported conditions, to account for the observed depletion rate. This discrepancy immediately raised the possibility that other processes occurring during the ion deceleration in matter, rather than NEEC, were responsible for the measured signal.
The experimental interpretation was further questioned in a subsequent correspondence by Guo \textit{et al.}, who pointed out possible contamination and background-related issues in the original analysis, followed by a reply from Chiara \textit{et al.} \cite{Guo2021}. More importantly, an independent low-background isomer-beam experiment by Guo \textit{et al.} did not observe any statistically significant NEEC signature and established an upper limit of $2\times10^{-5}$ for the excitation probability \cite{PhysRevLett.128.242502}. This upper limit strongly disfavors the large probability reported in Ref.~\cite{chiara2018isomer} and is instead consistent with much smaller theoretical estimates.
Several later studies attempted to refine the atomic-physics description of the beam-based scenario. In particular, Rzadkiewicz \textit{et al.} proposed a resonant-transfer treatment that includes the momentum distribution of target electrons, showing that the Compton profile can noticeably enhance some capture channels \cite{PhysRevLett.127.042501}. They later extended the analysis to highly excited open-shell electronic configurations, which provide additional enhancement relative to simpler closed-shell or ground-state descriptions \cite{PhysRevC.108.L031302}. Nevertheless, these refinements enhance selected capture channels but still do not bridge the many-orders-of-magnitude gap between standard theoretical estimates and the 2018 experimental claim.
Very recently, the long-standing controversy over $^{93m}$Mo was further clarified by a new high-precision beam-based measurement \cite{kbf5-6fcl}. Ding \textit{et al.} reported depletion probabilities of $2.0(2)\times10^{-5}$ in Pb and $4.7(13)\times10^{-6}$ in C during the slowing-down process of purified high-energy $^{93\mathrm{m}}\mathrm{Mo}$ ions. Their results agree with calculations for inelastic nuclear scattering and indicate that the observed isomer depletion is dominated by inelastic nuclear scattering rather than NEEC \cite{kbf5-6fcl}.
Taken together, the $^{93m}$Mo case is best described as a historically important but now substantially reinterpreted episode in the experimental search for NEEC.

In view of the unresolved status of the $^{93m}$Mo case, recent work has shifted attention toward alternative candidate nuclei and improved beam-based schemes. In particular, Slabkowska \textit{et al.} proposed a new experiment based on the $6^{-}$ isomer $^{84m}$Rb at 463.618 keV, which could be depleted through NEEC via a nearby $5^{-}$ triggering level at 467.116 keV \cite{PhysRevC.109.054327}. Their multiconfigurational Dirac--Fock calculations showed that the dominant contribution comes from the $M1$ channel, while the $E2$ component is much weaker, and predicted a total NEEC depletion probability nearly three orders of magnitude larger than that estimated for the $^{93m}$Mo case. They also discussed a feasible production route based on the fusion--evaporation reaction $^{82}$Se+$^{7}$Li. Although this proposal remains to be tested experimentally, it is important because it illustrates how a more favorable level structure and transition pattern may improve the prospects for future beam-based NEEC searches.

In parallel with the search for more favorable nuclear candidates, theoretical studies have also explored whether NEEC rates can be enhanced by relaxing standard assumptions about the initial electronic configuration. In particular, Gargiulo \textit{et al.} showed that lifting the conventional ground-state assumption can substantially increase the NEEC resonance strength by opening additional capture channels \cite{PhysRevLett.128.212502}. For partially filled ions such as ${ }^{73}\mathrm{Ge}$, electronically excited configurations allow K-shell capture channels that are energetically inaccessible under the ground-state assumption, leading in some cases to enhancements of more than three orders of magnitude relative to estimates based on the ground-state assumption. These results highlight the importance of atomic-state preparation and electronic-structure effects in realistic NEEC calculations. At the same time, such enhancements should be interpreted primarily as an improvement in the theoretical accessibility of NEEC, rather than as a direct resolution of the outstanding experimental difficulties associated with background suppression and unambiguous signal identification.

\begin{figure}[htp]
\captionsetup{justification=justified, singlelinecheck=false}
	\centering
	\includegraphics[width=12cm]{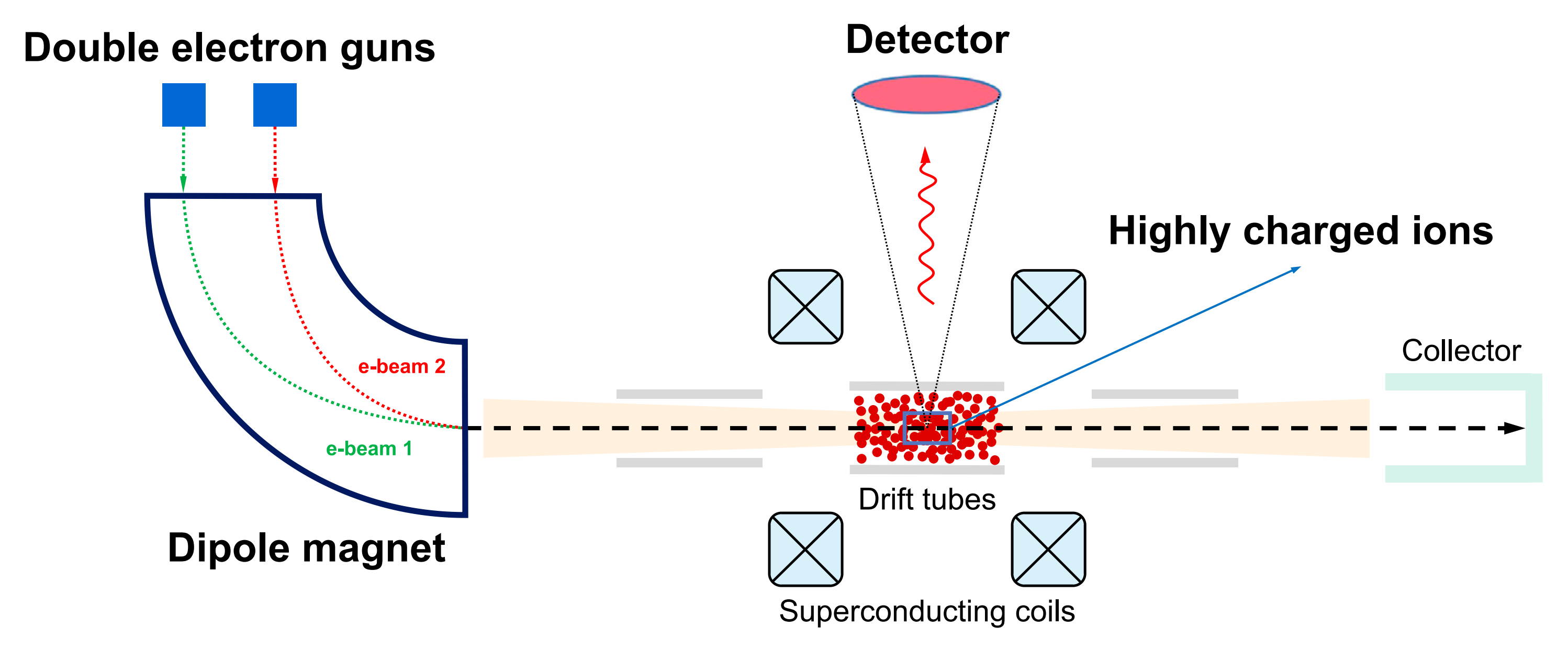}
	\caption{Schematic experimental setup: A dipole magnet converges two electron beams of different energies into a single beam. During interaction with trapped atoms, highly charged ions (HCIs) are formed, and NEEC may occur \cite{10.3389/fphy.2023.1203401}.}
	\label{fig6.3.1.2}
\end{figure}

Experimental efforts have also increasingly focused on more controllable platforms for future NEEC searches. In this context, both Electron Beam Ion Traps (EBITs) and storage rings have been proposed as promising environments for improving the resonance matching between electrons and highly charged ions while reducing competing backgrounds. Wang \textit{et al.} proposed an EBIT-based scheme in which monoenergetic electron beams interact with trapped highly charged ions under well-controlled charge-state and energy conditions \cite{10.3389/fphy.2023.1203401}. For several nuclei, including ${ }^{181}\mathrm{Ta}$, they evaluated the expected NEEC production rates and compared them with competing channels such as Coulomb excitation, showing that, for selected nuclei and charge states, the EBIT configuration may provide favorable conditions for NEEC studies because of its tunable electron energy and controlled ion environment. The schematic experimental setup is shown in figure \ref{fig6.3.1.2}, including the converging electron beams, ion confinement, and the corresponding detection geometry. Complementarily, Yang \textit{et al.} proposed a storage-ring scheme in which circulating highly charged ions interact with electron beams under well-defined energy conditions, with anti-coincidence techniques employed to suppress backgrounds from radiative recombination and Coulomb excitation \cite{10.3389/fphy.2024.1410076}. Calculations for candidate nuclei such as ${ }^{229}$Th suggest that storage-ring-based measurements may offer another promising route for NEEC searches. Taken together, these proposals indicate that highly controlled beam-based platforms may improve the prospects for NEEC detection; however, both approaches remain to be validated experimentally, and their practical usefulness will depend on achieving sufficient background suppression and unambiguous event identification under realistic operation conditions.

\begin{figure}[htp]
\captionsetup{justification=justified, singlelinecheck=false}
	\centering
	\includegraphics[width=12.5cm]{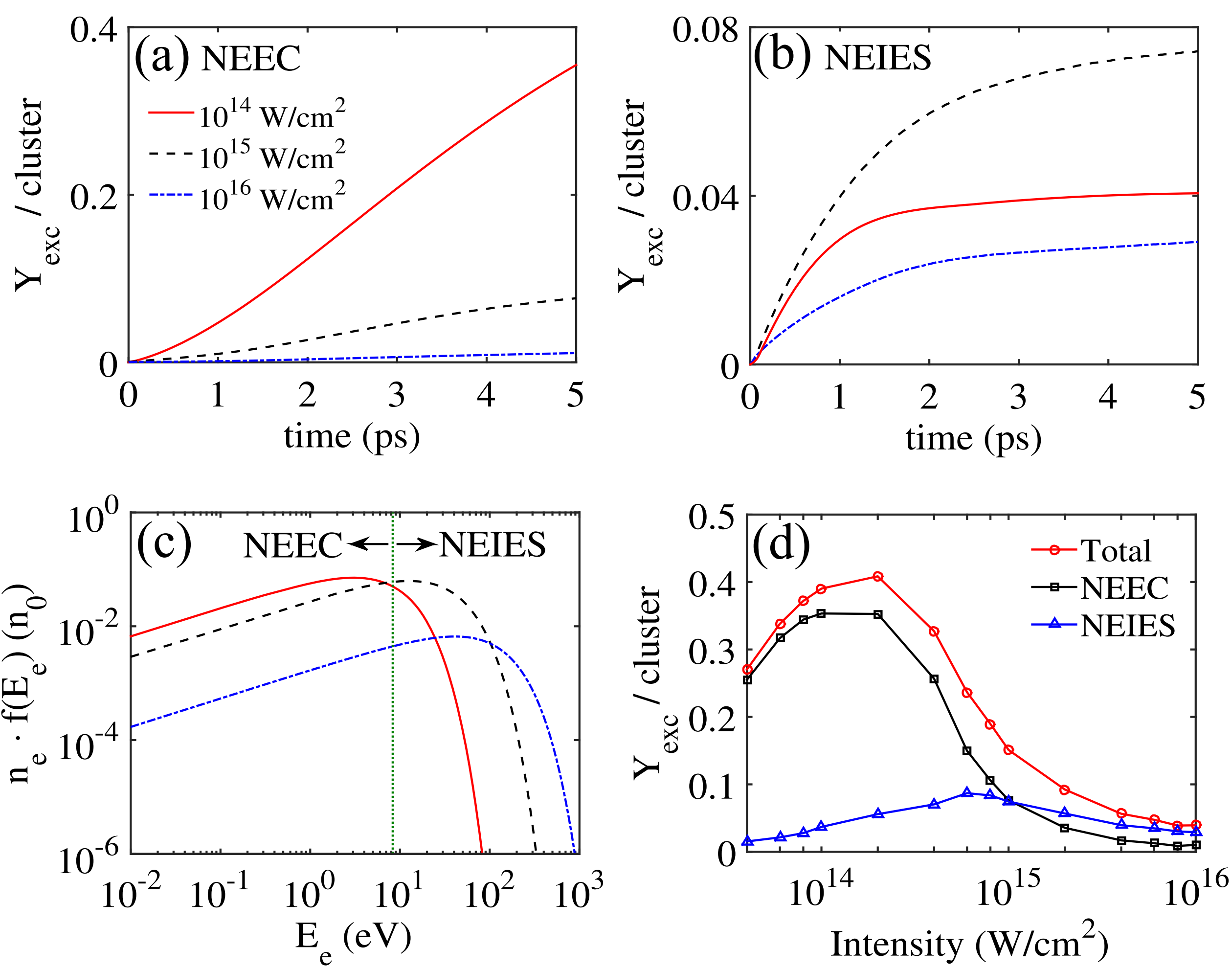}
	\caption{(a) Yield of ${ }^{229\mathrm{m}}$Th from NEEC under three different laser intensities $I=10^{14}\ \mathrm{W} / \mathrm{cm}^2$, $10^{15}\ \mathrm{W} / \mathrm{cm}^2$ and $10^{16}\ \mathrm{W} / \mathrm{cm}^2$. (b) Yield of ${ }^{229\mathrm{m}}$Th from NEIES under the same conditions. (c) Electron kinetic energy distributions at 1 ps during the cluster expansion. (d) Yield of ${ }^{229\mathrm{m}}$Th as a function of laser intensity at 5 ps \cite{PhysRevLett.130.112501}.}
	\label{fig6.3.1.3}
\end{figure}

Beyond beam-based platforms, laser-generated nanoplasmas have also been proposed as a possible environment for studying NEEC. Qi \textit{et al.} investigated, through numerical simulations, the excitation of ${ }^{229}$Th in laser-heated clusters irradiated by femtosecond pulses with intensities of $10^{14}$--$10^{16}\ \mathrm{W/cm^2}$ \cite{PhysRevLett.130.112501}. In this scenario, the laser first ionizes the cluster and creates a dense nanoplasma, in which confined electrons repeatedly interact with thorium ions and may drive nuclear excitation through free-bound processes. Their simulations indicated that NEEC gives the dominant contribution at moderate laser intensities around $10^{14}\ \mathrm{W/cm^2}$, whereas NEIES becomes more important as the laser intensity increases toward $10^{16}\ \mathrm{W/cm^2}$. As shown in figure \(\ref{fig6.3.1.3}\), the calculated yields and energy-dependent behavior of NEEC and NEIES differ markedly across the relevant intensity range.
This line of research was further developed in a subsequent laser-cluster proposal for ${ }^{235}$U \cite{PhysRevC.110.L051601}. In that work, Qi \textit{et al.} argued that, under suitable laser parameters, NEEC can overwhelmingly dominate the isomeric excitation yield in ${ }^{235}$U clusters, with contributions from competing mechanisms reduced to below the percent level. A particularly attractive feature of this proposal is that the long lifetime of the ${ }^{235m}$U isomer allows detection after the prompt atomic background has subsided. In addition, the distinct dependences of the predicted isomer yield on laser intensity, wavelength, and pulse duration were proposed as further signatures for distinguishing NEEC from competing excitation channels. These results suggest that laser-heated clusters may provide a useful theoretical testbed for comparing competing excitation mechanisms and for designing future verification strategies, although experimental validation under such plasma conditions remains an open challenge.

The potential of NEEC in laser-induced plasmas was also explored by Ma \textit{et al.} through particle-in-cell simulations of isomer production in Ag and Ge nanowire targets \cite{10.1063/5.0212163}. In their study, both Coulomb excitation and NEEC were evaluated under laser-driven plasma conditions. For the $^{73m}$Ge isomer, the calculated peak efficiency of NEEC was $1.65\times10^{11}\ \mathrm{particles\ s^{-1}\ J^{-1}}$, which is much smaller than the corresponding Coulomb-excitation contribution of $1.0\times10^{19}\ \mathrm{particles\ s^{-1}\ J^{-1}}$. Under these conditions, Coulomb excitation therefore remains the dominant production mechanism, making it challenging to isolate a NEEC contribution. The authors further suggested that lower-intensity laser conditions, which reduce the population of high-energy electrons while preserving a larger fraction of low-energy electrons, may enhance the relative importance of NEEC and thereby improve the prospects for probing its effects.

NEEC remains strongly limited by its narrow resonance condition and by competing processes such as Coulomb excitation and nuclear scattering, leading to excitation probabilities far below dominant channels in most scenarios. This explains the lack of a generally accepted experimental confirmation. Current efforts therefore focus on optimizing nuclear level structures, electronic configurations, and controlled environments to improve resonance matching and background suppression. In addition, NEEC is of interest for low-energy systems such as $^{229}$Th in the context of nuclear-clock-related studies, where it may provide a complementary pathway for probing electron–nucleus coupling in complex environments \cite{Piotrowski20}.

It is worth noting that while current studies of NEEC typically operate under conditions where the normalized vector potential remains below unity ($a_0 < 1$), the electron--laser interaction can still be accurately treated within the perturbative or semiclassical regime. In such scenarios, the standard factorized ansatz $\Psi(\mathbf{r}_e, \mathbf{r}_n, t) = \psi_e(\mathbf{r}_e, t) \otimes \psi_n(\mathbf{r}_n)$ remains valid, as the electron wavefunction evolves under external fields independently of nuclear feedback. This is justified in NEEC processes reported in low-temperature plasmas or XFEL-generated environments, where the field intensities typically correspond to $a_0 \sim 10^{-2}$--$10^{-1}$.

However, if future NEEC experiments can be extended to heavier nuclei, the regime $a_0 \gtrsim 1$ may become accessible. In this case, relativistic corrections and dynamic coupling between electrons and nuclei become non-negligible, requiring a fundamentally different theoretical treatment. Specifically, one must consider the breakdown of the product ansatz and adopt a fully correlated wavefunction framework.

To properly describe the entangled dynamics of electrons and nuclei in the relativistic strong-field regime, a generalized wavefunction expansion can be introduced as
\begin{equation}
    \Psi(\mathbf{r}_e, \mathbf{r}_n, t) = \sum_{i, \alpha} C_{i\alpha}(t)\, \phi_i^{\rm QED}(\mathbf{r}_e, t)\, \chi_\alpha(\mathbf{r}_n),
\end{equation}
where $\phi_i^{\rm QED}(\mathbf{r}_e, t)$ are electron wavefunctions dressed by the external field. In the simplest form, one may adopt Volkov states \cite{PhysRevA.22.1786}
\begin{equation}
    \phi_p^{\rm Volkov}(\mathbf{r}_e,t) = \exp\left[\frac{i}{\hbar }\left(\mathbf{p}\cdot \mathbf{r}_e - \int_0^t \frac{1}{2m} \left(\mathbf{p} + \frac{e}{c}\mathbf{A}(\tau)\right)^2 \mathrm{d}\tau\right)\right],
\end{equation}
which describe a free electron interacting with a plane-wave field via the vector potential $\mathbf{A}(t)$. The nuclear component \( \chi_\alpha(\mathbf{r}_n) \) is described by any of the nuclear models. In particular, the rigid triaxial even-even nuclei are described by superpositions of the following functions \cite{DAVYDOV1958237}

\begin{equation}
|IMK\rangle = \sqrt{\frac{2I + 1}{(1 + \delta_{K0})16\pi^2}} \left[ D^I_{MK}(\theta) + (-1)^I D^I_{M,-K}(\theta) \right],
\end{equation}
where \( K = 0, 2, 4\ldots \), \( D^I_{MK}(\theta) \) are the Wigner D-functions, depending on three Eulerian angles, \( I \) the nuclear spin, \( M \) and \( K \) its projections on the axes \( z \) of the lab. frame and \( \zeta \) of the body-fixed one, respectively.

Coupling coefficients $C_{i\alpha}(t)$ are governed by the time-dependent Schr\"odinger equation:
\begin{equation}
    i\hbar \frac{{\rm d}}{{\rm d}t} C_{i\alpha}(t) = \sum_{j,\beta} \left\langle \phi_i^{\rm QED} \chi_\alpha \middle| H(t) \middle| \phi_j^{\rm QED} \chi_\beta \right\rangle C_{j\beta}(t),
\end{equation}
where $H(t)$ includes both the field interaction and Coulomb electron--nucleus coupling. Such a correlated basis provides a flexible framework to capture electron–nucleus entanglement in strong fields and may serve as a viable approach to future NEEC modeling, particularly if experimental or theoretical developments extend into the relativistic regime with $a_0 \gtrsim 1$.

\subsubsection{Nuclear excitation by inelastic electron scattering (NEIES)}

{\ }

{\ }

\noindent NEIES is a process where an electron transfers a portion of its kinetic energy to a nucleus, exciting it to a higher energy state without emitting photons. Similar to Coulomb excitation, NEIES is mediated by the electromagnetic interaction and can be described in terms of virtual-photon exchange. Its distinguishing feature is that the projectile is a free electron, which undergoes inelastic scattering while transferring energy to the nucleus. Thus, NEIES may be regarded as electron-induced Coulomb excitation or inelastic electron-nucleus scattering, with the interaction treated using relativistic electron wavefunctions and nuclear electromagnetic transition operators. The unique role of free electrons in NEIES makes it a complementary mechanism to traditional Coulomb excitation. Electron-induced nuclear excitation through inelastic scattering can occur both in hot plasmas \cite{PhysRevLett.124.242501,CPC10.1088/1674-1137/ac9f0a,PhysRevC.106.044604} and in electron-accelerator experiments \cite{PhysRevC.106.064604,PhysRev.92.978,PhysRevC.79.014604,PhysRevC.85.044612,PhysRevC.105.014608,Ya.Dzyublik2013}. The former process may play a significant role in nuclear transformations in stars.

For NEIES, the electron transitions between free scattering states, while the nucleus transitions from its ground state $\left|I_i M_i\right\rangle$ to an excited state $\left|I_f M_f\right\rangle$. Using Fermi's Golden Rule, the differential cross section for NEIES is obtained by summing over the final states of the system

\begin{equation}
	\frac{\mathrm{d} \sigma}{\mathrm{d} \Omega}=\frac{2 \pi}{v_i} \rho\left(\mathcal{E}_f\right)\left|V_{f i}\right|^2,
\end{equation}
where $v_i=p_i c^2 / \mathcal{E}_i$ is the velocity of the incoming electron, $\rho\left(\mathcal{E}_f\right)=p_f \mathcal{E}_f /\left(8 \pi^3 c^2\right)$ is the density of final states, and $\mathcal{E}_i$ and $\mathcal{E}_f$ are the initial and final state energies of the electron, respectively, with $E_i= \mathcal{E}_i$ and $E_f=\mathcal{E}_f+E_{ is }$. Here, $E_{\mathrm{is}}$ is the nuclear excitation energy.

The transition probability for NEIES depends on the matrix element $V_{f i}$, which incorporates the nuclear and electronic components of the interaction. The total cross section for NEIES is obtained by integrating over the scattering angles

\begin{equation} 
	\sigma_{\mathrm{NEIES}}\left(\mathcal{E}_i\right) = \frac{8 \pi^2}{c^4} \frac{p_f}{p_i} \frac{\mathcal{E}_f+m_e c^2}{p_f^2} \frac{\mathcal{E}_i+m_e c^2}{p_i^2} \cdot \mathcal{S}, 
\end{equation}

\begin{equation} 
	\mathcal{S} = \sum_{\mathcal{T} \lambda} B\left(\mathcal{T} \lambda ; I_i \rightarrow I_f\right) \frac{k^{2 \lambda+2}}{(2 \lambda+1)!!^2} \sum_{\eta_i \eta_f} \left(2 j_i+1\right) \left(C_{j_f 1 / 2 \lambda 0}^{j_i 1 / 2}\right)^2 \left|M_{f i}^{\mathcal{T} \lambda}\right|^2. 
\end{equation}
The NEIES efficiency depends on the energy and flux of the incoming electrons, as well as the overlap between the nuclear transition energy and the energy transfer from the scattered electron. The excitation rate is given by

\begin{equation}
	\lambda_{\mathrm{NEIES}}= n_e \int \sigma_{\mathrm{NEIES}}\left(\mathcal{E}_i\right) v_i f\left(\mathcal{E}_i\right) \mathrm{d} \mathcal{E}_i,
\end{equation}
where $f\left(\mathcal{E}_i\right)$ is the electron energy distribution.

Unlike NEEC, NEIES does not require a narrow resonant capture condition; instead, excitation becomes possible once the incident electron can transfer sufficient energy to the nucleus. This characteristic is particularly advantageous in systems like \(^{229}\)Th, where the isomeric state at 8.36 eV poses significant experimental challenges for resonant photon excitation. A central challenge in NEIES lies in quantitatively assessing its efficiency and optimizing the conditions for practical applications. Over the past years, a series of theoretical and computational studies have addressed this challenge, providing valuable insights into the potential of NEIES as an effective pathway for nuclear excitation. A contribution comes from Tkalya, who investigated the excitation of \(^{229}\)Th to its isomeric state using low-energy electrons (9-12 eV) \cite{PhysRevLett.124.242501}. By solving the Dirac equation, Tkalya demonstrated that NEIES can achieve excitation cross sections on the order of \(10^{-26}-10^{-25} \mathrm{cm}^2\) under the optimal conditions. These cross sections significantly surpass those of photon-driven excitation mechanisms, indicating the efficiency of NEIES for this nuclear system. 

Building on these foundations, Liu and Wang extended the investigation of NEIES to ${}^{235}$U, which possesses an isomeric state at 76.7 eV \cite{PhysRevC.106.064604}. Their work highlighted the crucial importance of using realistic relativistic scattering states in NEIES calculations. Within the Dirac distorted-wave Born approximation, they demonstrated that, for low-energy electrons with kinetic energies around 100 eV, the excitation cross sections obtained using distorted Dirac continuum waves are enhanced by about six to seven orders of magnitude compared with the Dirac plane-wave Born approximation. This dramatic enhancement originates from the strong distortion of the electron wavefunction by the Coulomb and atomic potentials. They further showed that relativistic effects remain significant even at such low energies: replacing Dirac distorted waves with Schr\"odinger distorted waves reduces the cross sections by roughly one order of magnitude. These findings clearly indicate that an accurate treatment of relativistic wavefunction distortion is essential for modeling electron--nucleus interactions in heavy nuclei such as uranium. For completeness, it should be noted that finite-nuclear-mass (recoil) effects, which can be included via the Foldy–Wouthuysen transformation of the two-body Dirac Hamiltonian, introduce corrections of order $E_e/(M_N c^2)$ \cite{PhysRev.78.29,Bethe2013Quantum} (where $M_N$ denotes the nuclear mass). For typical NEIES conditions ($E_e = E_i \lesssim$ 100 eV) and heavy nuclei such as $^{229}$Th ($M_N c^2 \approx 2.1\times10^{11}$~eV), the fractional change in the excitation cross section is below $10^{-9}$. Such a minute correction is many orders of magnitude smaller than the relativistic and QED effects discussed above, and can thus be safely neglected in current theoretical treatments.

\begin{figure}[htp]
\captionsetup{justification=justified, singlelinecheck=false}
	\centering
	\includegraphics[width=10cm]{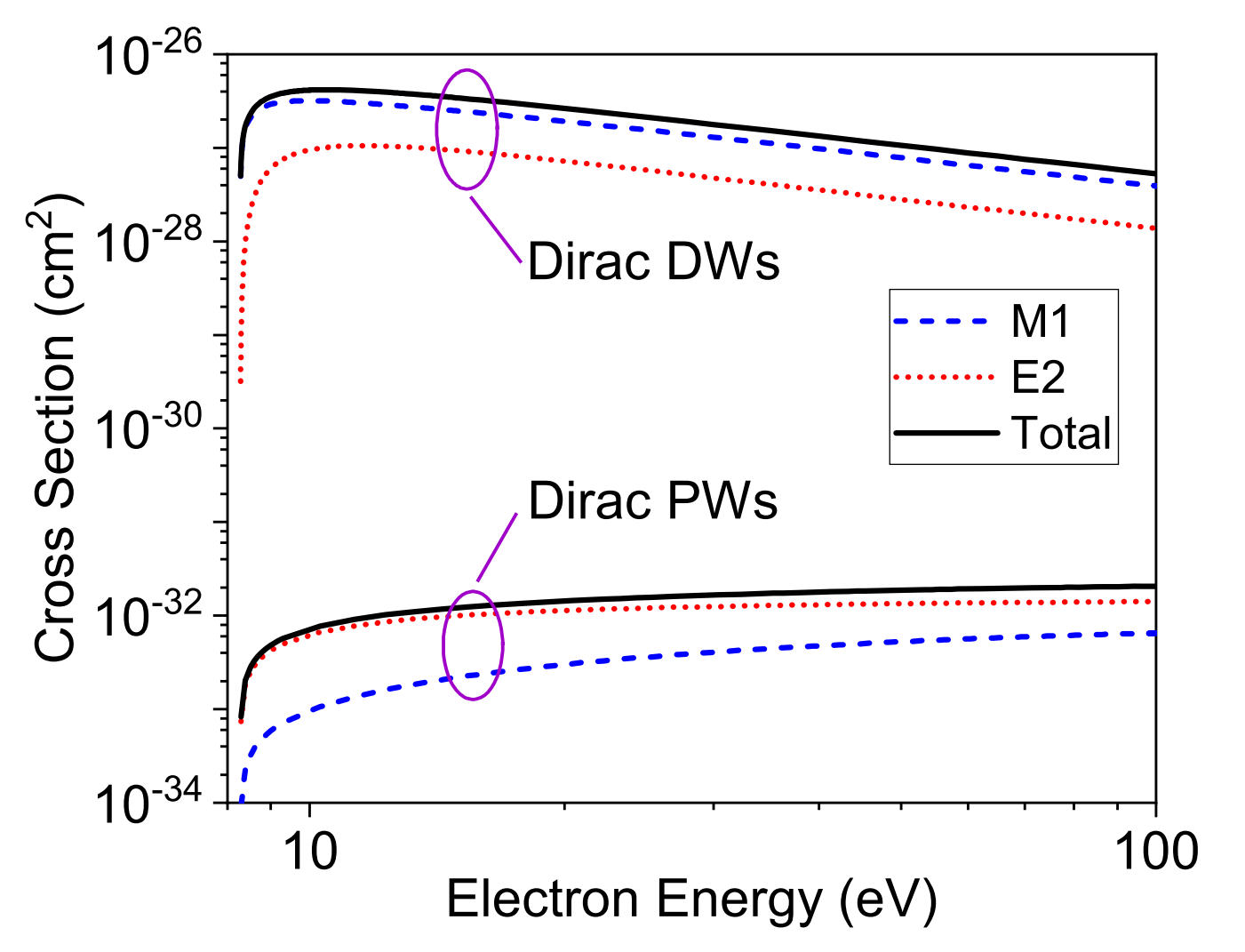}
	\caption{The isomeric excitation cross section of ${ }^{229}$Th calculated using Dirac DWs and Dirac PWs. For each case, the total cross section and the separate contributions from the M1 and E2 channels are presented \cite{PhysRevC.106.044604}.}
	\label{fig7.3.2}
\end{figure}

Zhang \textit{et al.} further expanded the theoretical framework by calculating NEIES cross sections for ${ }^{229}$Th across electron energies below 100 eV \cite{PhysRevC.106.044604}. As shown in figure \(\ref{fig7.3.2}\), their results confirmed the high efficiency of NEIES for isomeric excitation, with cross sections ranging from $10^{-27}$ to $10^{-26}\ \mathrm{cm}^2$. Their study also delved into how the factors such as ion-core potentials, relativistic effects, and nuclear transition probabilities influence the excitation efficiency. These comprehensive analyses established a robust theoretical foundation for guiding future experimental designs targeting nuclear excitation in thorium.

Recently, Xu \textit{et al.} offered a detailed examination of NEIES, specifically focusing on ${ }^{229}$Th \cite{PhysRevC.110.064621}. Their work extended the scope of NEIES studies by analyzing excitation cross sections for both the isomeric state (8.36 eV) and the second-excited state (29.19 keV) across a broad range of electron energies and ionic charge states. Using the DHFS method, they improved the accuracy in their calculations by incorporating detailed electron-nucleus interactions, exchange potentials, and relativistic effects. Notably, their findings demonstrated that the indirect excitation via the second-excited state can enhance the population of nuclei in the isomeric state, offering an efficient alternative to direct excitation of the isomeric state by low-energy electrons. They also revealed that up to $10\%$ of ${ }^{229}$Th$^{4+}$ ions could be accumulated in the isomeric state under optimal electron flux and energy conditions. Extending NEIES beyond $^{229}$Th, recent DHFS-plus-PIC calculations for laser-irradiated $^{235}$U clusters also showed that multichannel indirect excitation can enhance the isomer accumulation by up to 11 orders of magnitude over direct excitation and is strongly sensitive to laser polarization \cite{s6r5-5y5v}. More recently, low-energy NEIES has also been investigated as a possible mechanism for nuclear isomer depletion~\cite{slrl-9yfr}.

The above studies demonstrate NEIES as a versatile and effective mechanism for nuclear excitation, particularly in systems where traditional photon-mediated excitation is challenging. By combining precise theoretical modeling with experimental proposals, these works have provided opportunities for exploring nuclear excitation in complex environments, such as hot plasma and laser-driven systems \cite{PhysRevLett.128.052501,pnas2413221121,s6r5-5y5v}. For instance, the ability to control the population of nuclear isomers through NEIES may provide useful input for nuclear-clock-related studies, where the \(^{229}\)Th isomer transition could serve as a highly precise frequency standard.

\subsubsection{Nuclear excitation by electron transition (NEET)}

{\ }

{\ }

\noindent NEET is a resonant process where an electron transitions from a higher atomic energy level to a lower one, transferring the excess energy directly to the nucleus and inducing a nuclear transition. It was first introduced by Morita and collaborators in the 1970s as a mechanism that could link atomic and nuclear processes \cite{10.1143/PTP.49.1574}. Unlike photon-mediated processes, NEET operates via virtual photon exchange between the electron and nucleus, making it a photon-less yet highly specific mechanism for coupling the electronic and nuclear states. This process is significant in systems where the energy of electronic transitions closely matches the nuclear excitation energy, enabling resonant energy transfer within a confined atomic environment  \cite{Dzyublik2011,PhysRevC.88.054616,PhysRevC.74.031301,Fujioka_1985,TKALYA1992209,PhysRevLett.85.1831,SAKABE20051,PhysRevC.95.034310,KARPESHIN1999579,HARSTON2001447}.

The NEET process begins with the system in an initial state consisting of the nuclear ground state, the electrons in a higher atomic orbital, and the radiation field in its vacuum state. The system evolves under the interaction Hamiltonian $V$, transitioning to a final state where the nucleus is in an excited state, and the electrons move to a lower orbital. The system's time evolution is described in the interaction picture, with the interaction Hamiltonian $V_I(t)$ driving this transition. The total transition rate is given by \cite{10.3389/fphy.2023.1166566}

\begin{equation} 
	\omega_{\mathrm{NEET}} = 4 \pi \cdot \mathcal{S} \cdot \frac{\Gamma_{\mathrm{NEET}}}{\left(E_f-E_i\right)^2 + \Gamma_{\mathrm{NEET}}^2 / 4}, 
\end{equation}

\begin{equation} 
	\mathcal{S} = \sum_{\mathcal{T} \lambda} \left[B\left(\mathcal{T} \lambda ; I_i \rightarrow I_f\right) \frac{\kappa^{2 \lambda+2}}{(2 \lambda+1)!!^2} \left(C_{j_f 1 / 2 \lambda 0}^{j_i 1 / 2}\right)^2 \left|M_{f i}^{\mathcal{T} \lambda}\right|^2 \right].
\end{equation}
Here, $\Gamma_{\mathrm{NEET}}$ denotes the effective total resonance width of the coupled electronic--nuclear state, accounting for the relevant electronic and nuclear decay channels. This width can be expressed as $\Gamma_{\mathrm{NEET}}=\Gamma_i+\Gamma_f+\Gamma_n$, where $\Gamma_i$ and $\Gamma_f$ are the widths of the initial and final electronic states, respectively, and $\Gamma_n$ is the nuclear decay width, including contributions from radiative decay and internal conversion.

The resonance condition for NEET requires that the energy released by the electronic transition precisely matches the energy required for the nuclear excitation. Mathematically, this is expressed as $\mathcal{E}_i=\mathcal{E}_f+E_{\mathrm{is}}$.
For ${ }^{229} \mathrm{Th}$, with an isomeric energy of approximately 8.36 eV , this condition enables efficient energy transfer under specific electronic configurations. On resonance, the Lorentzian factor in the transition rate peaks sharply, leading to an enhanced transition probability proportional to $2 / \Gamma_{\mathrm {NEET}}$, underscoring the importance of narrow intermediate-state widths for maximizing the NEET efficiency.

Unlike NEEC or NEIES, NEET involves bound electronic states in both the initial and final atomic configurations. This bound--bound character makes the process highly sensitive to the atomic electronic structure and to the energy mismatch between the electronic and nuclear transitions. Although the localized nature of bound electrons can provide a favorable environment for electron--nucleus coupling, it does not necessarily lead to a larger transition probability than free-electron processes such as NEEC or NEIES. In practice, the strict resonance requirement significantly limits the range of applicable nuclei and electronic configurations. Moreover, competition between ordinary electronic relaxation, such as spontaneous photon emission, and nuclear excitation further reduces the effective NEET probability. Despite these challenges, NEET remains a unique mechanism for studying atomic--nuclear coupling, particularly for low-energy nuclear transitions, and may provide useful insights for nuclear-clock-related studies and precision measurements.

Early work by Sakabe \textit{et al.} identified isotopes and transitions feasible for NEET \cite{SAKABE20051}. By analyzing the energy levels of over 30 isotopes, they highlighted ${ }^{235}$U, ${ }^{237}$Np, and ${ }^{229}$Th as promising candidates. Their study emphasized the importance of the energy alignment, focusing on cases where the electronic and nuclear transition energies differ by less than 5 keV . This work guided experimental efforts toward isotopes with favorable nuclear transition multipolarities and electronic configurations.

To address the challenge of energy mismatch between electronic and nuclear transitions, Izosimov introduced the concept of utilizing autoionization states (AS) to enhance the NEET efficiency \cite{Izosimov01092008}. These states, characterized by their short lifetimes and high-energy densities, act as intermediates that bridge mismatches between the nuclear and electronic transition energies. Izosimov demonstrated that AS states could increase NEET probabilities in nuclei like ${ }^{237}$Np, facilitating energy transfer under less restrictive conditions. This approach also showed potential for triggering nuclear isomer decay, making NEET a viable mechanism for controlled energy release in nuclear systems. Further progress was made by Chodash \textit{et al.}, who investigated NEET in ${ }^{235}$U using high-power laser pulses to generate uranium plasma \cite{PhysRevC.93.034610}. Although no definitive NEET signal was detected, their work established an upper limit for NEET rates ($\lambda_\mathrm{NEET}<1.8\times 10^{-4}\ \mathrm{s}^{-1}$), refining theoretical predictions and experimental methods. This study showed the difficulty of isolating NEET signals in environments with competing atomic and nuclear relaxation processes.

\begin{figure*}[htb]
\captionsetup{justification=justified, singlelinecheck=false}
	\centering
	\includegraphics[width=10cm]{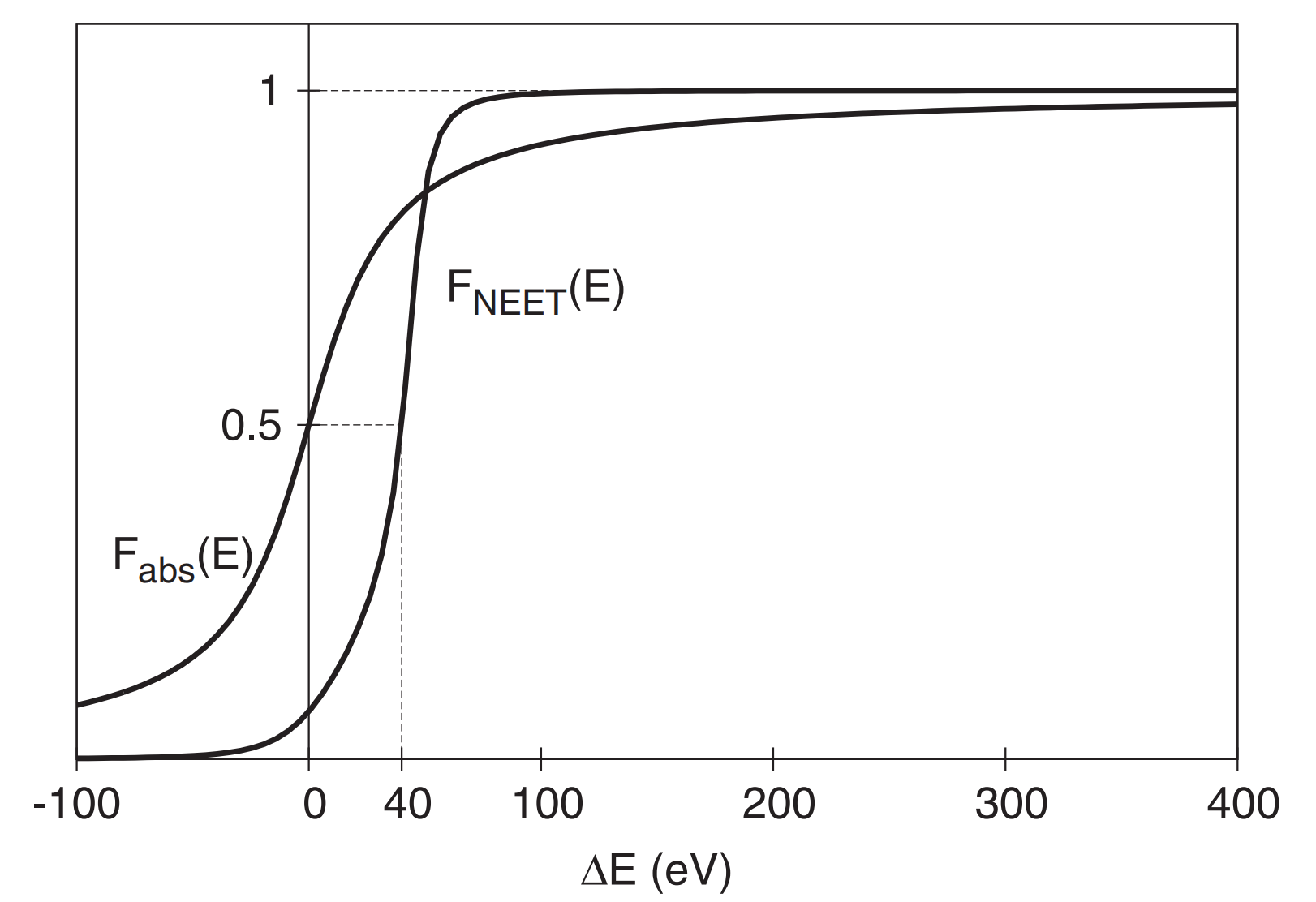}
	\caption{Calculated NEET edge function $F_{\mathrm{NEET}}(E)$ and $K$-absorption function $F_{\mathrm{abs}}(E)$ for $^{197}\mathrm{Au}$ as functions of the X-ray photon energy $E=B_K+\Delta E$, where $B_K$ is the $K$-shell binding energy \cite{PhysRevC.88.054616}.}
	\label{fig733}
\end{figure*}

The NEET process in the vicinity of the $K$-shell ionization threshold induced by incident X-rays has been theoretically studied in Refs.~\cite{PhysRevA.75.022509,PhysRevC.88.054616}. A detailed treatment was given by Dzyublik, who combined scattering theory with quantum electrodynamics. Figure \ref{fig733} illustrates the calculated NEET edge function $F_{\mathrm {NEET }}(E)$ and the $K$-absorption function $F_{\mathrm {abs }}(E)$ for ${}^{197}$Au as a function of X-ray photon energy. The edge function $F_{\mathrm {NEET}}(E)$ is defined as the NEET cross section $\sigma_{\mathrm {NEET}}(E)$, measured in units $\sigma_{\mathrm {NEET}}(\infty)$, corresponding to energies much exceeding the $K$-ionization threshold $B_K$. The comparison between these two functions provides valuable insights into the energy transfer processes involved in NEET and the conditions under which NEET is most likely to occur.

Despite the challenges posed by NEET's strict resonance conditions, recent advancements have provided new insights into its potential. Zhang and Wang conducted a comparative analysis of NEET, NEEC, and NEIES in ${ }^{229}$Th \cite{10.3389/fphy.2023.1166566}. Their study demonstrated that NEET cross sections are several orders of magnitude smaller than those of NEEC and NEIES under typical experimental conditions. This disparity arises from NEET's strict dependence on precise resonance conditions and the limited number of suitable bound-bound electronic transitions that can match the nuclear excitation energy.

For $^{229}\mathrm{Th}$, NEET provides a possible pathway for isomeric excitation by exploiting near-resonant electronic transitions, thereby linking atomic and nuclear dynamics. While achieving the precise electronic conditions required for NEET remains challenging, advancements in ion trapping, laser cooling, and atomic manipulation technologies are paving the way for more controlled experimental investigations. These developments, coupled with deeper theoretical understanding, may help clarify the practical potential of NEET for exploring nuclear structure and low-energy nuclear transitions.

\section{Experimental advances in laser-assisted nuclear excitation}\label{Seventh}

In the past years, with the improvement of experimental conditions and the efforts in the theoretical study of laser-nucleus interaction, several breakthroughs have been made experimentally in laser modulation of nuclei \cite{PhysRevLett.128.052501,pnas2413221121, Shvydko2023,Kraemer2023,PhysRevLett.132.182501,PhysRevLett.133.013201,Zhang2024,Zhang20242,PhysRevLett.134.113801,Yamaguchi2024,Xiao2026}. This is of great significance for the full utilization of nuclei, especially in the preparation of nucleus clocks \cite{Shvydko2023,Kraemer2023,PhysRevLett.132.182501,PhysRevLett.133.013201,Zhang2024,Zhang20242,PhysRevLett.134.113801,Yamaguchi2024,PhysRevResearch.7.L022036,PhysRevResearch.7.013052} and nuclear batteries \cite{Walker1999,CARROLL2007960}. Currently reported experiment results have focused on the laser-induced nuclear excitation of nuclei, since the laser peak intensities required for directly impacting the nuclear decay are far away from the current available in the laboratories. By contrast, the laser or electron excitation of nuclei has relaxed requirements, which are readily satisfied by the vast majority of laboratory desktop-class lasers. In this chapter, some experimental results of laser-induced nuclear excitation of high interest are discussed. The great success of these experiments not only reveals the structural information of nuclei but also paves the way towards future experiments and applications in several fields such as advanced nuclear energy systems, medical isotope production, and quantum information processing.

\subsection{Coulomb excitation of $^{83}$Kr in laser-cluster interactions }

\begin{figure*}[h]\centering
\captionsetup{justification=justified, singlelinecheck=false}
	\includegraphics[width=12cm]{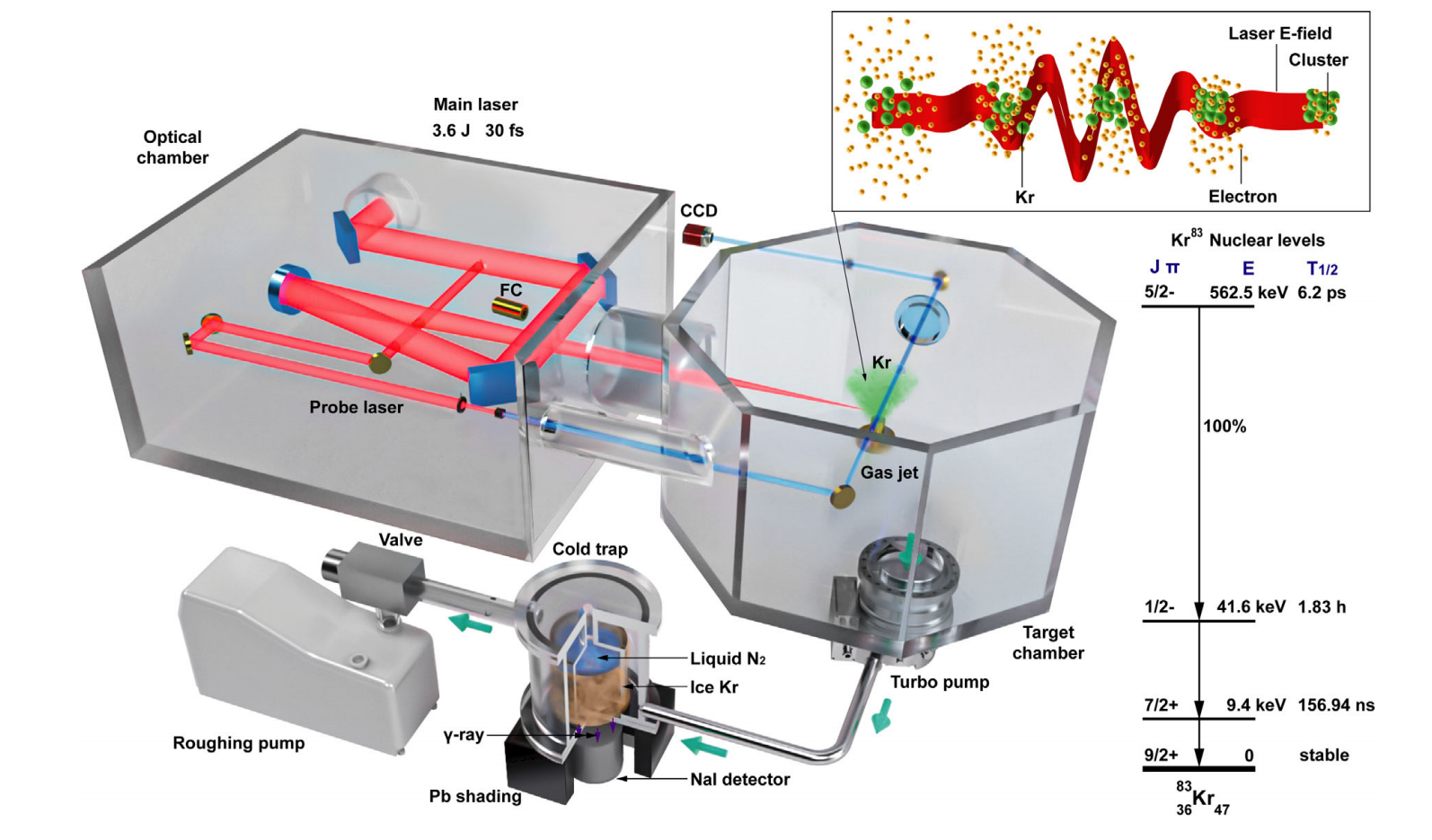}
	\caption{Schematic illustration of the experimental setup for the laser-induced nuclear excitation of $^{83}$Kr. The right inset shows the decay scheme of $^{83}$Kr, providing details of the energy levels and decay pathways involved in the process \cite{PhysRevLett.128.052501}.}
	\label{fig7.3.3.1.1}
\end{figure*}

One of the earliest experiments on laser-induced nuclear excitation was reported by Feng \textit{et al} \cite{PhysRevLett.128.052501}. They achieved femtosecond pumping of nuclear isomeric states for the first time through Coulomb excitation of ions and quivering electrons in laser-cluster interactions. In the experiment, they employed a Ti: sapphire laser system with an intensity of $10^{19}\ \mathrm{W/cm^{2}}$ as shown in figure \ref{fig7.3.3.1.1}. The laser pulse was focused on nanoparticles or clusters ejected from natural $^{83}$Kr gas. By monitoring crystal fluorescence, the second ($E$ = 41.6 keV) and third ($E$ = 562.5 keV) excited states of $^{83}$Kr were discovered. Irradiating Kr clusters with a 30 fs, 120 TW laser pulse, the peak excitation efficiency of the second and third excited states are $5.07\times10^{14}$ and $2.34\times10^{15}$ p/s, respectively, and the average efficiencies are 290 and 390 p/s, respectively. Simulations show that highly efficient Kr nuclear excitation arises from the nonlinear resonant interaction between high-density ions and high-energy electrons.

Another significant advancement in laser-based nuclear cross-section measurement experiments came from the same group \cite{pnas2413221121}. They successfully measured extremely small nuclear isomeric excitation cross-sections (10-100 pb, $1\,\mathrm{pb}=10^{-36}\,\mathrm{cm}^2$) of $^{83}$Kr through the laser-cluster interaction. The experiment was conducted at the Laboratory of Laser Plasmas, Shanghai Jiao Tong University, with a $3$ J, $800$ nm $p$-polarized laser pulse focused on a Kr gas jet containing $11.5\%$ of the $^{83}$Kr isotope. Kr atoms formed clusters due to the supersonic adiabatic expansion, and the laser pulse ionized the cluster atoms within femtoseconds, generating a nanometer-sized plasma ball with high-density (about $10$ times the solid-state atomic density) and high-temperature (100 keV-1 MeV). This facilitated nuclear excitations of $^{83}$Kr from the ground state to the $41.56$-keV isomeric state (half-life $1.83$ h) via inelastic electron scattering. The decay signals were detected by a NaI crystal X-ray detector, a curved image plate, and magnetic spectrometers at different angles. By stretching the laser pulse to different durations while maintaining its energy, different plasma temperatures were created, and the corresponding isomeric excitation cross-sections were extracted. The experimentally retrieved cross-sections were on the order of 10-100 pb, showing a similar trend with temperature as the theoretical curve, although with some discrepancies within a factor of two. This indicates that the laser-cluster interaction method has the potential to measure the small cross-sections of various nuclear reactions and provides a new approach for studying the nuclear transition mechanisms in stellar evolution \cite{Misch_2021,sym13101831}.

\subsection{X-ray resonance excitation $^{45}$Sc}

The ultra-narrow nuclear resonance transition of \(^{45}\)Sc, with a lifetime of up to 0.47 seconds between the ground state and the 12.4-keV isomeric state, positions this isotope as a promising candidate for high-precision timekeeping applications. Although the scientific potential of \(^{45}\)Sc has been recognized for some time, the practical realization of its applications necessitates resonant excitation of \(^{45}\)Sc, which in turn relies on high-brightness, accelerator-driven X-ray sources that have recently emerged. 

\begin{figure*}[htb]\centering
\captionsetup{justification=justified, singlelinecheck=false}
	\includegraphics[width=12cm]{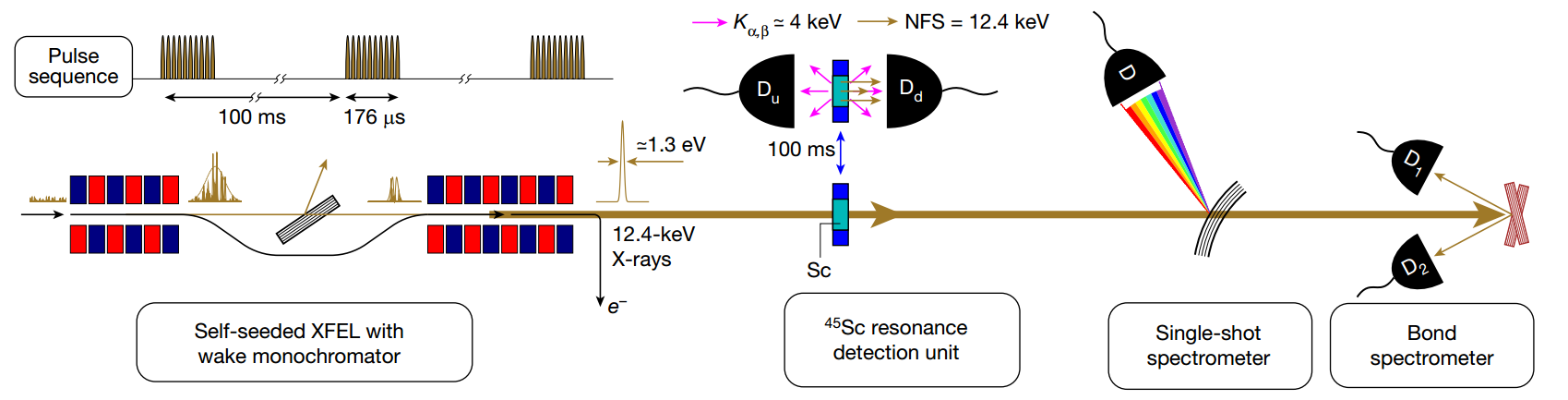}
	\caption{ Experimental setup designed for the resonant X-ray excitation of the nuclear resonant state of $^{45}$Sc using X-ray free electron laser (XFEL) pulses. It shows the key components and their arrangements, along with the process of exciting the $^{45}$Sc nucleus and detecting the resonance \cite{Shvydko2023}. }
	\label{fig7.3.4.1.1}
\end{figure*}

For a notable advancement, Shvyd\ensuremath{'}ko \textit{et al.} \cite{Shvydko2023} successfully achieved resonant X-ray excitation of the nuclear resonant state of \(^{45}\)Sc for the first time, utilizing an X-ray free electron laser (XFEL) \cite{Decking2020,Liu2023}. The experiments were conducted using the Materials Imaging and Dynamics instrument at the European XFEL. A schematic representation of the experimental setup is shown in figure \ref{fig7.3.4.1.1}. The experiment used 12.4-keV X-ray pulses generated by the EuXFEL to irradiate Sc-metal foil, exciting \(^{45}\)Sc nuclei from the ground state to the 12.4-keV isomeric state. The excited nuclei mainly decayed via internal conversion, emitting characteristic \(K_{\alpha}\) and \(K_{\beta}\) X-rays that were detected to confirm the resonance excitation. Through the careful measurement of nuclear decay products, the transition energy of \(^{45}\)Sc was determined to be 12,389.59 ± 0.12 (systematic error) eV ± 0.15 (statistical error), with an uncertainty two orders of magnitude smaller than previously reported \cite{BURROWS2008171}. Noteworthy the production efficiencies were reported, with the third excited state (\(E = 562.5\) keV) exhibiting a production rate of \(2.34 \times 10^{15}\) particles/s. In comparison, the production efficiency for the second excited state (\(E = 41.6\) keV) reached \(5.07 \times 10^{14}\) particles/s. These discoveries and results pave the way for applying \(^{45}\)Sc in extreme metrology, nuclear clock technology, and ultra-high-precision spectroscopy, among others.

\subsection{Excitation of $^{229}$Th and preparation of nuclear clocks}

The first experiment that has garnered significant attention in recent years for observing the radiative decay of $^{229m}$Th was conducted by Kraemer \textit{et al.} at the ISOLDE facility of the European Center for Nuclear Research (CERN) in 2023 \cite{Kraemer2023}. This experiment involved vacuum-ultraviolet spectroscopic measurements of $^{229m}$Th doped in large bandgap crystals such as CaF$_{2}$ and MgF$_{2}$. In the experiment, they measured a photon energy of 8.338 (24) eV, and determined the half-life of $^{229m}$Th embedded in MgF$_{2}$ to be 670 (102) seconds. The isomer ${ }^{229\mathrm{m}}$Th was populated through the $\beta$ decay of \(^{229}\)Ac, which was generated via the $\beta$ decay chain of \(^{229}\)Fr and \(^{229}\)Ra ions implanted into large-bandgap CaF$_{2}$ and MgF$_{2}$ crystals. The lattice environment suppressed non-radiative decay channels (e.g., electron conversion) by ensuring the electron binding energy exceeded the isomer’s decay energy, allowing radiative decay to dominate. Vacuum-ultraviolet spectroscopy detected 148.7 nm photons emitted during the isomer’s radiative decay, confirming the transition energy and half-life.This study bears significant implications for developing nuclear clocks, as $^{229}$Th features a notably low excitation energy, positioning it as a leading candidate for such applications.

Following this breakthrough, the potential of $^{229}$Th for nuclear clock applications has been thoroughly investigated \cite{PhysRevLett.132.182501,PhysRevLett.133.013201,Zhang2024,Ooi2026,Zhang20242,Elwell2025,PhysRevLett.134.113801}. Recent studies have focused on various aspects of $^{229}$Th, including its excitation in different host materials and the implications for nuclear clock applications. With the development of laser-nucleus interaction experiments in recent years, the experiments on laser-induced nucleus excitation become more and more practical, the relevant theoretical model has been steadily improved, and the excited target nucleus has progressively been clarified. The recent experiments have not only led to confidence in the preparation of nucleus clocks but also provided invaluable experience for using lasers to manipulate other behaviors of nuclei.

\subsubsection{Laser excitation in $^{229}$Th-doped CaF$_2$ crystals}

{\ }

{\ }

\noindent Tiedau \textit{et al.} addressed the long-standing challenge of directly exciting the 8.4 eV nuclear transition in \(^{229}\)Th by developing a tabletop VUV laser system and optimizing Th-doped CaF$_{2}$ crystals to suppress the background signals \cite{PhysRevLett.132.182501}. This approach overcame previous limitations in spectral precision and radioluminescence noise, enabling the first observation of resonant nuclear fluorescence. The experimental setup featured a spectroscopy vacuum chamber with a crystal mount, photomultiplier tubes, and a pneumatic shutter system. To avoid optical damage, the Th-doped CaF$_2$ crystal was cooled to a low temperature. 

The isomer ${ }^{229\mathrm{m}}$Th in Th-doped CaF$_{2}$ crystals was resonantly excited using a VUV laser system generating 10 ns pulses at 148.38 nm. The laser frequency was precisely stabilized using four-wave mixing in Xe gas and calibrated against an optical frequency standard. The Th ions, doped into the CaF$_{2}$ lattice as Th$^{4+}$ with charge compensation by two interstitial F$^{-}$ ions, were confined in the Lamb-Dicke regime, suppressing Doppler shifts and recoil effects. Resonant excitation was confirmed by observing delayed fluorescence decay with a lifetime of 630(15) s, corresponding to a vacuum half-life of 1740(50) s after correcting for the refractive index enhancement \(n^{3}\). This observation demonstrates a remarkably long lifetime for radiative decay on the nuclear transition, suggesting limited decoherence in the specific experimental setup. Nevertheless, it does not fully constrain all potential sources of dephasing that may arise from host-lattice interactions. The precise dephasing rate thus remains an open theoretical and experimental question. This includes contributions from magnetic-dipole interactions between the $^{229}\rm{Th}$ nuclear spin and the surrounding $^{19}\rm{F}$ nuclear spins in the Th:CaF$_2$ host, which have been investigated in nuclear magnetic resonance fields \cite{h2sn-wvzm}. Predictions such as those by Kazakov \textit{et al.} \cite{Kazakov_2012} suggest coherence times below 1 s due to fluorine spin interactions. However, more recent studies \cite{hiraki2025laser} indicate that the local fluorine environment may be more variable than previously assumed, potentially mitigating spin-induced dephasing at specific lattice sites.

This evolving understanding of the local fluorine configuration suggests that nuclear coherence times in $^{229}$Th:CaF$_2$ may be highly site-dependent. In particular, lattice environments where fluorine ions are spatially more distant from Th$^{4+}$ centers could exhibit reduced magnetic-dipole interactions, leading to prolonged coherence lifetimes. Further research could use advanced spectroscopy and isotopic methods to identify the best crystal structures for reducing spin dephasing in thorium-doped nuclear materials, improving their overall performance.

Direct X-ray absorption fine-structure measurements further identified the dominant defect motif in Th:CaF$_2$ as Th$^{4+}$ substituting Ca$^{2+}$ with two nearby interstitial F$^{-}$ ions, thereby placing the microscopic interpretation of crystal-field splitting and dephasing on firmer structural grounds \cite{Takatori_2025}. Terhune \textit{et al.} additionally showed that off-resonant VUV illumination can photoinduce quenching of the $^{229}$Th isomer by exciting in-gap electronic states that reopen internal-conversion decay channels, highlighting defect control as a practical clock-design parameter \cite{glzr-thyw}. Guan \textit{et al.} further observed X-ray-induced quenching of the $^{229}$Th clock isomer in CaF$_2$, with a temperature dependence that tracks crystal afterglow and supports a carrier-migration-mediated non-radiative deexcitation pathway \cite{75bb-thn7}.

In the experiment by Tiedau \textit{et al.}, the VUV photons were detected using a Cs-I photomultiplier tube behind dielectric mirrors to filter out radioluminescence background. The signal was absent in Th-232 control crystals, confirming isotope specificity. This breakthrough establishes the feasibility of laser-based nuclear spectroscopy for \(^{229}\)Th, opening avenues for precision tests of fundamental physics, development of optical nuclear clocks with highly stability, and investigations into nuclear coherence and lattice-nucleus interactions. 

\subsubsection{Investigation in $^{229}$Th-doped LiSrAlF$_6$ crystals}

{\ }

{\ }

\noindent Elwell \textit{et al.} addressed the challenge of achieving laser excitation of the \(^{229}\)Th nuclear transition in a chemically distinct host material by investigating \(^{229}\)Th-doped LiSrAlF$_6$ crystals \cite{PhysRevLett.133.013201}. This work aimed to explore how lattice environments affect nuclear coherence and decay dynamics, providing critical insights into the feasibility of solid-state nuclear clocks across different materials.  The experimental procedure involved irradiating the crystals with a VUV laser and monitoring the resulting fluorescence.

The \(^{229}\)Th-doped LiSrAlF$_6$ crystals were irradiated with VUV laser pulses generated via resonance-enhanced four-wave mixing in Xe gas \cite{PhysRevLett.133.013201}. The laser system delivered 30 pulses per second at 148.38 nm, with frequency stabilized to Xe atomic transitions. The \(^{229}\)Th ions, embedded as Th$^{4+}$ with charge compensation by interstitial F$^{-}$ ions, experienced an electric field gradient (EFG) from the crystal lattice. This EFG induced quadrupole splitting of nuclear energy levels, enabling resonant excitation of the \(I=5/2 \to I=3/2\) transition. A narrow spectral feature at 148.38219(4) nm (\(2020407.3(5)\) GHz) was observed, with a fluorescence decay lifetime of 568(13) s. Density functional theory (DFT) calculations revealed that Th \(5f\) states within the crystal bandgap (6.8 eV) mediated electron-hole pair formation, leading to non-radiative decay channels and redshifted fluorescence (182 nm). Only a fraction (1\%) of \(^{229}\)Th nuclei avoided such quenching, contributing to the observed narrow resonance. The observed fluorescence exhibited multiple timescales, and a narrow spectral feature, limited by the laser linewidth, was identified at $148.38219(4)_{stat}(20)_{sys}$ nm, which decayed with a lifetime of $568(13)_{stat}(20)_{sys}$ s. Through density functional theory calculations, they suggested potential mechanisms underlying the observed phenomena, including coupling the $^{229}$Th nucleus to the crystal's electronic and phononic degrees of freedom. 

This study advances the field by demonstrating nuclear laser spectroscopy in an alternative host material, emphasizing the role of crystal chemistry in nuclear coherence. The identification of quenching mechanisms through DFT also opens new avenues for engineering defect-free environments, potentially unlocking the full potential of \(^{229}\)Th as a next-generation frequency reference.

\subsubsection{Frequency ratio measurement between $^{229}$Th and $^{87}$Sr}

{\ }

{\ }

\noindent Recently, Zhang \textit{et al.} addressed the challenge of establishing a direct frequency link between nuclear and atomic clocks by developing a vacuum ultraviolet (VUV) frequency comb system referenced to the JILA \(^{87}\)Sr optical lattice clock \cite{Zhang2024}. This approach overcame previous limitations in spectral precision and enabled the first precise measurement of the \(^{229}\)Th nuclear transition frequency relative to an atomic clock. They utilized a VUV frequency comb to directly excite the narrow $^{229}$Th nuclear clock transition within a solid-state CaF$_2$ host material.

The ${ }^{229\mathrm{m}}$Th isomer in a CaF$_{2}$ crystal was excited using a VUV frequency comb generated via seventh harmonic generation in a femtosecond enhancement cavity \cite{Zhang2024}. The comb was phase-locked to the \(^{87}\)Sr clock laser at 698 nm, achieving a repetition rate of 75 MHz and a spectral linewidth of $\leq$10 GHz. The VUV comb (148.3 nm) resonantly excited the nuclear transition in Th$^{4+}$ ions embedded in the crystal lattice, which experiences an electric field gradient leading to quadrupole splitting. The excitation process involved populating the excited state from the ground state, with transitions observed at 2,020,407,283.847(4) MHz and other quadrupole-split components. Fluorescence photons emitted during the 641(4) s decay were detected using a PMT behind VUV-reflective filters, yielding a frequency ratio \(v_{Th}/v_{Sr} = 4.707072615078(5)\). This work demonstrates coherent control of nuclear transitions and enables precision tests of fundamental physics, representing a notable advancement in nuclear metrology. The measured frequency ratio and quadrupole structure provide valuable insights into nuclear properties and crystal-field interactions, offering a foundation for the development of portable solid-state nuclear clocks and investigations into temporal variations of fundamental constants with improved sensitivity.

Building on the first frequency-comb measurement, Higgins \textit{et al.} showed that the $m=\pm5/2\rightarrow\pm3/2$ line in $^{229}$Th:CaF$_2$ shifts by only 62(6) kHz between 150 and 293 K, thereby identifying an especially robust transition for subsequent reproducibility studies \cite{PhysRevLett.134.113801}. Ooi \textit{et al.} subsequently condensed the clock-performance aspect into a reproducibility benchmark, reporting a 220 Hz transition-frequency reproducibility in $^{229}$Th:CaF$_2$ and identifying dopant-induced inhomogeneous broadening and thermal co-thermometry as key constraints for solid-state nuclear clocks \cite{Ooi2026}.

\subsubsection{$^{229}$ThF$_4$ thin films for solid-state nuclear clocks}

{\ }

{\ }

\noindent To tackle the challenges posed by material scarcity and radioactivity in solid-state nuclear clock development, Zhang \textit{et al.} developed \(^{229}\)ThF\(_4\) thin films through a miniaturized radioactive physical vapour deposition (PVD) process  \cite{Zhang20242}. They observed the nuclear clock transition by irradiating the films with a VUV laser. Spectroscopic lines were detected at specific frequencies for films on different substrates, and the lifetimes of the excited states were measured. The results showed that solid-state nuclear clocks could use the $^{229}$ThF$_4$ thin films.

The \(^{229}\)ThF\(_4\) thin films (30–100 nm thickness) were irradiated with VUV laser pulses generated via resonance-enhanced four-wave mixing in Xe gas \cite{Zhang20242}. The laser system delivered 30 pulses per second at ~148.38 nm, with frequency stabilized to Xe atomic transitions. The \(^{229}\)Th ions, embedded as Th$^{4+}$ in the film, experienced an electric field gradient (EFG) from the crystalline environment, leading to quadrupole splitting of nuclear energy levels. Resonant excitation of the \(I=5/2 \to I=3/2\) transition was observed at 2020406.8(4) GHz (MgF$_{2}$ substrate) and 2020409.1(7) GHz (Al$_{2}$O$_{3}$ substrate). Fluorescence decay lifetimes of 150(15) s (Al$_{2}$O$_{3}$) and 153(9) s (MgF$_{2}$) were measured, attributed to Purcell-enhanced decay in the high refractive index (\(n \approx 1.95\)) host and potential quenching from lattice defects. This approach enabled laser spectroscopy of the \(^{229}\)Th nuclear transition with microgram quantities of material, reducing radioactivity by three orders of magnitude compared to bulk crystals. This breakthrough establishes \(^{229}\)ThF\(_4\) thin films as a scalable platform for solid-state nuclear clocks, enabling integration with photonic devices and reducing radiation hazards. The measured lifetimes, shortened by defect-induced nonradiative decay channels like electron-hole pair formation, quantify nuclear-solid interaction dynamics, such as phonon-mediated coherence reduction and laser-driven transition efficiency. These results provide empirical data for modeling nuclear spin-phonon interactions in crystalline matrices and guide experimental designs for coherent control of nuclear states in solid-state environments.

In a complementary low-bandgap platform, Elwell \textit{et al.} demonstrated laser-based conversion-electron M\"ossbauer spectroscopy of $^{229}$ThO$_2$, recording the nuclear resonance through internal-conversion electrons with a $12.3(3)\ \mu\mathrm{s}$ lifetime and thereby extending laser-driven $^{229}$Th spectroscopy beyond fluorescence-based wide-bandgap hosts \cite{Elwell2025}. These emerging research progress on the nuclear excitation of $^{229}$Th in various systems has important implications for developing nuclear clocks. Accurate measurements of transition frequencies and lifetimes in different host materials provide essential data for designing and improving these clocks. The ability to control and manipulate the nuclear excitation in $^{229}$Th is vital for achieving more accurate and stable timekeeping devices. On the trapped-ion side, Zitzer \textit{et al.} measured the hyperfine structure and isotope shift of sympathetically cooled $^{229}$Th$^{3+}$, refining nuclear moments and strengthening the spectroscopy basis for ion-based nuclear clocks \cite{PhysRevA.111.L050802}. Host-dependent isomer-shift calculations by Perera \textit{et al.} further showed that different solid hosts can span an approximately 80 MHz clock-frequency window, making host-specific electronic relaxation an essential consideration in cross-platform comparisons of $^{229}$Th clocks \cite{mhwc-4m14}. Further work could focus on optimizing the host materials, improving the detection techniques, and exploring the potential of $^{229}$Th in other applications such as quantum optics \cite{doi10.1126/science.1187770,PhysRevLett.109.262502}. The insights gained from these studies also open up possibilities for investigating new physical phenomena and testing fundamental theories of physics.

\section{Summary and outlook}\label{eighth}

Over the past decade, remarkable progress has been made in laser-nucleus interactions field, not only deepening our knowledge of the fundamental properties of nuclei but also opening up novel avenues for nuclear energy utilization and technological applications. While the interplay between high-intensity lasers and nuclei has emerged as a frontier for probing sub-attosecond decay dynamics and novel nuclear excitations, experimental progress remains constrained by the extreme field requirements necessary to observe measurable decay modifications. The most convincing experimental advances are concentrated in laser-assisted nuclear excitation, whereas direct laser-induced modifications of radioactive charged-particle decay remain weak under accessible optical and near-infrared laser conditions.

From a theoretical perspective, the rapid advancement of laser technology has spurred the development of diverse models and methods to evaluate the impact of high-intensity lasers on the nuclear decay half-lives and excitation rates. Based on the different interaction mechanisms and treatments of lasers as either fields or particles, these approaches have laid a solid theoretical foundation for subsequent investigations. The evolution of laser technology, from its early stages to the current state-of-the-art high-power high-intensity lasers employing chirped pulse amplification technology and optical parametric chirped pulse amplification technology, has enabled the exploration of laser-nucleus interactions under extreme conditions. This has, in turn, propelled research into traditional nuclear decay processes such as $\alpha$ decay and has also paved way for new laser-induced nuclear excitation mechanisms.

For $\alpha$ decay under extreme laser fields, theoretical studies grounded in the time-dependent Schrödinger equation and employing approximations like the quasistatic approximation and the Kramers-Henneberger transformation. It has been discovered that the laser fields can modify the energy spectrum of $\alpha$ particles and, in certain cases, lead to substantial alterations in the penetration probability and half-life of $\alpha$ decay. In future experimental studies, through systematic calculations of $\alpha$ decay characteristics in different nuclei, the factors such as nuclear deformation, shell effects, and odd-even staggering effects should be carefully considered which play a vital role in the behavior of $\alpha$ decay in laser fields. Additionally, the $\alpha$-decay rate, influenced by laser-driven electron screening and by laser-induced nuclear excitation of isomeric states, offers scalable solutions for modulating the $\alpha$-decay rate through laser fields. These findings not only enhance our understanding of the microscopic mechanisms underlying the $\alpha$ decay but also offer valuable insights for further exploration in experiments. 

Regarding proton radioactivity, although relatively fewer studies have been conducted compared to the $\alpha$ decay, existing research indicates that the rates of change in the half-life and penetration probability of proton radioactivity exhibit symmetry in extreme laser field environments. This provides important clues for understanding the behavior of proton radioactivity in the laser fields. Two-proton radioactivity, a relatively rare yet significant nuclear decay mode, is still in the early stages of research in laser fields. Current research primarily uses linearly polarized Gaussian lasers to assist the two-proton radioactivity. It has been found that high-intensity laser fields can affect the tunneling probability and the formation of two-proton pairs in two-proton radioactivity. Although the impact of lasers on the preformation probability is relatively small, the preformation probability contains crucial information about the nuclear structure. These studies are therefore best viewed as theoretical maps of possible strong-field trends, since rare-nucleus production, low statistics, and intense laser-driven backgrounds currently limit direct measurements.

The study of nuclear excitation in laser fields has a distinct theoretical basis from that of nuclear decay, and several mechanisms have been identified. Direct laser excitation is a fundamental process involving photons directly interacting with nuclei to induce transitions between nuclear energy levels. However, the efficiency of direct laser excitation is typically low due to the small interaction cross-section and the challenge of achieving the required photon energies. Photoexcitation via blackbody radiation and bremsstrahlung offers an alternative excitation pathway. Although less efficient than direct excitation mechanisms, it remains relevant in specific contexts such as laser-heated clusters. On the contrary, electron-coupled nuclear excitation mechanisms, including NEEC, NEIES, and NEET, play a crucial role in the nuclear excitation. First, NEEC occurs when a free electron is captured into a bound atomic state, transferring its energy to the nucleus. NEIES involves the transfer of energy from a electron to the nucleus through direct interaction, while NEET consists of the transfer of energy from an electron transitioning between atomic energy levels to the nucleus. These mechanisms operate under different physical conditions and provide a rich theoretical framework for studying the nuclear excitation. For nuclear optical clocks, understanding these nuclear excitation mechanisms is crucial as it enables more precise control over nuclear transitions, which is essential for improving the accuracy and stability of nuclear optical clocks. In contrast to direct charged-particle decay control, this direction already has a firmer experimental foundation, particularly through the rapidly developing $^{229}$Th nuclear-clock programme.

Significant breakthroughs have been achieved experimentally in recent years in the study of laser-nucleus interactions. The number of laser-induced nuclear excitation experiments has steadily increased, and the associated difficulties have been gradually overcome. Theoretical models and numerical approaches have been refined, and the target nuclei for excitation have become more clearly defined. The experimental results have not only instilled ones' confidence in the development of nuclear clocks but have also provided invaluable experience for future investigations into other aspects of laser-nucleus interactions. Specifically, the precise measurement of nuclear excitation frequencies and lifetimes in experiments related to nuclear optical clocks has allowed for more accurate calibration and optimization of these nuclear optical clocks, bringing us closer to realizing their full potential in practical applications. Furthermore, these schemes of nuclear excitation by external means have also led to the realization that controlling nuclear decay can also be accomplished indirectly, for example by laser-accelerated electrons to manipulate the nuclear decay.

As we look ahead, although there are still many problems and controversies in this field today, such as the disconnect between theory and experiment including the gap between the current intensity required for nuclear decay and the experimental capabilities, as well as the lack of experimental validation for indirect nuclear excitation mechanisms—the landscape of future research in this field is abundant with potential. On the theoretical side, there is a pressing need to develop a more comprehensive and unified theoretical framework that can accurately represent the laser-nucleus interactions. Advancing more precise computational models and simulation techniques will be essential for predicting and clarifying the experimental outcomes. On the experimental front, the focus will likely shift toward more exact measurements of nuclear excitation and decay processes under various laser conditions. This will require the use of advanced laser technologies featuring higher intensities, shorter pulse durations, and enhanced control over laser parameters. In the realm of applications, the potential of laser-nucleus interactions in fields such as nuclear energy, nuclear medicine, and fundamental physics research is virtually boundless. For instance, the ability to modulate nuclear decay rates and excitation processes could enable revolutionary nuclear reactors with enhanced safety and efficiency, as well as novel diagnostic and therapeutic modalities in nuclear medicine. Moreover, the study of laser-nucleus interactions might offer novel avenues for testing fundamental physical theories and exploring physics beyond the Standard Model, and also provide effective methods of the nuclear isomer population and their depletion governing, which will lead to creation of compact superpower energy sources.

The field of laser-nucleus interactions has witnessed transformative advancements over the past decade, marked by breakthroughs in theoretical models, experimental techniques, and technological applications. Despite the challenges posed by the current energy level mismatch between the laser intensity and nuclear decay, continued research into this interdisciplinary field continues to yield transformative insights into nuclear manipulation at extreme electromagnetic fields. In the future, as lasers intensify beyond the current capabilities, the interplay between light and matter promises to unlock new regimes of nuclear control, transforming our ability to manipulate energy, time, and matter at the most fundamental scales.

\section*{Acknowledgments}
This work was supported by the National Natural Science Foundation of China (Grant
Nos. 12375244, 12035011, 12135009, 13535009), the National Key R\&D Program of China (Contract No. 2023YFA1606503), and the Natural Science Foundation of Hunan Province of China (Grant No. 2025JJ30002).

\section*{References}
\addcontentsline{toc}{section}{References}

\bibliography{IOPLaTeXGuidelines}

\end{document}